\documentclass[a4paper,12pt]{article}
\usepackage{amsmath, amssymb, amsthm}
\usepackage{fullpage}
\usepackage{hyperref}
\usepackage{graphicx}
\usepackage{parskip}
\usepackage{color}
\usepackage{appendix}

\numberwithin{equation}{section}

\begin{document}
\title{Broken planar Skyrmions -- statics and dynamics}
\author{Paul Jennings$^*$ and Thomas Winyard$^\dagger$\\[10pt]
{\em \normalsize Department of Mathematical Sciences, }\\{\em \normalsize Durham University, Durham, DH1 3LE, U.K.}\\[10pt]
{\normalsize $^*$ paul.jennings@durham.ac.uk \quad $^\dagger$ t.s.winyard@durham.ac.uk}}

\maketitle
\vspace{15pt}
\begin{abstract}
The broken planar Skyrme model is a theory that breaks global $O\left(3\right)$ symmetry to the dihedral group $D_N$. It has been shown that the single soliton solution is formed of $N$ constituent parts, named partons, that are topologically confined. The multi-soliton solutions have already been computed for $N = 3$ and were shown to be related to polyiamonds. We extend this for larger $N$ and demonstrate that this polyform structure continues (planar figures formed by regular $N$-gons joined along their edges, of which polyiamonds are the $N=3$ subset). Furthermore, we numerically simulate the dynamics of this model for the first time. It will be demonstrated that the time dependent behaviour of these solutions can be broken down into the interactions of its constituent partons. The results are then compared with those of the standard planar Skyrme model. 
\end{abstract}

\pagebreak

\section{Introduction}
The Skyrme model \cite{Skyrme:1961vq} is a (3+1)-dimensional theory that admits soliton solutions, called Skyrmions, which represent baryons. This has been well studied \cite{bible} with solutions calculated for a large range of charges \cite{Battye:2001qn}. The Skyrme model has been derived from Quantum Chromodynamics (QCD) \cite{Witten,Witten2}, and then more recently from holographic QCD, as a low-energy effective theory in the large colour limit \cite{Sakai:2004cn}.

The planar (or baby) Skyrme model \cite{Piette:1994ug} is the (2+1)-dimensional analogue of the Skyrme model. Planar Skyrmions are also manifest in their own right in condensed matter physics, such as in ferromagnetic quantum Hall systems \cite{Sondhi:1993zz}, and more recently observed in chiral ferromagnets \cite{Yu}. 

Although seen as a model of QCD, these models do not exhibit any classical colour dependent behaviour. The number of colours, $N$, appears only when the models are quantised (as a coefficient of the Wess-Zumino term). In this paper we are interested in a model that has a classical colour dependence, which has been proposed by J\"{a}ykk\"{a} \emph{et al.} \cite{Jaykka:2011ic}. For the resulting solitons of the three-colour theory, it was found that the energy density was arranged in lumps, called partons. Links were also identified between the structure of the higher charge solitons and polyiamonds. This paper left interesting open questions as to how this would generalise for systems with a greater number of colours.

In this paper we consider this planar Skyrme model with discrete symmetry, and examine static soliton field configurations for a range of $N$-colour systems. By examining the structure of the static solutions, we consider how the connection to polyiamonds generalises for higher-colour systems to polyforms. Finally we go on to consider the dynamics of these solitons and ascertain whether their structure impacts upon the scattering behaviour.

\section{The Model}
The planar Skyrme model has the form of a non-linear modified sigma model, described by the Lagrangian density
\begin{equation}
\mathcal{L}=\frac{1}{2}\partial_\mu \boldsymbol{\phi}\cdot\partial^\mu \boldsymbol{\phi}-\frac{\kappa^2}{4}\left(\partial_\mu\boldsymbol{\phi}\times\partial_\nu\boldsymbol{\phi}\right)\cdot \left(\partial^\mu \boldsymbol{\phi} \times\partial^\nu\boldsymbol{\phi} \right)-m^2 V[\boldsymbol{\phi}],
\end{equation}
where greek indices run over time and spatial dimensions ($\mu = 0,1,2$) and $\boldsymbol{\phi}(\mathbf{x},t)$ is a unit vector field, $\boldsymbol{\phi}=(\phi_1,\phi_2,\phi_3)$. Applying Derrick's theorem \cite{Derrick:1964} to the static energy, we can see that for non-zero potential energy, static soliton solutions are possible. This is due to the addition of the second term in the Lagrangian, stabilising the sigma model. This term is referred to as the Skyrme term, in accordance with its relation to the 3-dimensional Skyrme model. Hence soliton solutions of the theory are referred to as planar Skyrmions. We can also obtain from Derrick's theorem that the scale of the soliton is proportional to the constant $\sqrt{\kappa/m}$. The energy of this planar Skyrme model has the form,

\begin{multline}\label{eq : energy}
E=\int \left(\frac{1}{2}\dot{\boldsymbol{\phi}}\cdot\dot{\boldsymbol{\phi}}+\frac{\kappa^2}{2}( \dot{\boldsymbol{\phi}}\times \partial_i\boldsymbol{\phi})\cdot( \dot{\boldsymbol{\phi}}\times \partial_i\boldsymbol{\phi})\right)\,\mathrm{d}^2\mathbf{x}\\
+\int \left( \frac{1}{2}\partial_i\boldsymbol{\phi}\cdot\partial_i\boldsymbol{\phi}+\frac{\kappa^2}{4}\left(\partial_i\boldsymbol{\phi}\times \partial_j\boldsymbol{\phi}\right)\cdot\left(\partial_i\boldsymbol{\phi}\times \partial_j\boldsymbol{\phi}\right)+m^2V[\boldsymbol{\phi}]\right)\,\mathrm{d}^2\mathbf{x},
\end{multline}
where latin indices run over spatial dimensions ($i = 1,2$). For finite energy we require $\boldsymbol{\phi}$ to be a vacuum at spatial infinity, hence it can be viewed as a map from the compactified physical space, $\mathbb{R}^2 \cup \{\infty\}=\textit{S}\,^2$, to the target space $\textit{S}\,^2$. Since the second homotopy group $\pi_2(\textit{S}\,^2)=\mathbb{Z}$ there is a winding number associated to the map, which is characterised by the topological charge
\begin{equation}
B=-\frac{1}{4\pi}\int \boldsymbol{\phi}\cdot(\partial_1 \boldsymbol{\phi}\times \partial_2 \boldsymbol{\phi})\,\mathrm{d}^2\mathbf{x}.
\end{equation}

The field equation that follows from the Lagrangian is,
\begin{multline}
-m^2  \frac{\delta V}{\delta \boldsymbol{\phi}}-\partial_\mu\partial^\mu\boldsymbol{\phi}
+\kappa^2\left[ \partial_\mu\partial^\mu \boldsymbol{\phi}(\partial_\nu\boldsymbol{\phi}\cdot\partial^\nu\boldsymbol{\phi}) + \partial_\mu \boldsymbol{\phi}( \partial_\nu\boldsymbol{\phi}\cdot \partial^\mu\partial^\nu\boldsymbol{\phi})\right. \\
\left. -\partial_\mu\partial_\nu\boldsymbol{\phi} (\partial^\mu \boldsymbol{\phi}\cdot \partial^\nu \boldsymbol{\phi})-\partial_\mu \boldsymbol{\phi} (\partial^\mu \boldsymbol{\phi} \cdot \partial_\nu\partial^\nu \boldsymbol{\phi})\right] +\lambda\boldsymbol{\phi}=0,
\end{multline}
where $\lambda$ is a suitable Lagrange multiplier to enforce the condition that $\boldsymbol{\phi}\cdot\boldsymbol{\phi}=1$. The field equation is highly non-linear, and to study the behaviour of the system we must resort to numerical techniques.

A variety of different potentials have been proposed \cite{0951-7715-3-2-007,0951-7715-12-6-303,0951-7715-17-3-014,Jaykka:2010bq,nitta}, the standard potential term \cite{Piette:1994ug} is the analogue of the pion mass term in the Skyrme model, $V[\boldsymbol{\phi}]= 1-\phi_3$ . In this paper we consider the potential 
\begin{equation}
V[\boldsymbol{\phi}]=\left|1-(\phi_1+i\phi_2)^N\right|^2(1-\phi_3),
\end{equation}
for some integer $N\ge2$, which was considered for the $N=3$ case by J\"{a}kk\"{a} \emph{et al.} \cite{Jaykka:2011ic}. Note that up to quadratic order in $\phi_1$ and $\phi_2$ this reduces to the pion mass potential. Hence physically the fields $\phi_1$ and $\phi_2$ are massive fields with mass given by the constant $m$, as with the standard potential. This choice of potential breaks the $\textit{O}(3)$ symmetry of the system to the dihedral group $\textit{D}_N$, generated by rotation $(\phi_1+i\phi_2)\to(\phi_1+i\phi_2)e^{2\pi i/N}$ and reflection $\phi_2\to-\phi_2$. This choice of potential has vacuua at $\boldsymbol{\phi}=(0,0,1)$ and at the $N$th roots of unity on the $\phi_3=0$ equatorial circle. The vacuum at spatial infinity is chosen to be
\begin{equation}
\boldsymbol{\phi}_\infty=\lim_{|{\mathbf{x}|}\to\infty}\boldsymbol{\phi}(\mathbf{x},t)=(0,0,1).
\end{equation}
This choice does not further restrict the symmetry of the model since the generators of the dihedral group are independent of $\phi_3$. We will follow the notation of paper \cite{Jaykka:2011ic} and hence refer, somewhat suggestively, to the system for a particular choice of $N$ as the $N$-colour system.

\section{Static Planar Skyrmions}
In this section we specialise to the static case and examine the structure of (local) minimal energy solutions. The only work to date is for the three-colour system \cite{Jaykka:2011ic}. We shall recreate and then extend these findings, as well as examining the static solutions for higher-colour systems.

To find these soliton solutions we use an energy-minimising gradient flow algorithm, choosing to set $\kappa=m=1$ on a square grid with $(501)^2$ grid points and lattice spacing $\Delta x=0.04$. Spatial derivatives are approximated using fourth-order finite difference methods. We also fixed the boundary of our grid to be the vacuum at spatial infinity $\boldsymbol{\phi}_\infty = (0,0,1)$. For all our simulations the topological charge, when computed numerically, gives an integer value to five significant figures, indicating the accuracy of the results.

The gradient-flow algorithm requires an initial approximation to the static soliton. Consider the field configuration
\begin{equation}
\boldsymbol{\phi}=(\sin(f)\cos(B\theta),\sin(f)\sin(B\theta),\cos(f)),
\label{initialconditions}
\end{equation} 
for polar coordinates $r$ and $\theta$, and where $f$ is a monotonically decreasing function of $r$. The boundary conditions on $f$ are $f(0)=\pi$ and $f(R) = 0$, where the circle of radius $r = R$ lies inside the grid. Outside this radius the rest of the grid is set to the vacuum $\boldsymbol{\phi}_\infty$. We can see that this describes a field on the grid with topological charge $B$, and so for a suitable choice of $f$ this gives us our initial approximation.

We note that this initial approximation has the maximal symmetry $D_{NB}$, in the sense that the spatial rotation $\theta\to\theta+2\pi/NB$ can be compensated for by global rotation symmetry, while the reflection $\theta\to-\theta$ can be balanced by a global reflection. 

To find solutions with lower symmetry we also considered similar initial conditions but with a symmetry breaking perturbation. Once a pattern was discernible for these lower symmetry forms, we also used a product ansatz for our initial conditions. In other words we placed single solitons about our grid and then performed our gradient flow procedure.   

\subsection{Single Soliton Solutions}

\begin{figure}
\begin{center}
\begin{tabular}{c c c}
\includegraphics[scale=0.4,natwidth=1000,natheight=1000]{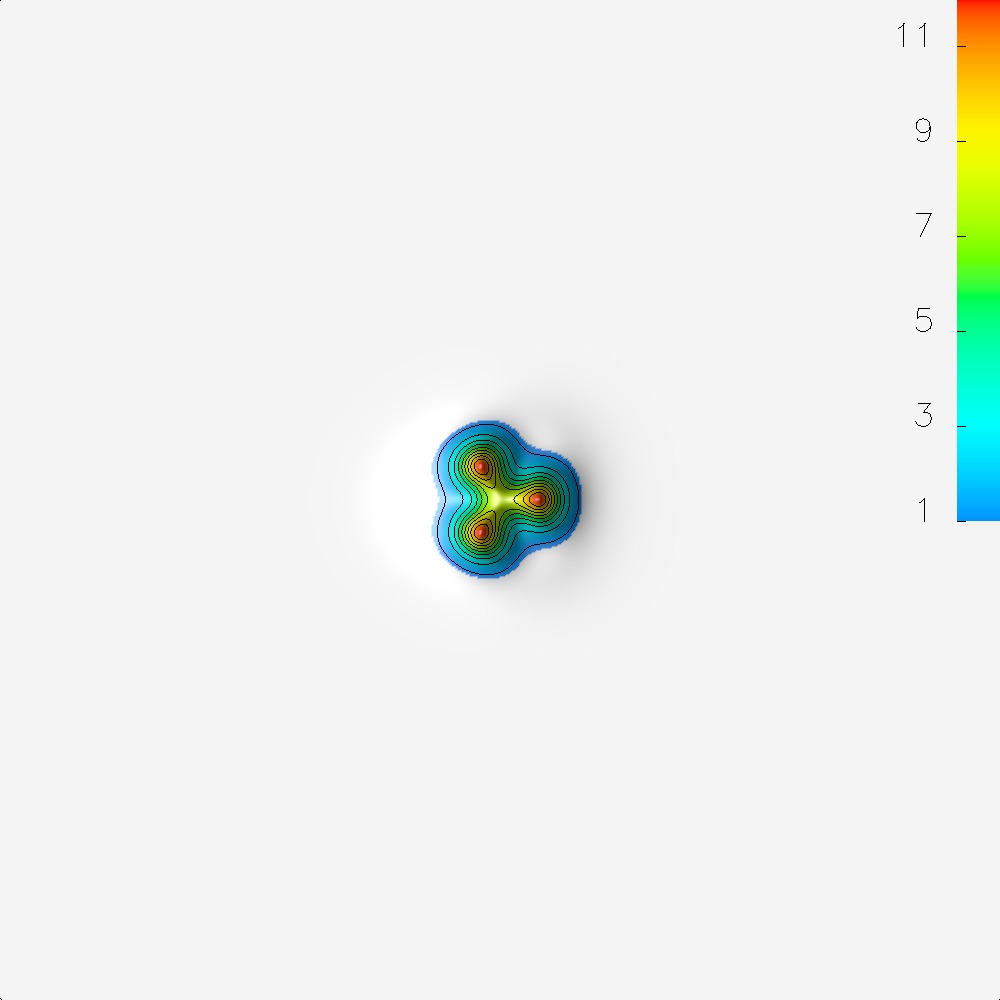} & \includegraphics[scale=0.4,natwidth=1000,natheight=1000]{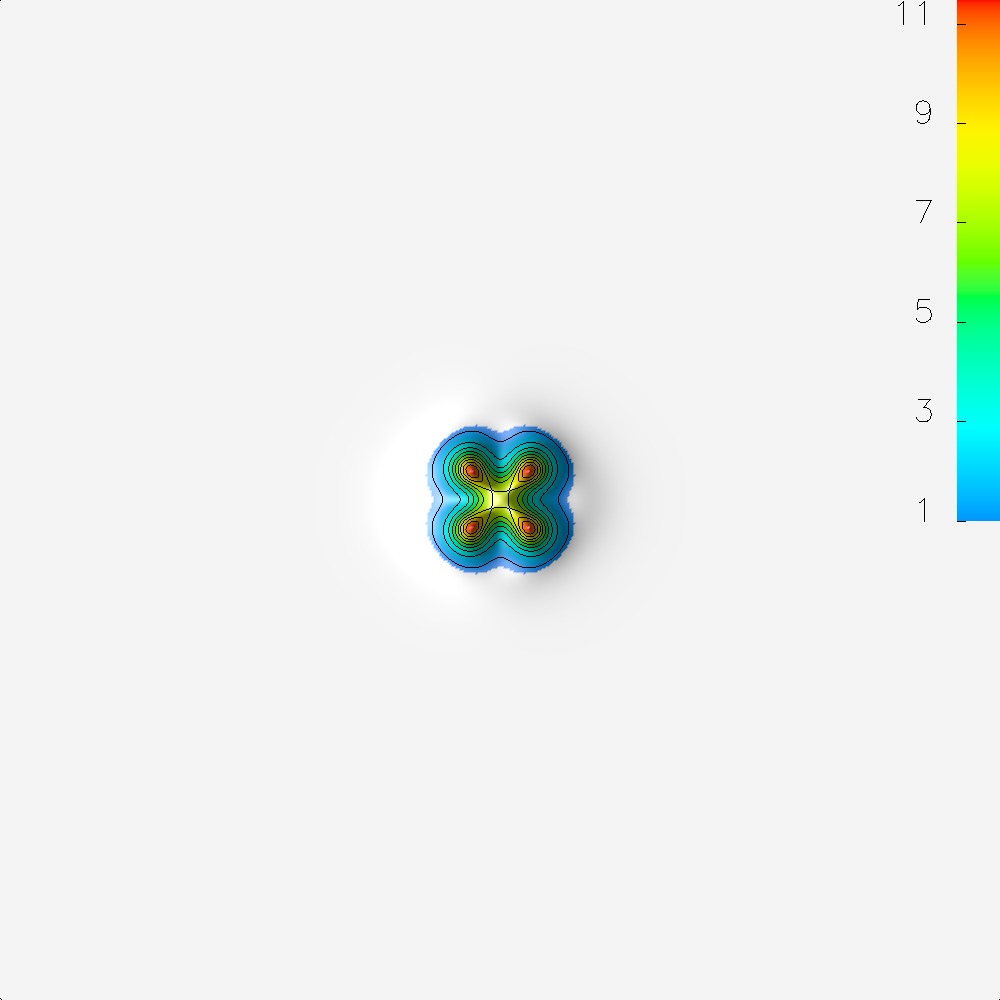} & \includegraphics[scale=0.4,natwidth=1000,natheight=1000]{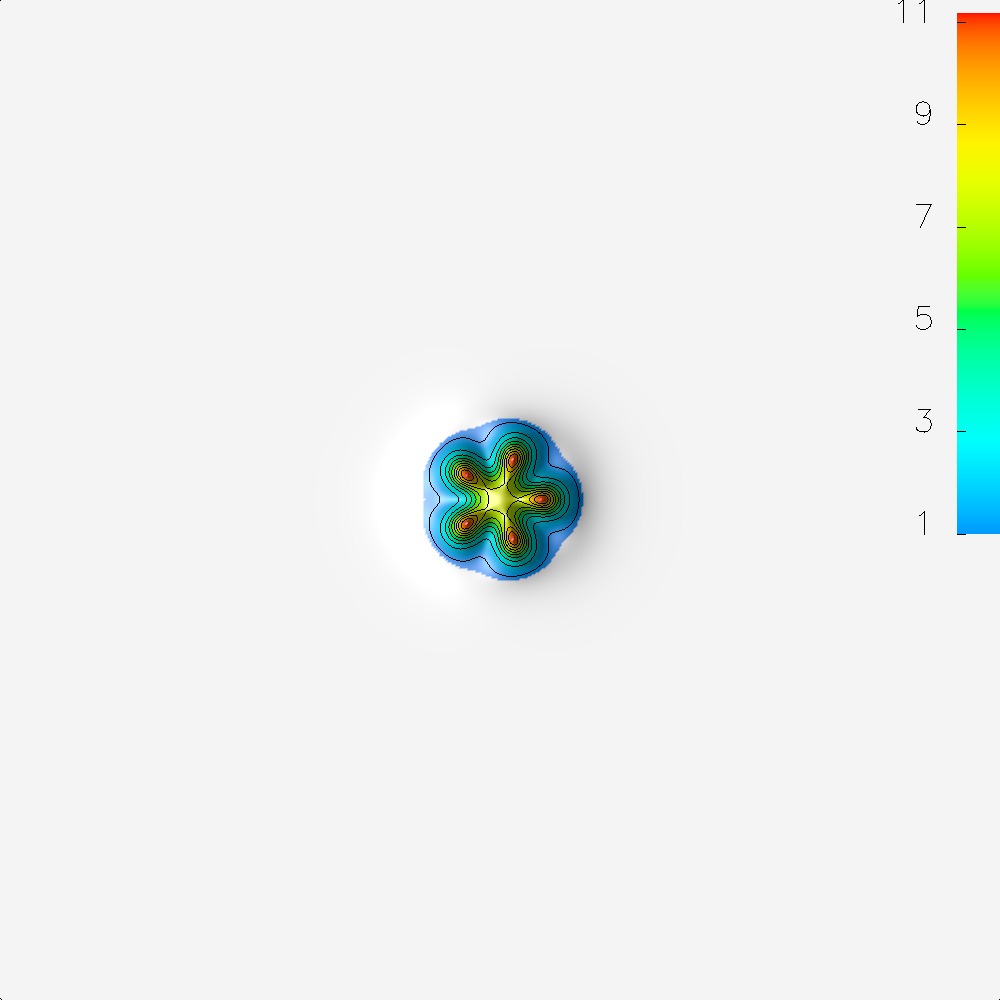}\\
\includegraphics[scale=0.4,natwidth=1000,natheight=1000]{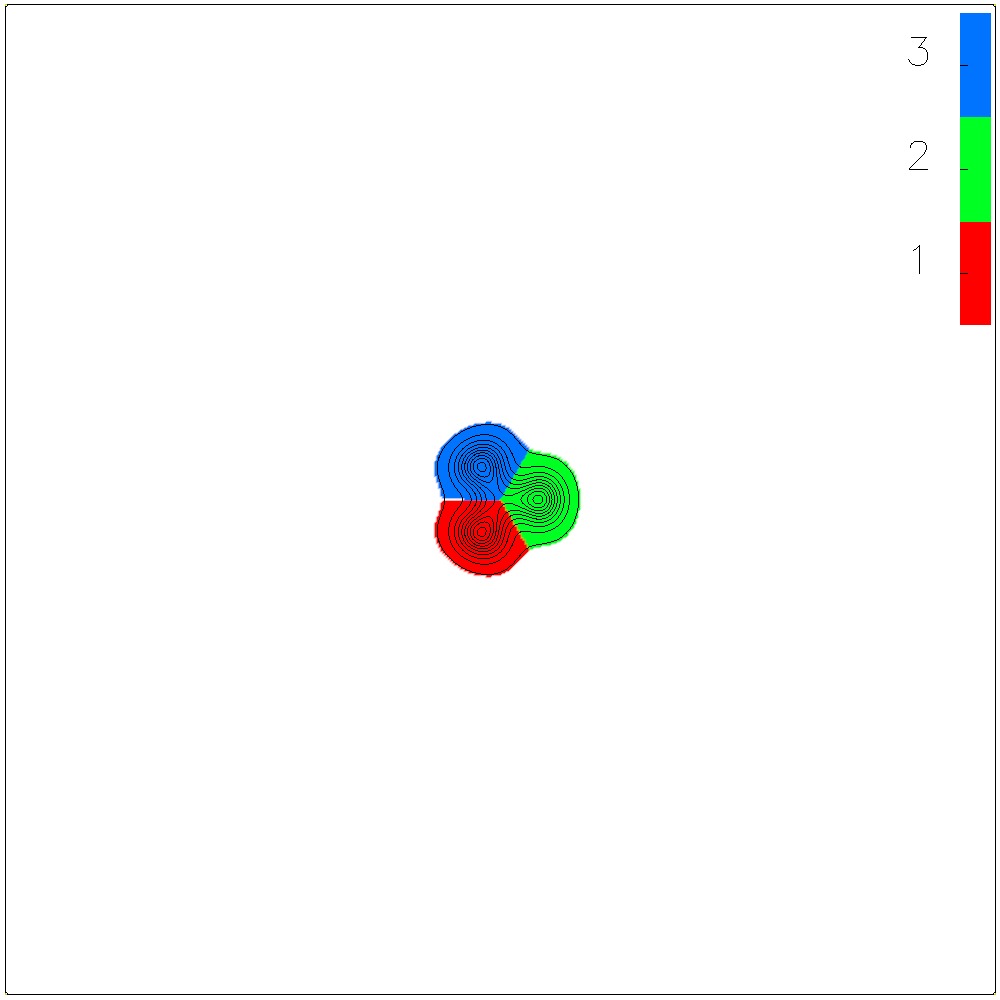} & \includegraphics[scale=0.4,natwidth=1000,natheight=1000]{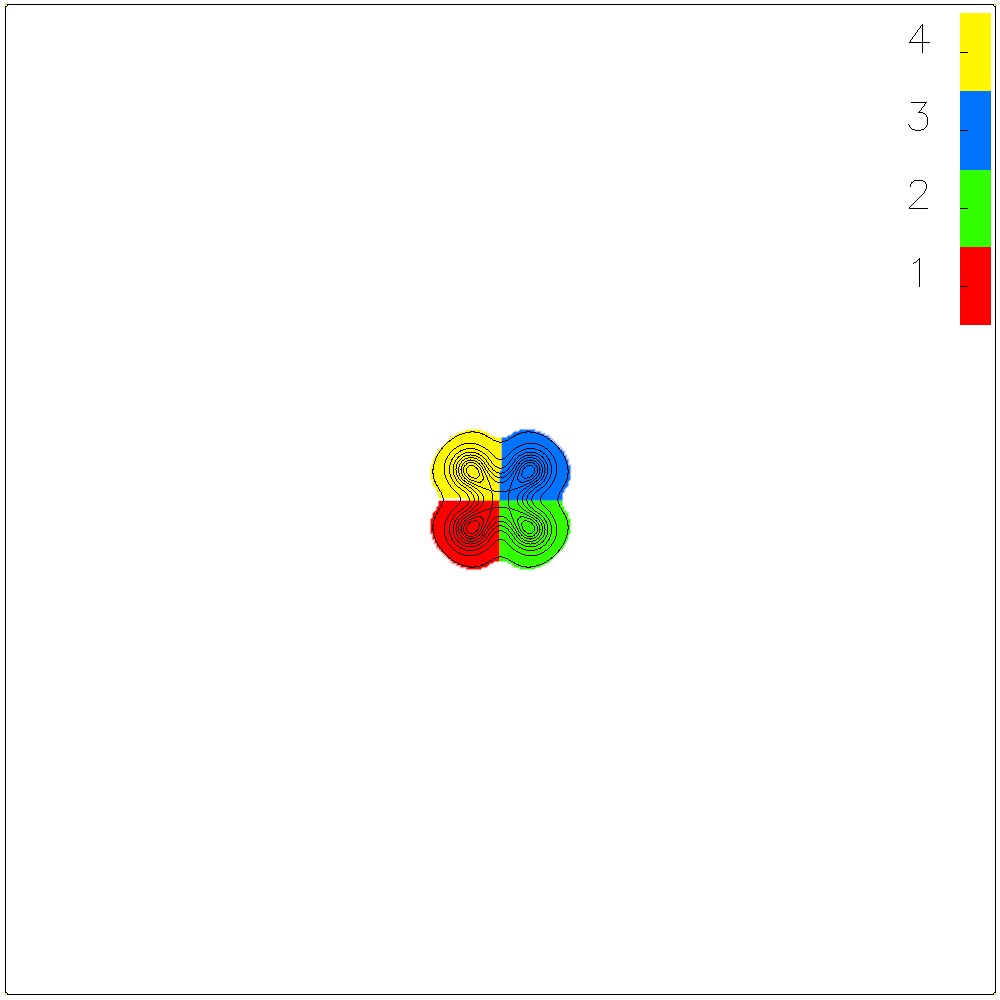} & \includegraphics[scale=0.4,natwidth=1000,natheight=1000]{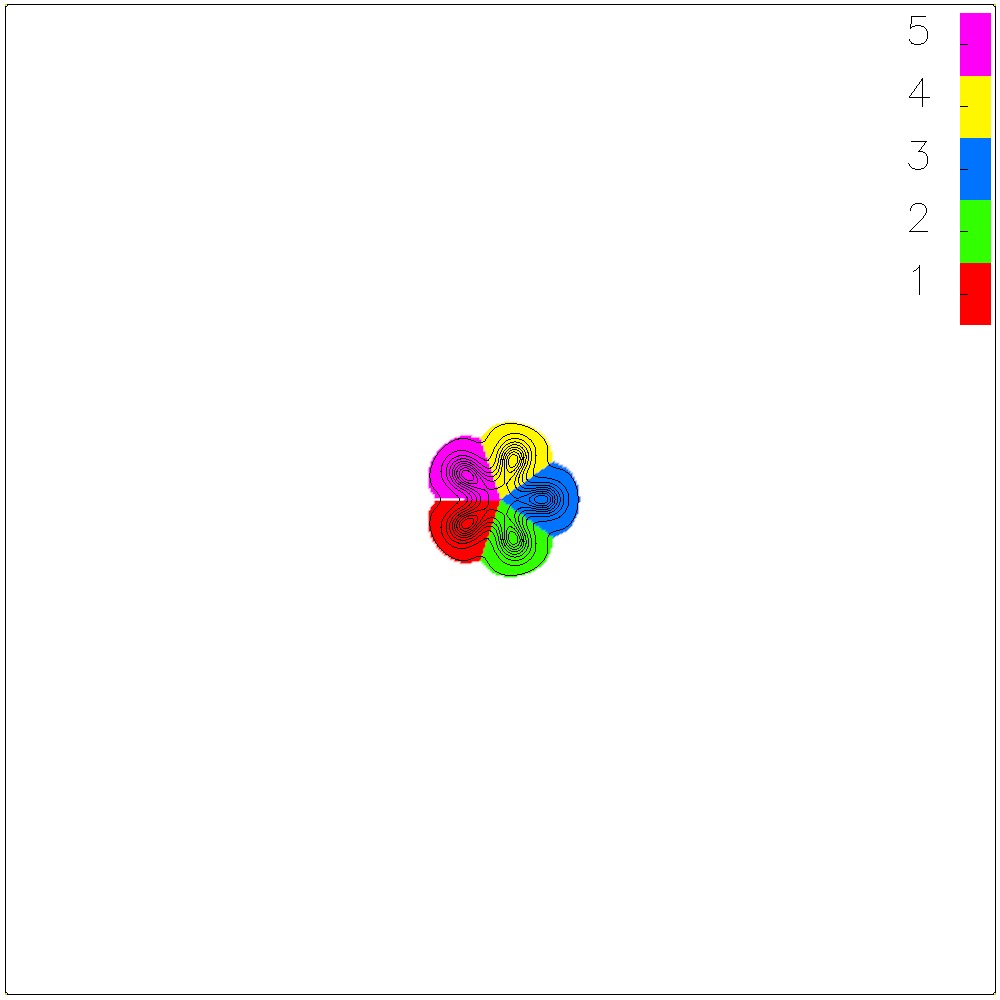}\\
(a) $N=3$, $B=1$ & (b) $N=4$, $B=1$ & (c) $N=5$, $B=1$
\end{tabular}
\caption{Energy density plots of the single soliton solutions for a)$N=3$, b)$N=4$ and c)$N=5$. The top image is coloured based on the energy density and the bottom image is coloured based on the segment in which the point lies in the target space.}
\label{singlesolitons}
\end{center}
\end{figure}

Applying our energy minimizing code on the initial conditions in equation (\ref{initialconditions}) for $N=3,4$ and $5$, and $B=1$, we obtain the contour plots in the top half of figure \ref{singlesolitons} (note that all images in this section show the entire grid and hence are to scale). These energy density plots exhibit the maximal symmetry group $D_N$, giving the predicted $N$ parton structure. Note that a plot of topological charge density will yield a similar result. The energy is given to be $E=34.79,34.58$ and $34.41$ respectively. 

We can further embellish the parton interpretation by introducing colour into our visualisation. Each peak of the energy density will have an associated colour, derived from the segment of the target 2-sphere in which the parton lies. These segments are formed by taking the angle in the $\phi_1$, $\phi_2$ plane (phase), and splitting the plane into $N$ segments using the phases of the $N$ vacuua on the $\phi_3=0$ equator.  

Each of these segments, or partons, contributes $1/N$ to the topological charge. Naturally this means that the combination of the vacuua structure and the requirement of integer topological charge, forces these partons to be topologically confined. If we add this additional structure to our figures, we obtain the results given in the lower half of figure \ref{singlesolitons}. 

\subsection{Multi-soliton Solutions}
For higher values of topological charge, we observe two prominent types of solution. These are shown in figure \ref{N3plots} for $N=3$, figure \ref{N4plots} for $N=4$ and in the appendices for $N=5,6$. The maximal symmetry solutions, shown in figures \ref{N3plots}(a,c,e) and \ref{N4plots}(a,b,e), are composed of $NB$ partons, situated on the vertices of a regular $NB$-gon. They retain the maximal symmetry of the initial conditions, namely $D_{NB}$. The $B>2$ maximally symmetric solutions have energies higher than that of the lower symmetry solutions, forming local minima. However for $B \geq 5$ the maximally symmetric solution could not be found.

For $B=2$ the hexagonal $N=3$ solution (\ref{N3plots}(b)) has an energy comparable with the lower symmetry solution (\ref{N3plots}(a)). Due to our expected numerical accuracy, we cannot determine which of the solutions is the global minimum of our model. However for all $N>3$, the lower symmetry solution appears to be an unstable saddle point and could not be attained via gradient flow. Hence the maximal $D_{2N}$ solution is the only solution found for $B=2$, $N\geq4$ and is hence identified as the global minimum.

The lower symmetry solution for $N=3$, $B=2$, shown in figure \ref{N3plots}(b), is formed by two $B=1$ single solitons, with a relative spatial rotation by $\pi$. This is as expected due to the form of the asymptotic fields being the same as for the standard planar Skyrme model. The leading order result states that two single solitons are in the maximally attractive channel when rotated relative to each other by $\pi$ \cite{Piette:dbs}. Due to the potential breaking axial symmetry, beyond leading order the asymptotic forces will discriminate between various orientations of the two solitons.

The most energetically favorable orientations for $B\geq2$ appears to be that of polyforms  \cite{polyominoe}, planar figures formed by regular $N$-gons joined along their edges. For the $N=3$ case these are known as polyiamonds and for $N=4$ polyominoes. Polyforms have been studied for millennia, with the earliest reference from ancient masters of the strategy game Go.  We will represent each soliton as a regular $N$-gon, with $N$ different colours located at the vertices, which are then joined along common edges. We can then see that each of the solutions shown in figures \ref{N3plots} and \ref{N4plots} exhibit this polyform structure.

Studying the solutions for $B=3$, $N=4$ as an example, the initial conditions described in equation (\ref{initialconditions}) produces the unstable $D_{12}$ maximally symmetric solution with energy $E/B=33.43$. A slight perturbation of these initial conditions, breaking the maximal symmetry, forms either the  \includegraphics[scale=0.14,natwidth=121,natheight=44]{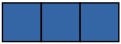} (line) solution in figure \ref{N4plots}(c) or the \includegraphics[scale=0.14,natwidth=82,natheight=82]{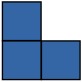} solution in figure \ref{N4plots}(d) with energies $E/B = 32.66$ and $32.77$ respectively, this pattern continues for all $N \geq 4$. In this example the line solution appears to be the global energy minima and this emerges to be the case for all $N$ and $B \geq 3$. This is not a surprise due to the standard potential giving the same result as shown in \cite{Foster:2009vk}. If we look at some of the results for higher $N$, the solutions are very difficult to find as they tend to want to relax to 
the line solution instead. Due to this we did not actually find solutions for \includegraphics[scale=0.14,natwidth=120,natheight=82]{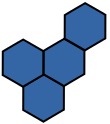}, \includegraphics[scale=0.14,natwidth=121,natheight=82]{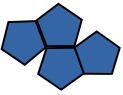} and \reflectbox{\includegraphics[scale=0.14,natwidth=121,natheight=82]{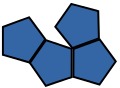}}.

The two key forms of solution discussed above continues for various $N$ and $B$. Some of these other solutions and energies can be seen in appendix A. There are however several caveats to the general forms discussed above. 

\begin{table}
\caption{The energy for soliton solutions and their symmetry group $G$ for $B \leq 4$ and (left) $N = 3$ (right) $N=4$}
\begin{center}
\begin{tabular}{c c c}
\begin{tabular}{c c c c c c}
$B$ & form & $E$ & $E/B$ & $G$ & figure \\
\hline
1 & \includegraphics[scale=0.15,natwidth=50,natheight=45]{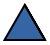} & 34.79 & 34.79 & $D_3$ &\ref{singlesolitons}(a)\\
2 & \includegraphics[scale=0.15,natwidth=45,natheight=45]{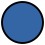} & 66.07 & 33.04 & $D_6$ &\ref{N3plots}(a)\\
2 & \includegraphics[scale=0.15,natwidth=70,natheight=45]{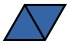} & 66.12 & 33.06 & $D_2$ &\ref{N3plots}(b)\\
3 & \includegraphics[scale=0.15,natwidth=55,natheight=55]{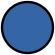} & 101.04 & 33.68 & $D_9$ &\ref{N3plots}(c)\\
3 & \includegraphics[scale=0.15,natwidth=90,natheight=45]{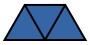} & 98.47 & 32.82 & $D_1$ &\ref{N3plots}(d)\\
3 & \includegraphics[scale=0.15,natwidth=90,natheight=80]{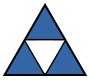} & 100.94 & 33.65 & $D_3$ &\ref{caveats}(a)\\
4 & \includegraphics[scale=0.15,natwidth=65,natheight=65]{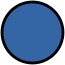} & 138.98 & 34.75 & $D_{12}$ &\ref{N3plots}(e)\\
4 & \includegraphics[scale=0.15,natwidth=110,natheight=45]{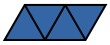} & 130.65 & 32.66 & $C_2$ &\ref{N3plots}(f)\\
4 & \includegraphics[scale=0.15,natwidth=90,natheight=80]{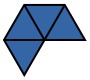} & 130.66 & 32.67 & $D_1$ &\ref{N3plots}(g)\\
4 & \includegraphics[scale=0.15,natwidth=90,natheight=80]{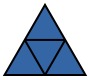} & 131.80 & 32.95 & $D_3$ &\ref{N3plots}(h)\\
4 & \includegraphics[scale=0.15,natwidth=90,natheight=80]{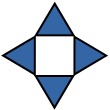} & 132.07 & 33.02 & $D_4$ &\ref{caveats}(b)\\
\hline

\label{energytable}
\end{tabular} & ~~ &
\begin{tabular}{c c c c c c}
$B$ & form & $E$ & $E/B$ & $G$ & figure \\
\hline
1 & \includegraphics[scale=0.14,natwidth=43,natheight=44]{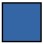} & 34.58 & 34.58 & $D_4$ & \ref{singlesolitons}(b)\\
2 & \includegraphics[scale=0.14,natwidth=45,natheight=45]{Images/forms/maximal.jpg} & 65.58 & 32.79 & $D_8$ & \ref{N4plots}(a)\\
3 & \includegraphics[scale=0.15,natwidth=55,natheight=55]{Images/forms/maximal2.jpg} & 100.28 & 33.43 & $D_{12}$ & \ref{N4plots}(b)\\
3 & \includegraphics[scale=0.14,natwidth=121,natheight=44]{Images/forms/poly-3-1.jpeg} & 97.97 & 32.66 & $D_2$ & \ref{N4plots}(c)\\
3 & \includegraphics[scale=0.14,natwidth=82,natheight=82]{Images/forms/poly-3-2.jpeg} & 98.32 & 32.77 & $D_1$ & \ref{N4plots}(d)\\
3 & \includegraphics[scale=0.15,natwidth=118,natheight=100]{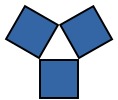} & 98.71 & 32.90 & $D_3$ &\ref{caveats}(c)\\
4 & \includegraphics[scale=0.15,natwidth=65,natheight=65]{Images/forms/maximal3.jpg} & 137.97 & 34.49 & $D_{16}$ &\ref{N4plots}(e)\\
4 & \includegraphics[scale=0.14,natwidth=160,natheight=43]{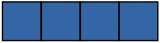}  & 129.94 & 32.49 & $D_2$ & \ref{N4plots}(f)\\
4 & \includegraphics[scale=0.14,natwidth=121,natheight=82]{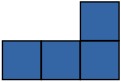} & 130.28 & 32.57 & $C_1$ & \ref{N4plots}(g)\\
4 & \includegraphics[scale=0.14,natwidth=121,natheight=82]{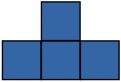} & 131.61 & 32.90 & $D_1$ & \ref{N4plots}(h)\\
4 & \includegraphics[scale=0.14,natwidth=82,natheight=82]{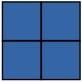} & 131.13 & 32.78 & $D_4$ & \ref{N4plots}(i)\\
4 & \includegraphics[scale=0.14,natwidth=120,natheight=82]{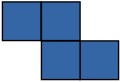} & 130.61 & 32.65 & $C_2$ & \ref{N4plots}(j)\\
4 & \includegraphics[scale=0.14,natwidth=130,natheight=125]{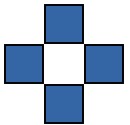} & 131.81 & 32.95 & $D_4$ & \ref{caveats}(d)\\
4 & \includegraphics[scale=0.14,natwidth=130,natheight=125]{Images/forms/poly-fourhole.jpeg} & 135.94 & 33.98 & $D_4$ & \ref{caveats}(e)\\
\hline
\label{N4energytable}
\end{tabular}
\end{tabular}
\end{center}
\end{table}

\subsection{Caveats to the Standard Solutions}

\begin{figure}
\begin{center}
\begin{tabular}{c c c}
\includegraphics[scale=0.4,natwidth=1000,natheight=1000]{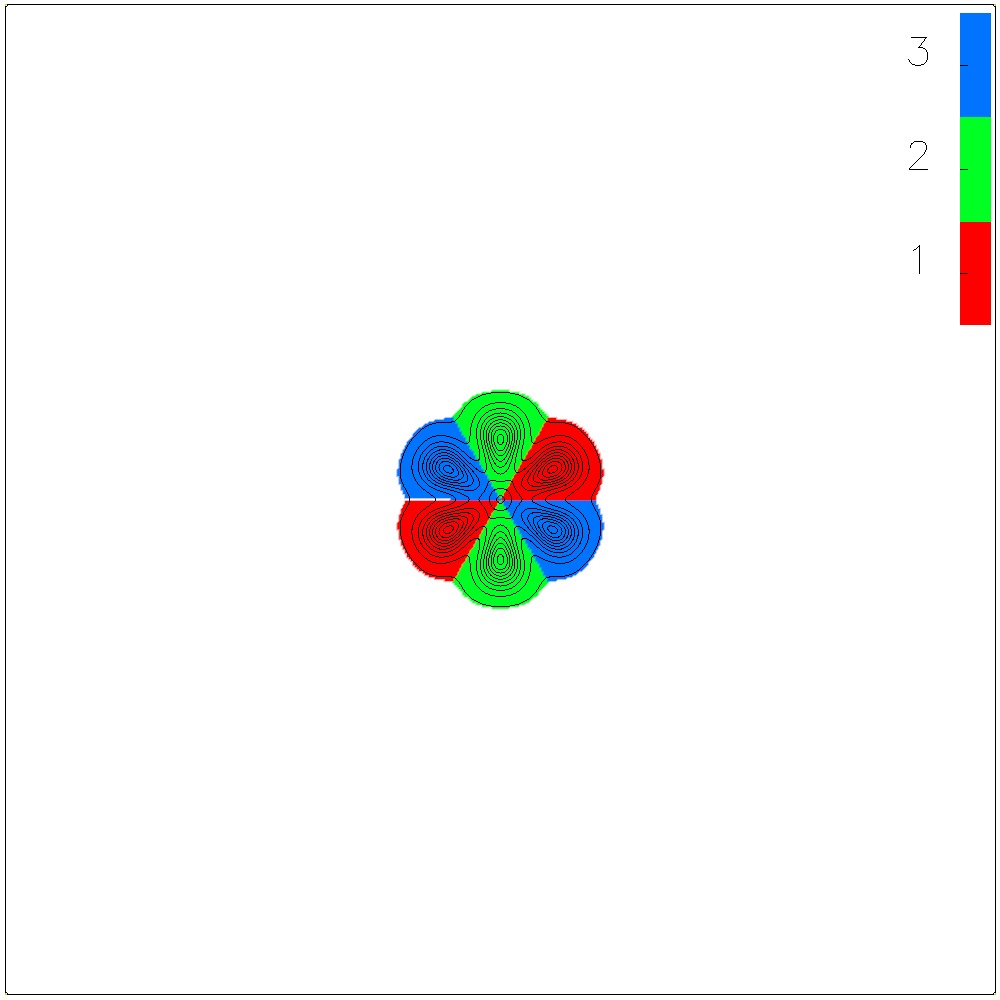} & \includegraphics[scale=0.4,natwidth=1000,natheight=1000]{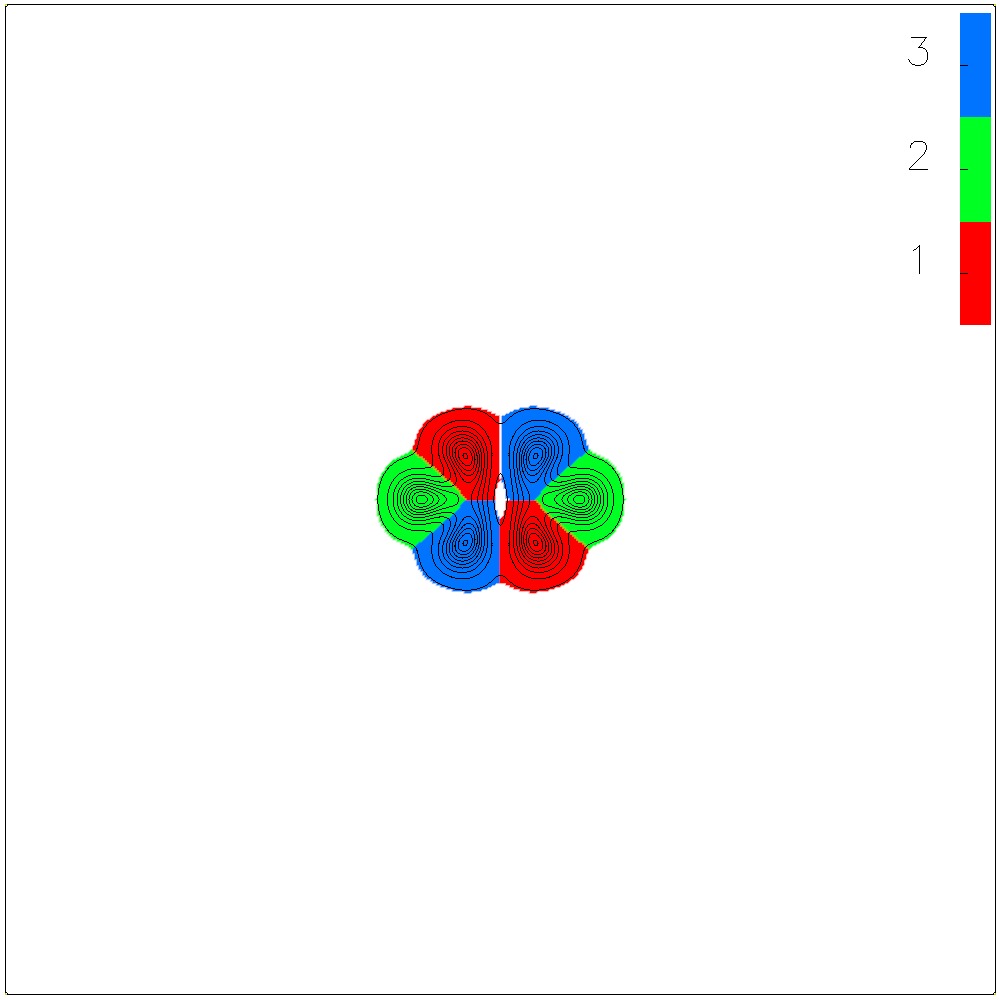} & \includegraphics[scale=0.4,natwidth=1000,natheight=1000]{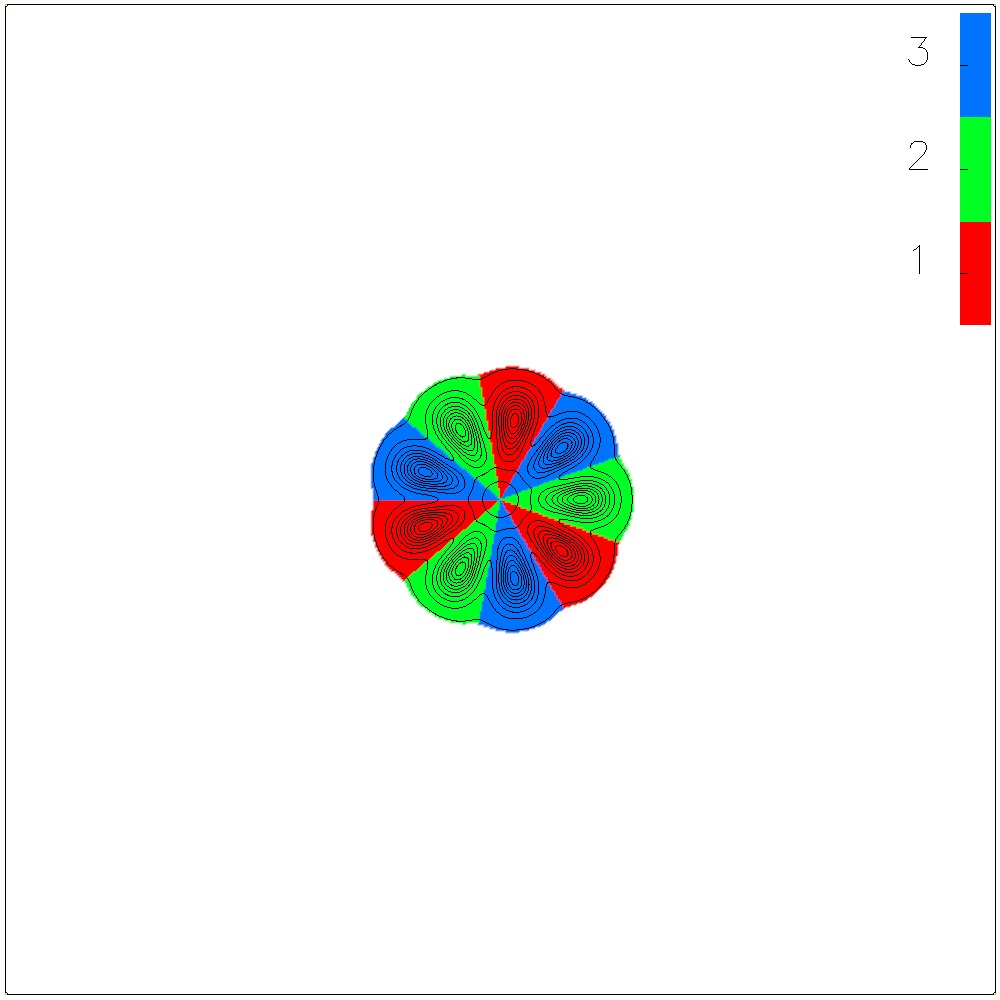}\\
(a) $B=2$ form = \includegraphics[scale=0.15,natwidth=45,natheight=45]{Images/forms/maximal.jpg} & (b) $B=2$ form = \includegraphics[scale=0.15,natwidth=70,natheight=45]{Images/forms/Polyiamond-2-1.jpg} & (c) $B=3$ form = \includegraphics[scale=0.15,natwidth=55,natheight=55]{Images/forms/maximal2.jpg}\\
\includegraphics[scale=0.4,natwidth=1000,natheight=1000]{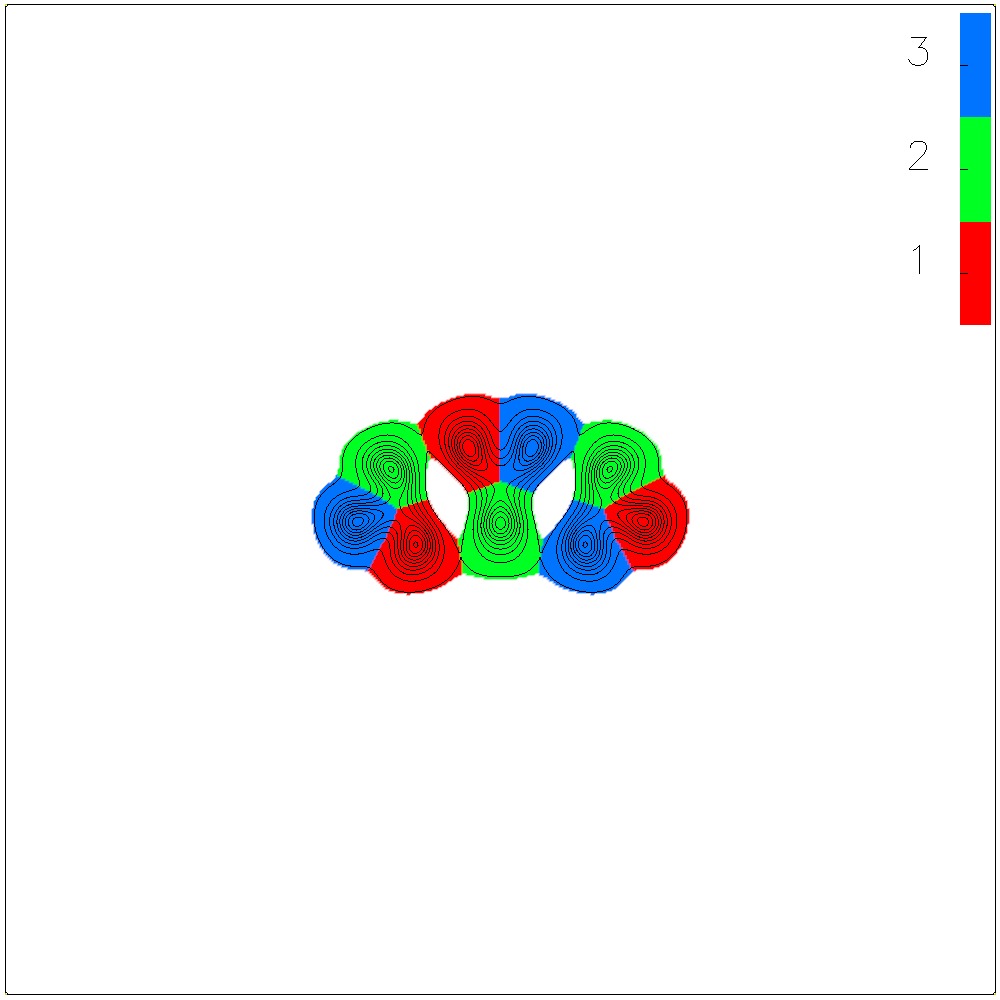} & \includegraphics[scale=0.4,natwidth=1000,natheight=1000]{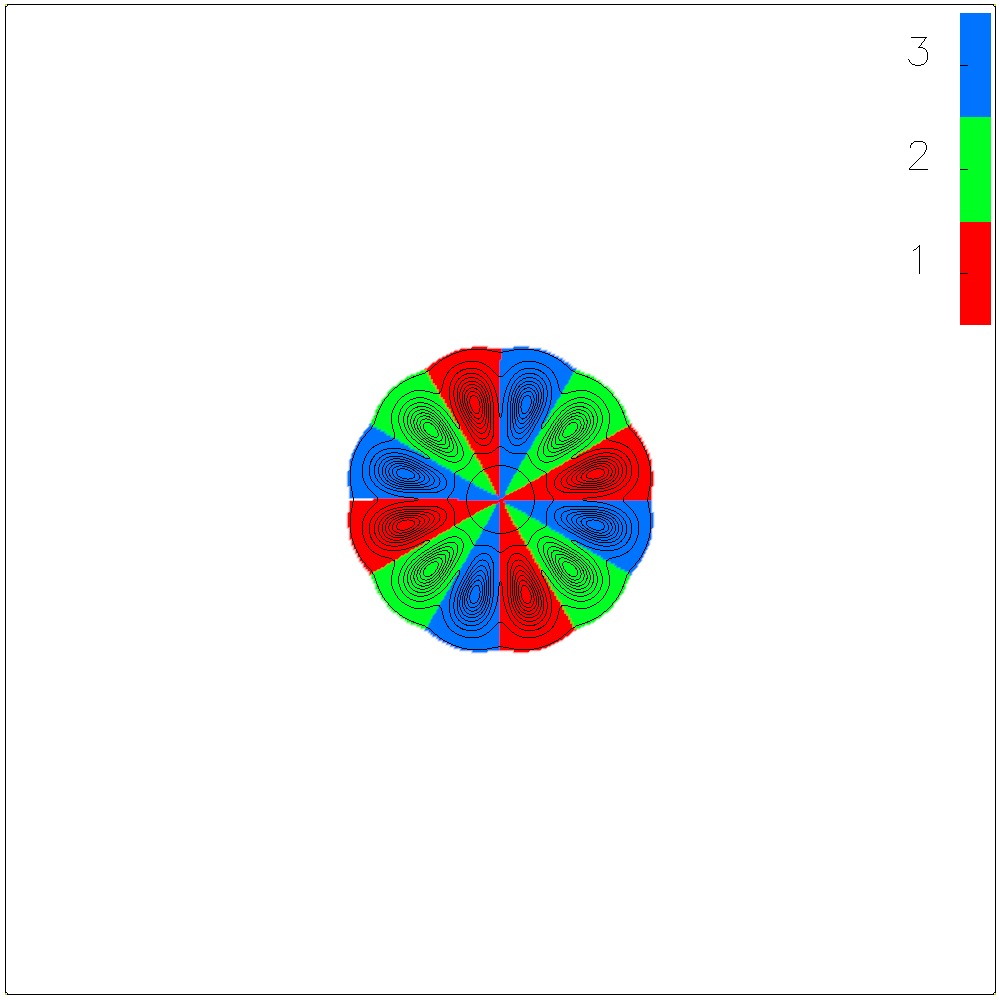} & \includegraphics[scale=0.4,natwidth=1000,natheight=1000]{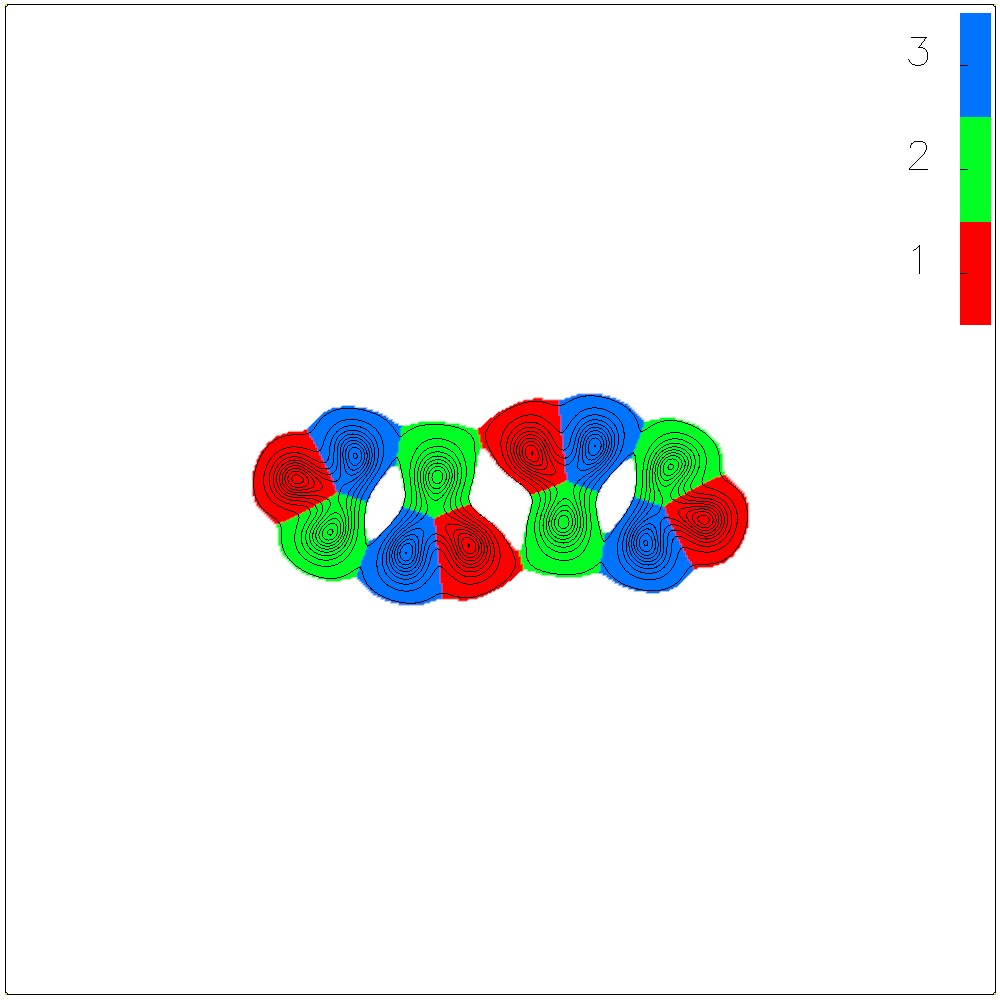}\\
(d) $B=3$ form = \includegraphics[scale=0.15,natwidth=90,natheight=45]{Images/forms/Polyiamond-3-1.jpg} & (e) $B=4$ form = \includegraphics[scale=0.15,natwidth=65,natheight=65]{Images/forms/maximal3.jpg} & (f) $B=4$ form = \includegraphics[scale=0.15,natwidth=110,natheight=45]{Images/forms/Polyiamond-4-2.jpg}
\end{tabular}
\begin{tabular}{c c}
\includegraphics[scale=0.4,natwidth=1000,natheight=1000]{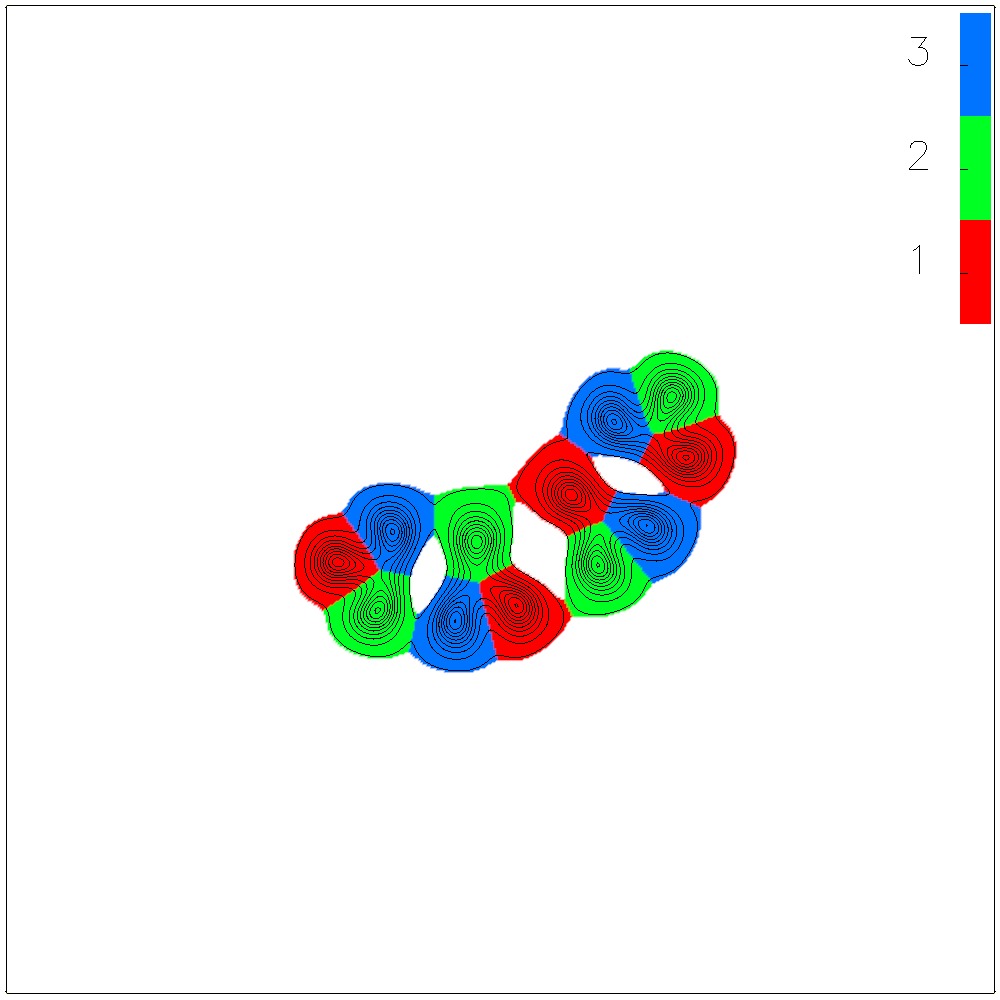} & \includegraphics[scale=0.4,natwidth=1000,natheight=1000]{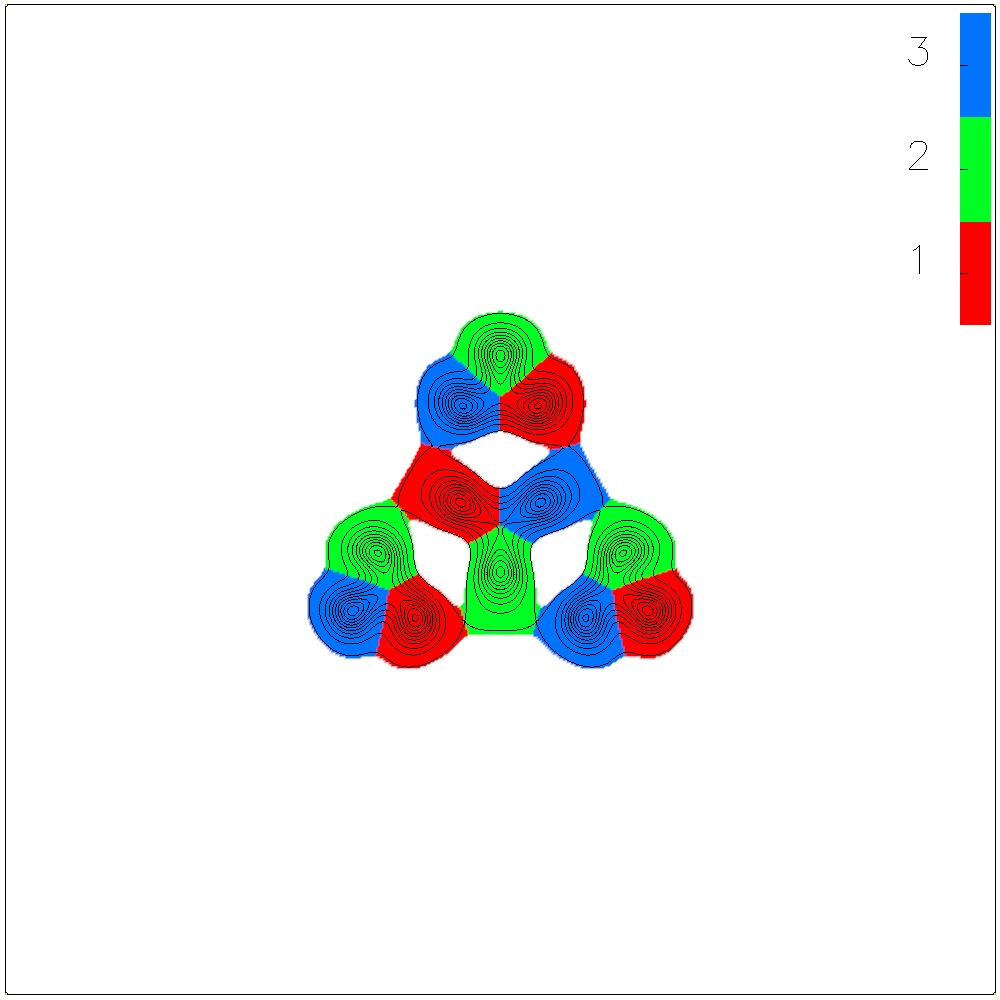}\\
(g) $B=4$ form = \includegraphics[scale=0.15,natwidth=90,natheight=80]{Images/forms/Polyiamond-4-1.jpg} & (h) $B=4$ form = \includegraphics[scale=0.15,natwidth=90,natheight=80]{Images/forms/Polyiamond-4-3.jpg}
\end{tabular}
\caption{Energy density plots of the multi-soliton solutions for $N=3$ and $B\leq4$ (colour is based on the segment in which the point lies in the target space).}
\label{N3plots}
\end{center}
\end{figure}

\begin{figure}
\begin{center}
\begin{tabular}{c c c}
\includegraphics[scale=0.4,natwidth=1000,natheight=1000]{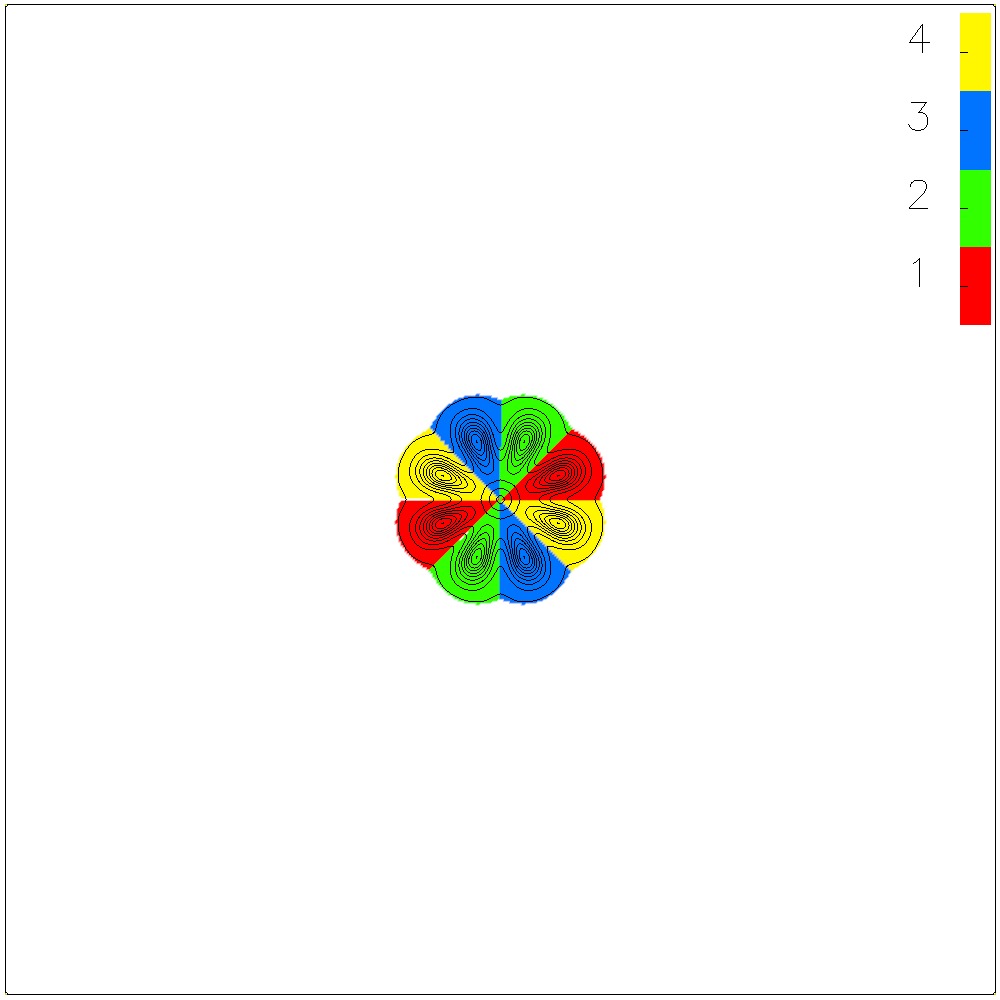} & \includegraphics[scale=0.4,natwidth=1000,natheight=1000]{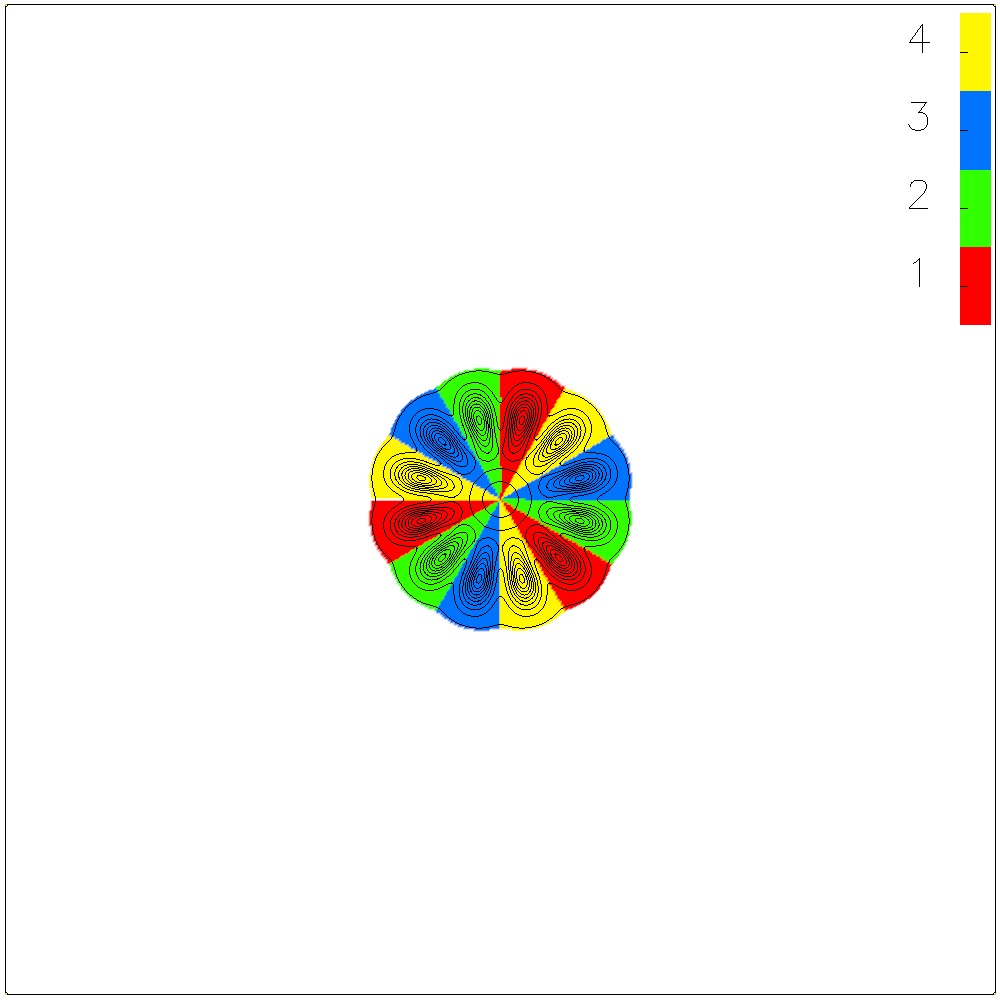} & \includegraphics[scale=0.4,natwidth=1000,natheight=1000]{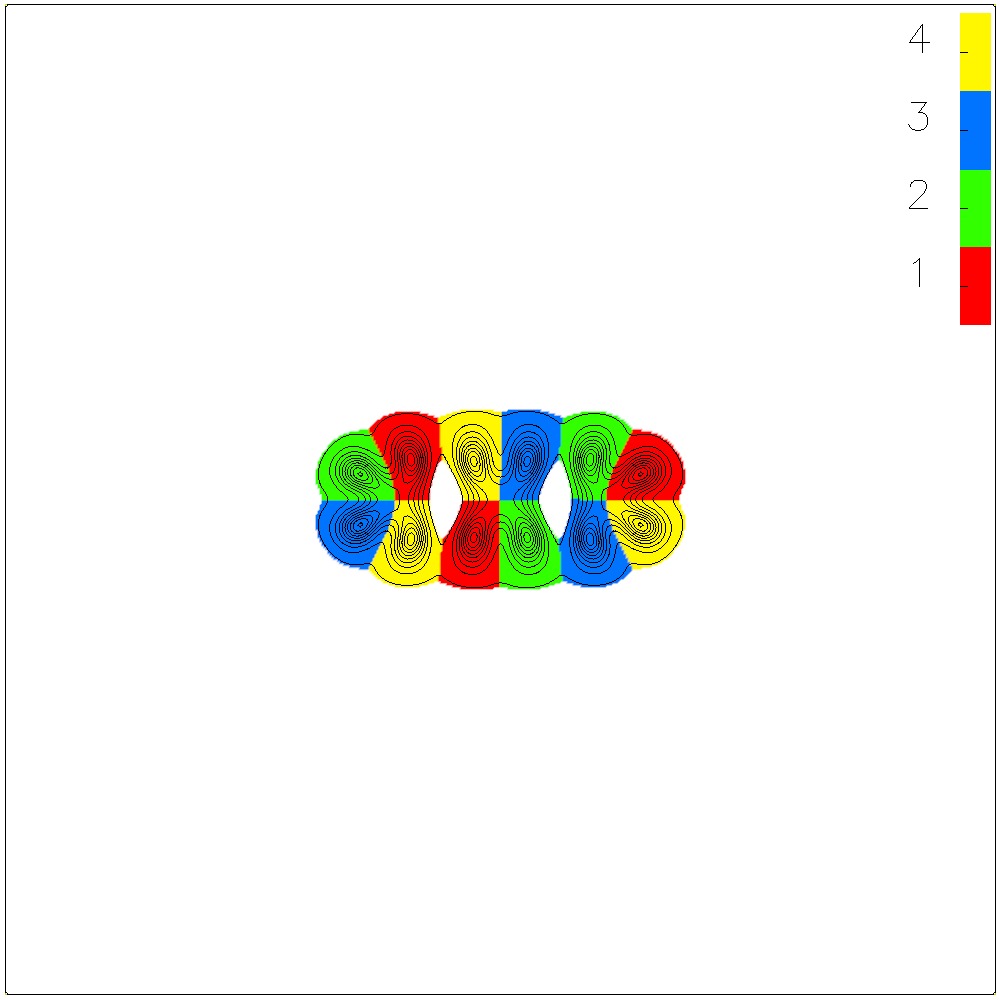}\\
(a) $B=2$ form = \includegraphics[scale=0.14,natwidth=45,natheight=45]{Images/forms/maximal.jpg} & (b) $B=3$ form = \includegraphics[scale=0.15,natwidth=55,natheight=55]{Images/forms/maximal2.jpg} & (c) $B=3$ form = \includegraphics[scale=0.14,natwidth=121,natheight=44]{Images/forms/poly-3-1.jpeg} \\
\includegraphics[scale=0.4,natwidth=1000,natheight=1000]{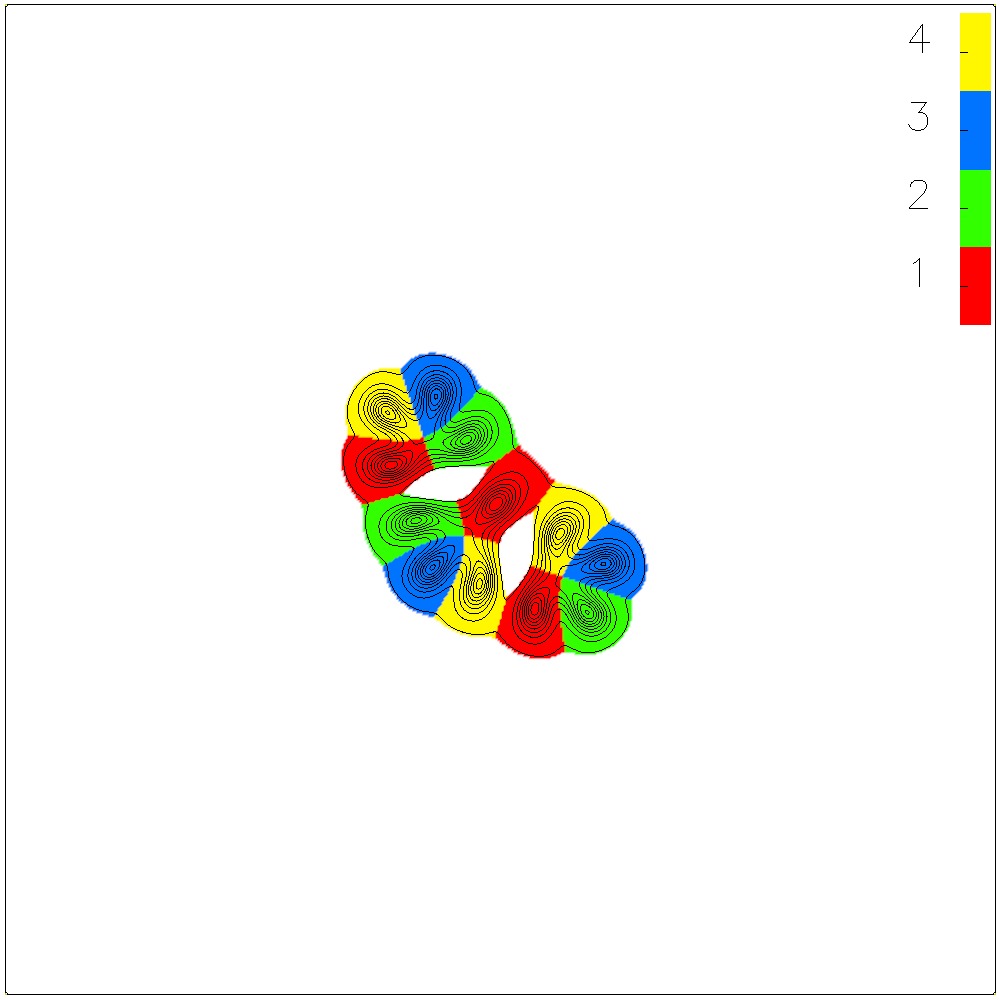} & \includegraphics[scale=0.4,natwidth=1000,natheight=1000]{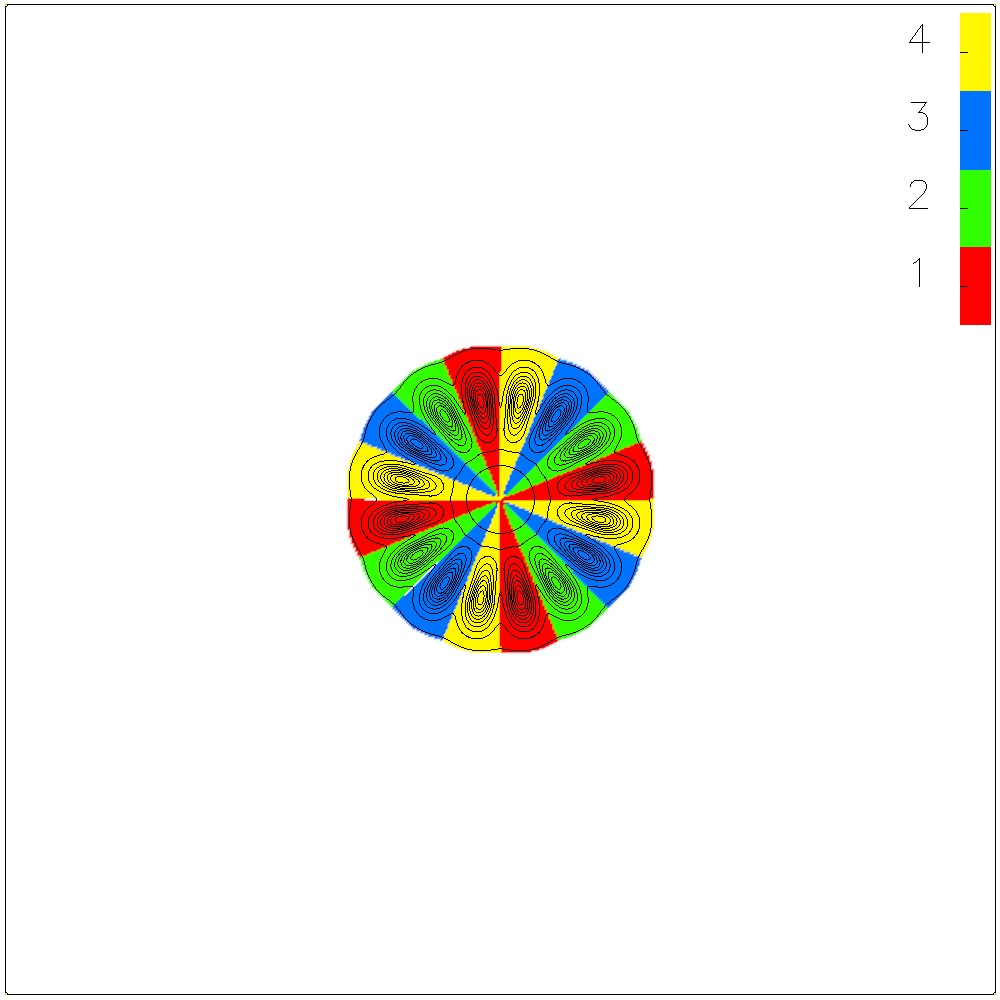} & \includegraphics[scale=0.4,natwidth=1000,natheight=1000]{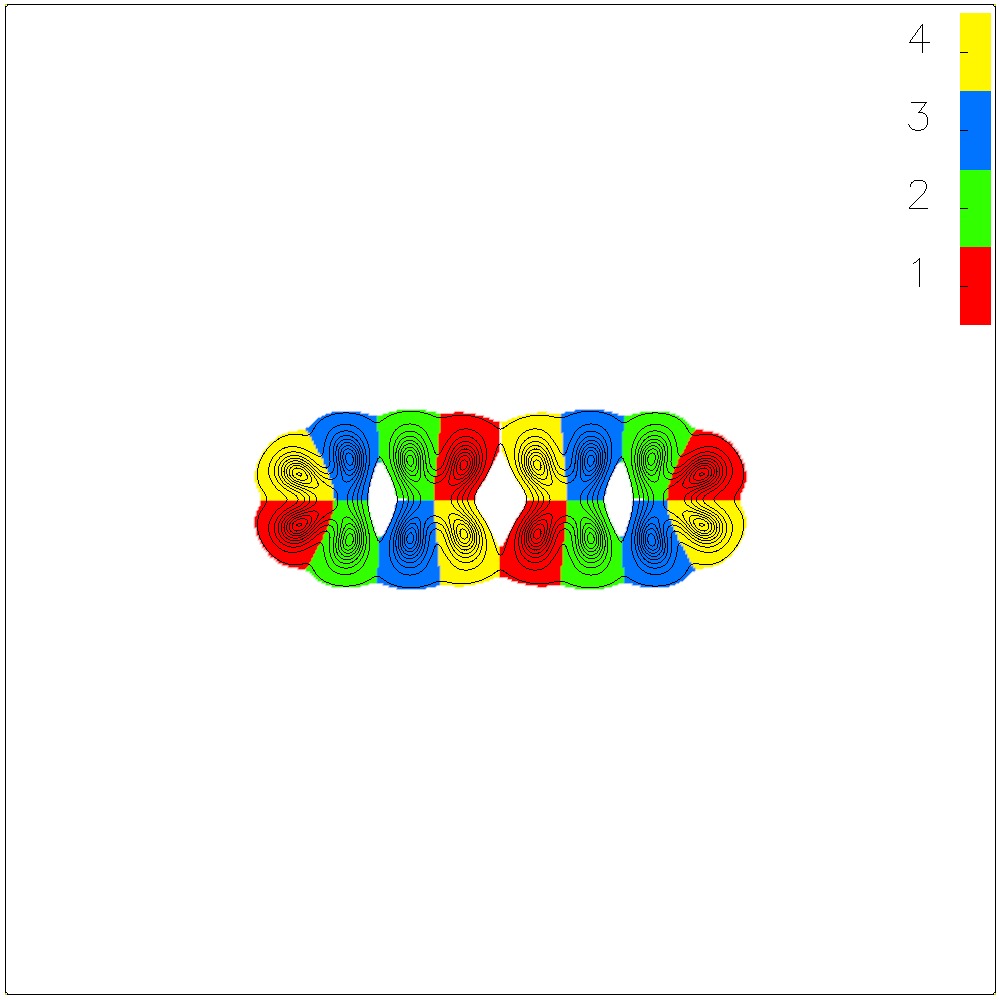}\\
(d) $B=3$ form = \includegraphics[scale=0.14,natwidth=82,natheight=82]{Images/forms/poly-3-2.jpeg} & (e) $B=4$ form = \includegraphics[scale=0.15,natwidth=65,natheight=65]{Images/forms/maximal3.jpg} & (f) $B=4$ form = \includegraphics[scale=0.14,natwidth=160,natheight=43]{Images/forms/poly-4-1.jpeg} \\
\includegraphics[scale=0.4,natwidth=1000,natheight=1000]{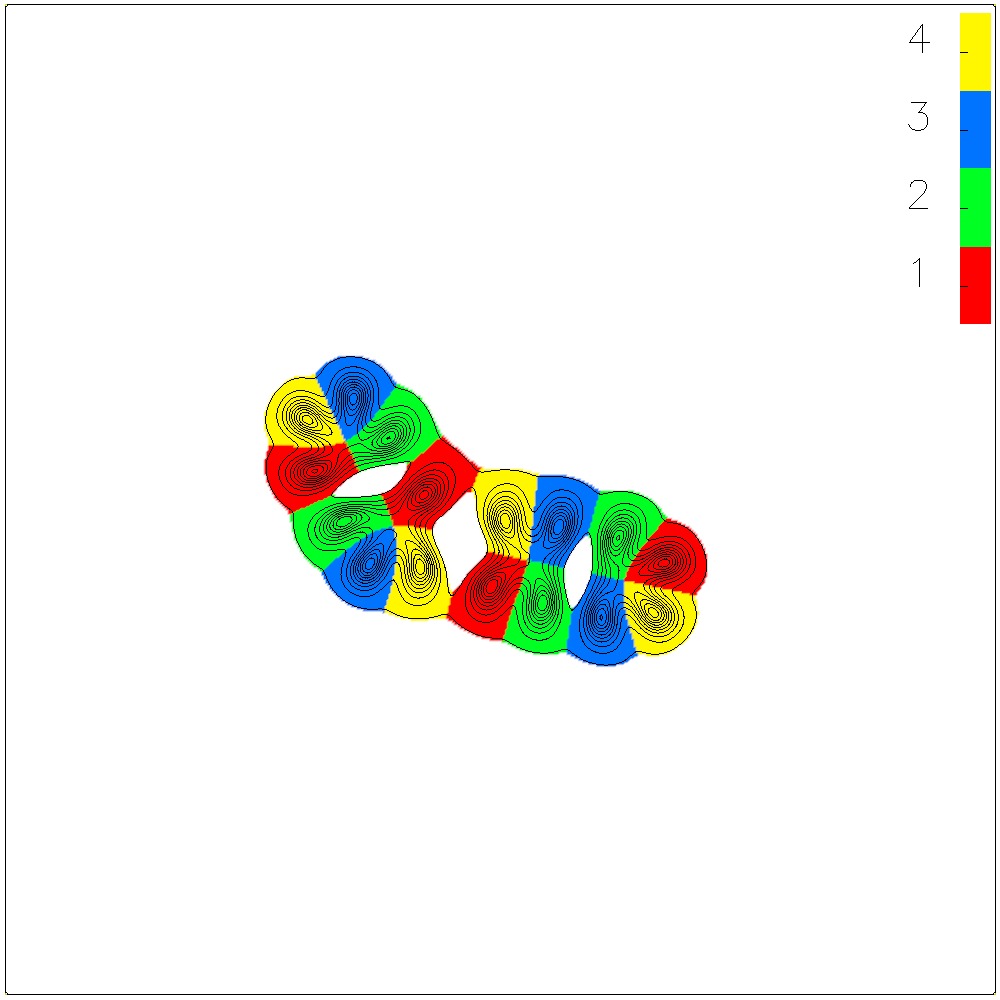} & \includegraphics[scale=0.4,natwidth=1000,natheight=1000]{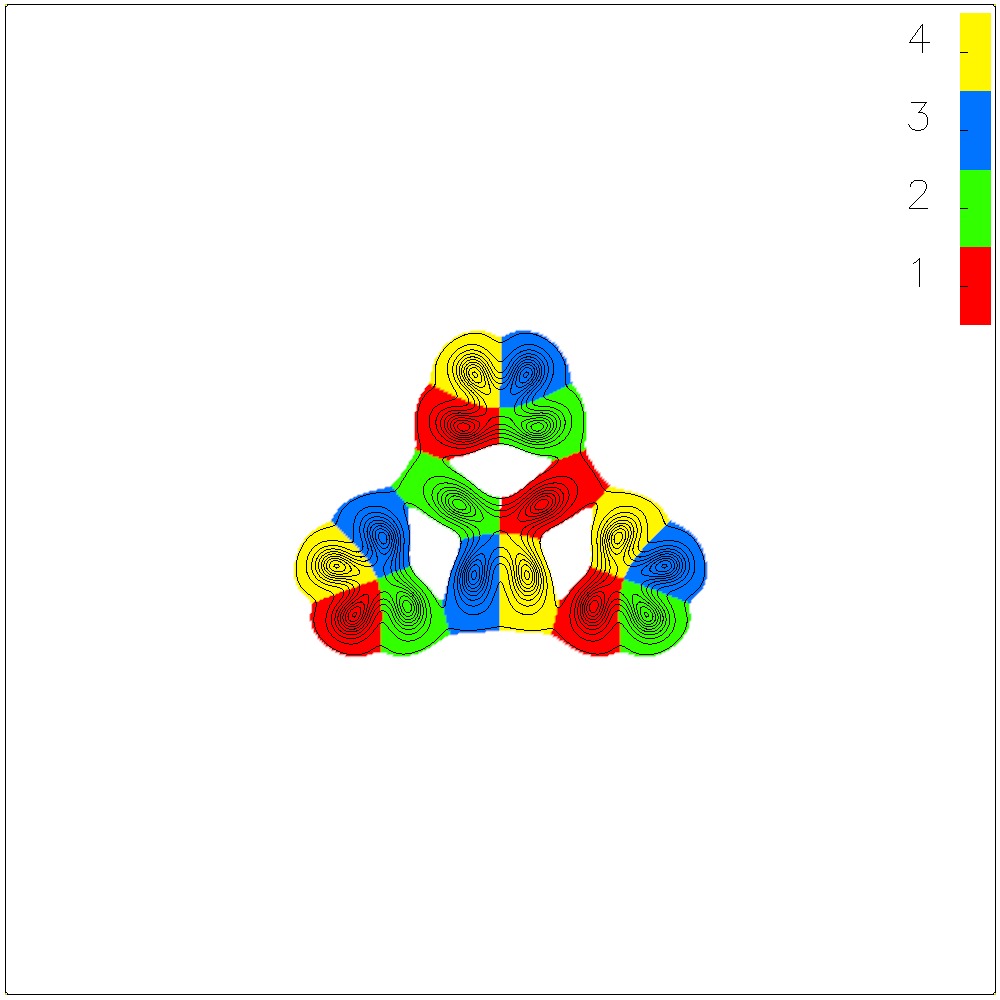} & \includegraphics[scale=0.4,natwidth=1000,natheight=1000]{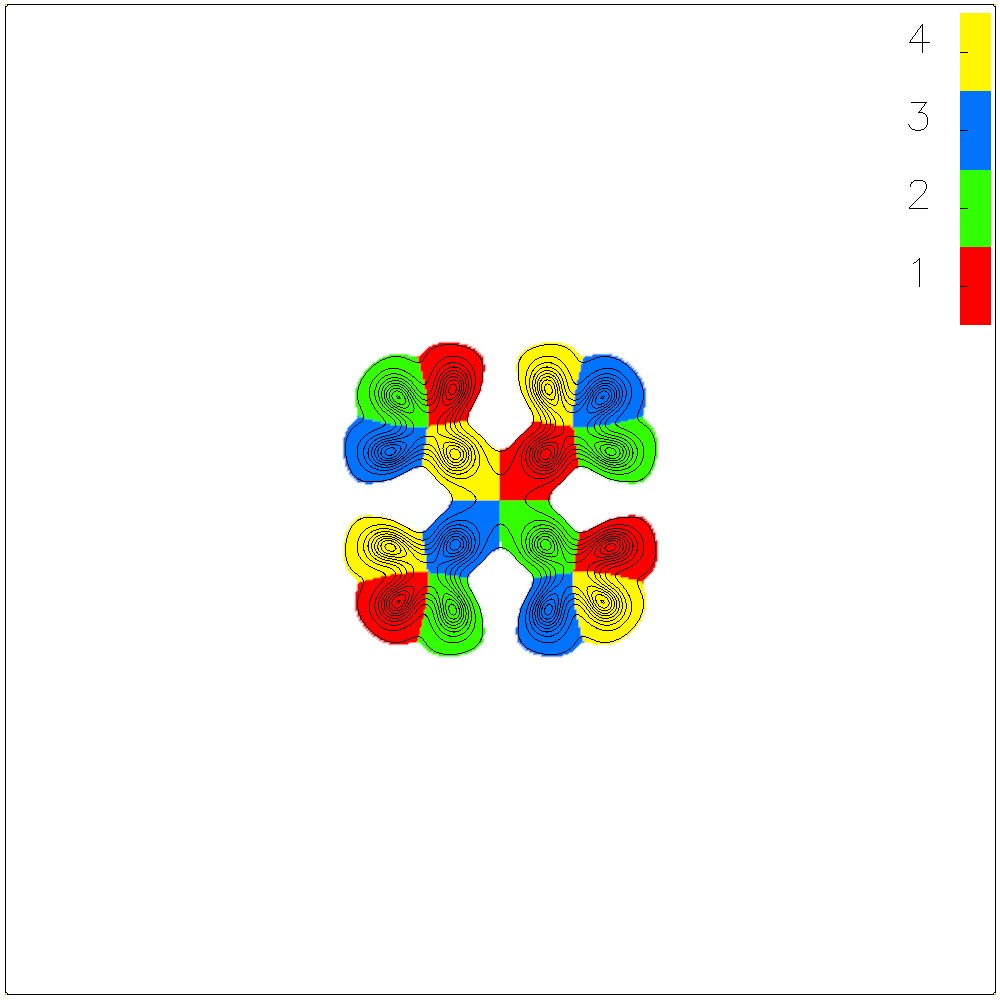}\\
(g) $B=4$ form = \reflectbox{\includegraphics[scale=0.14,natwidth=121,natheight=82]{Images/forms/poly-4-2.jpeg}} & (h) $B=4$ form = \includegraphics[scale=0.14,natwidth=121,natheight=82]{Images/forms/poly-4-3.jpeg} & (i) $B=4$ form = \includegraphics[scale=0.14,natwidth=82,natheight=82]{Images/forms/poly-4-4.jpeg}
\end{tabular}
\begin{tabular}{c}
\includegraphics[scale=0.4,natwidth=1000,natheight=1000]{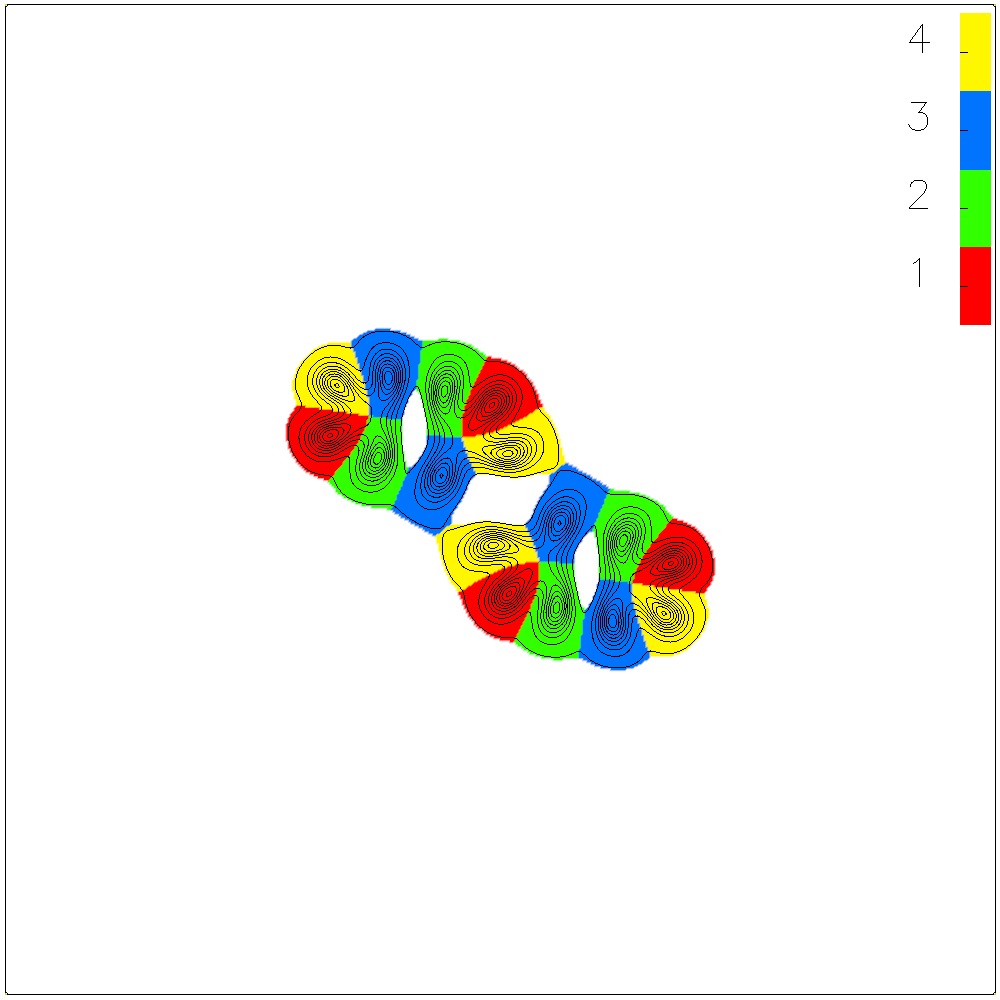}\\
(j) $B=4$ form = \includegraphics[scale=0.14,natwidth=120,natheight=82]{Images/forms/poly-4-5.jpeg}
\end{tabular}
\caption{Energy density plots of the multi-soliton solutions for $N=4$ and $B\leq4$ (colour is based on the segment in which the point lies in the target space).}
\label{N4plots}
\end{center}
\end{figure}

\begin{figure}
\begin{center}
\begin{tabular}{c c c}
\includegraphics[scale=0.4,natwidth=1000,natheight=1000]{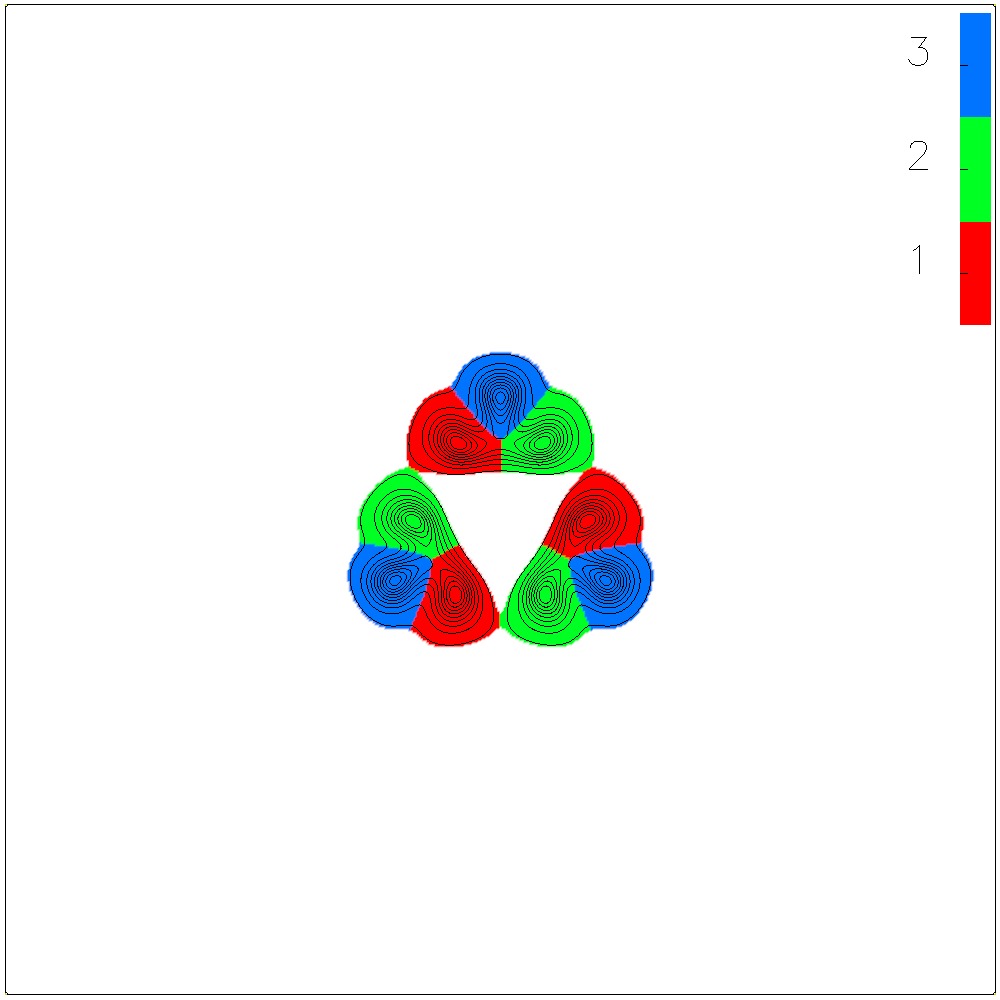} & \includegraphics[scale=0.4,natwidth=1000,natheight=1000]{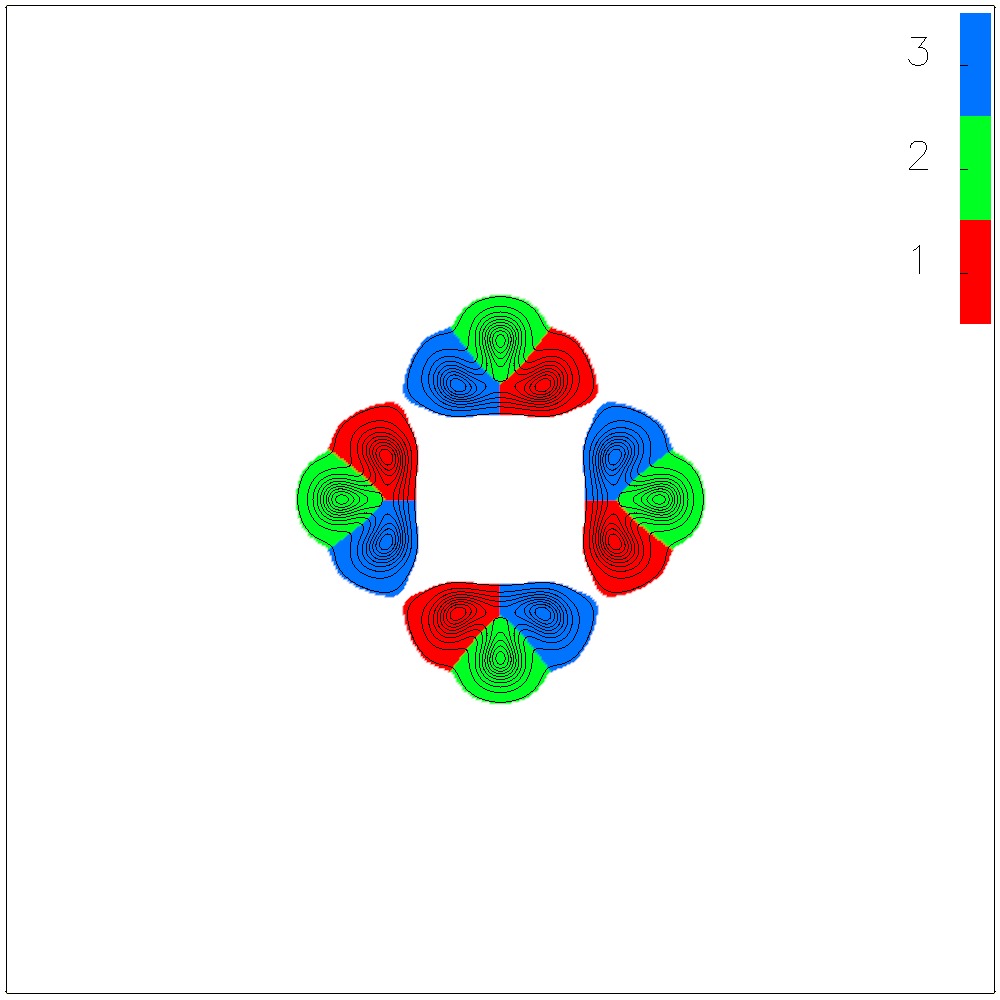} & \includegraphics[scale=0.4,natwidth=1000,natheight=1000]{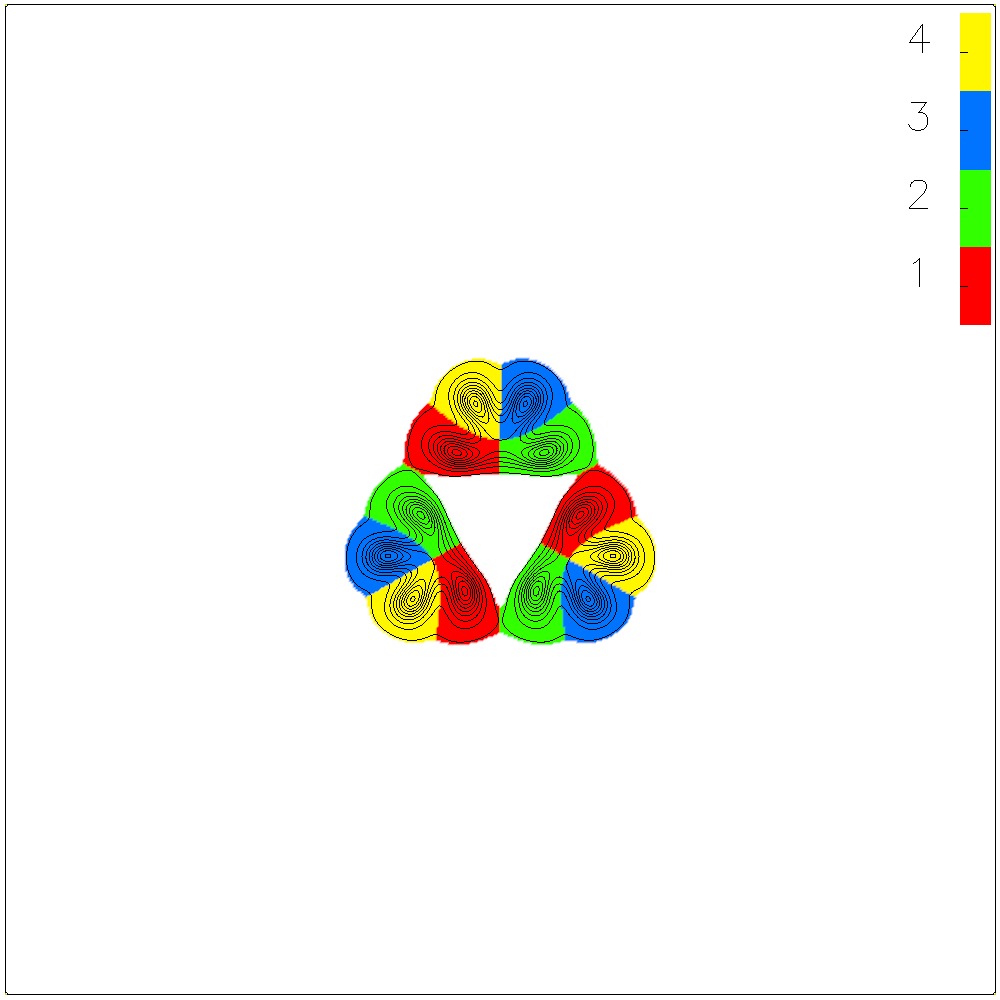}\\
(a) $N=3$, $B=3$ \includegraphics[scale=0.15,natwidth=90,natheight=80]{Images/forms/Polyiamond-3-3.jpg} & (b) $N=3$, $B=4$ \includegraphics[scale=0.15,natwidth=90,natheight=80]{Images/forms/Polyiamond-4-hole.jpeg} & (c) $N=4$, $B=3$ \includegraphics[scale=0.15,natwidth=118,natheight=100]{Images/forms/poly-tri.jpeg}\\
\includegraphics[scale=0.4,natwidth=1000,natheight=1000]{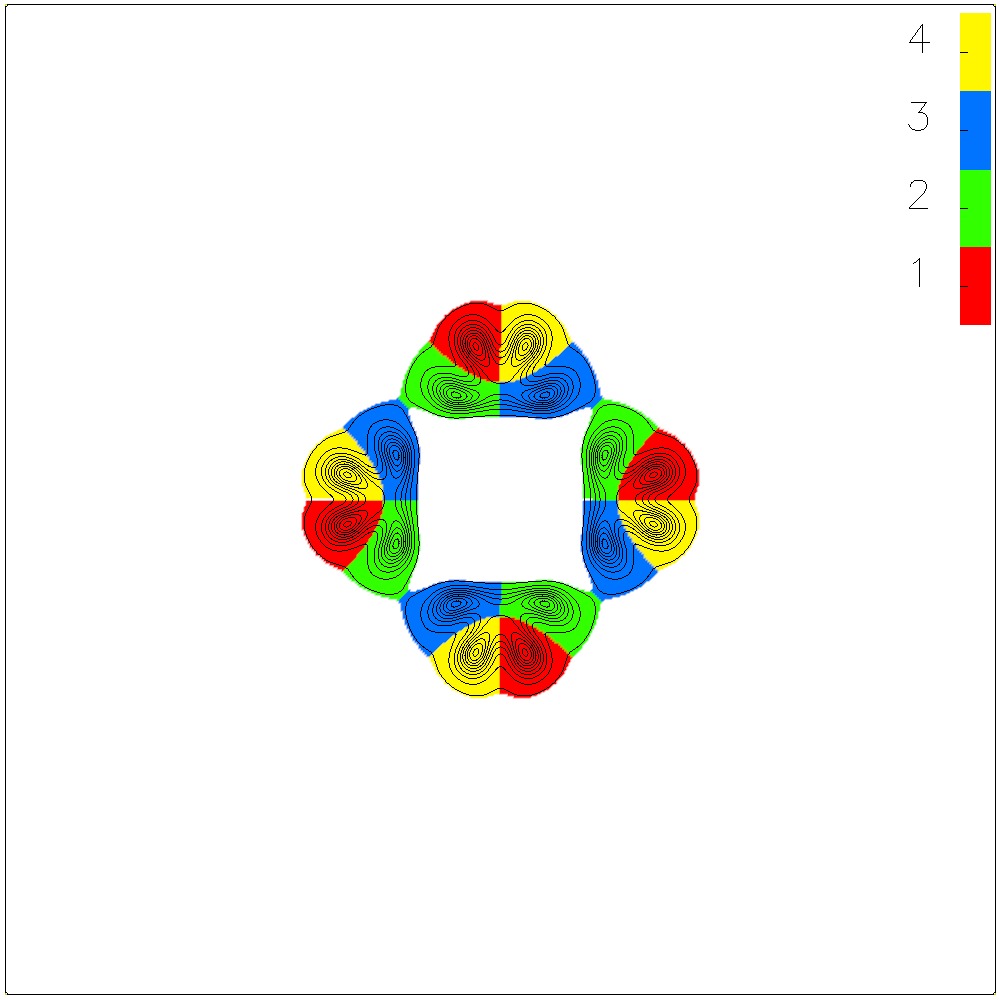} & \includegraphics[scale=0.4,natwidth=1000,natheight=1000]{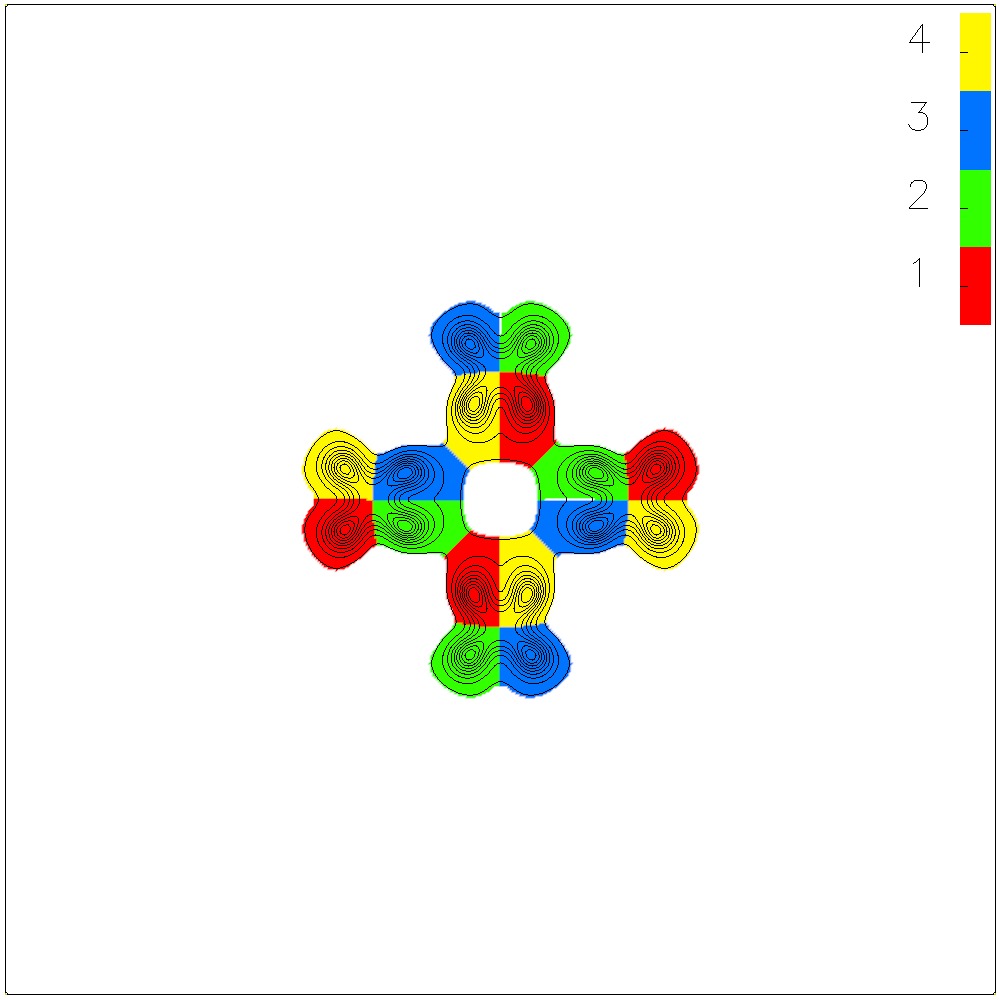} & \includegraphics[scale=0.4,natwidth=1000,natheight=1000]{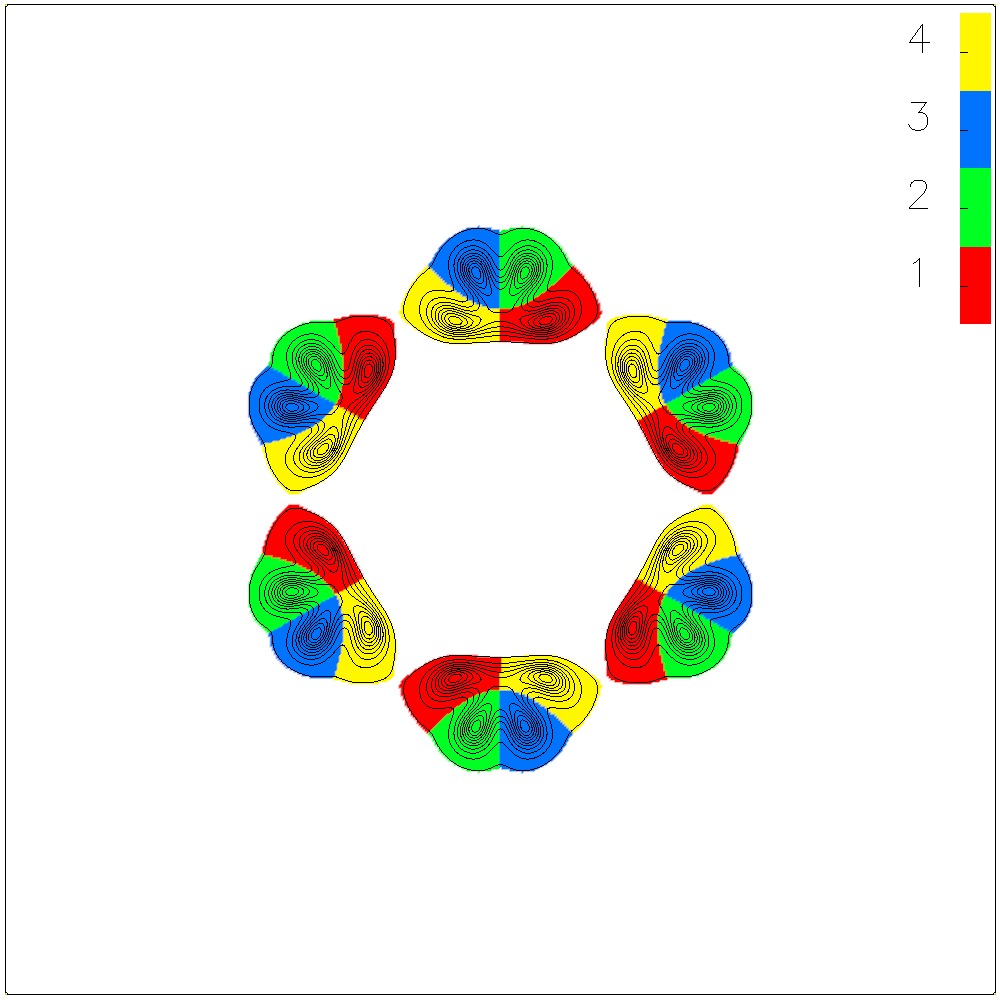}\\
(d) $N=4$, $B=4$ \includegraphics[scale=0.14,natwidth=130,natheight=125]{Images/forms/poly-fourhole.jpeg} & (e) $N=4$, $B=4$ \includegraphics[scale=0.14,natwidth=90,natheight=90]{Images/forms/poly-fourhole.jpeg} & (f) $N=4$, $B=6$ \includegraphics[scale=0.2,natwidth=90,natheight=90]{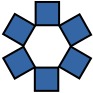}\\
\includegraphics[scale=0.4,natwidth=1000,natheight=1000]{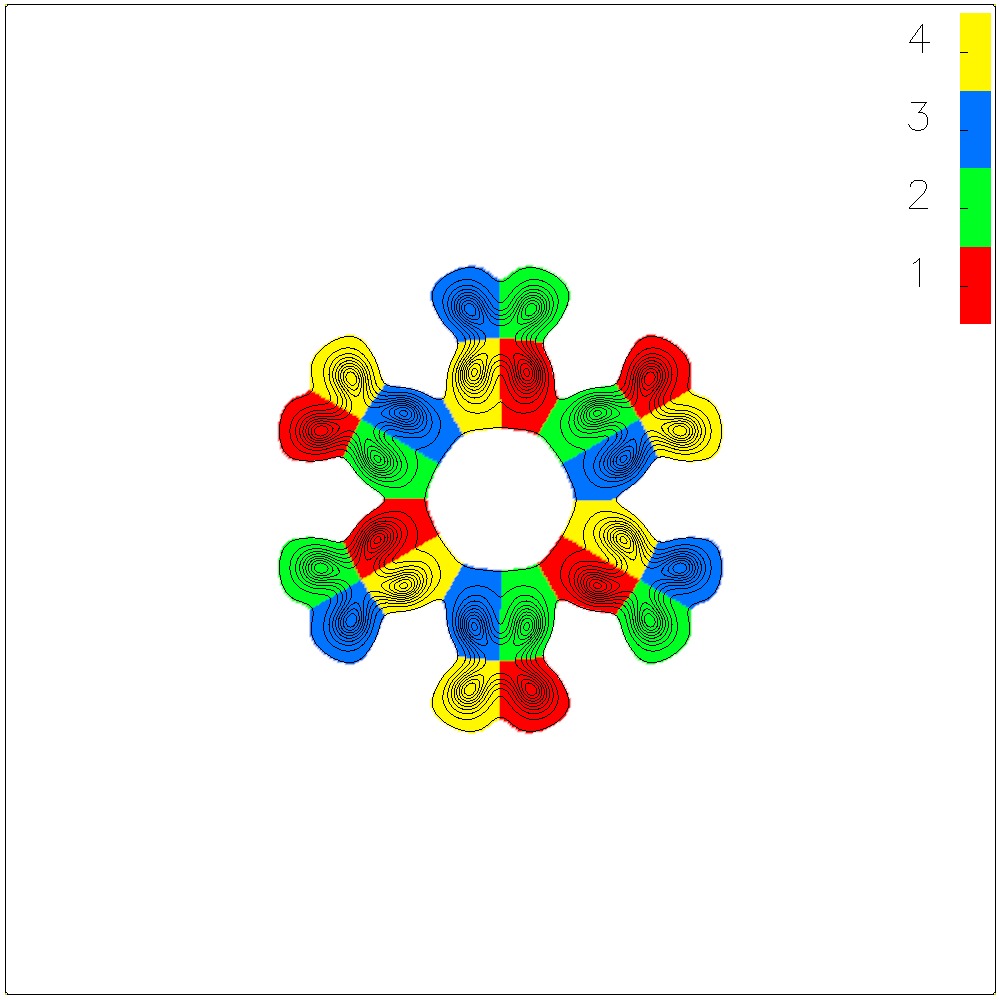} & \includegraphics[scale=0.4,natwidth=1000,natheight=1000]{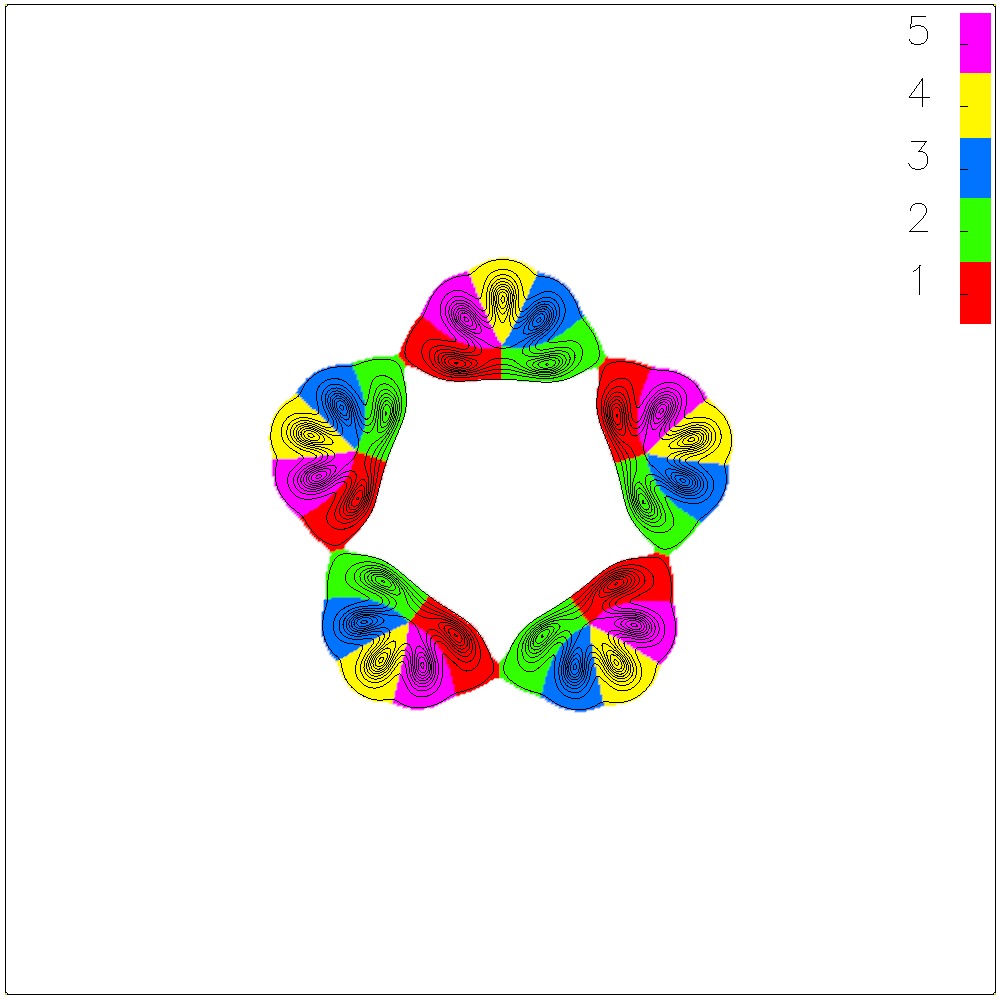} & \includegraphics[scale=0.4,natwidth=1000,natheight=1000]{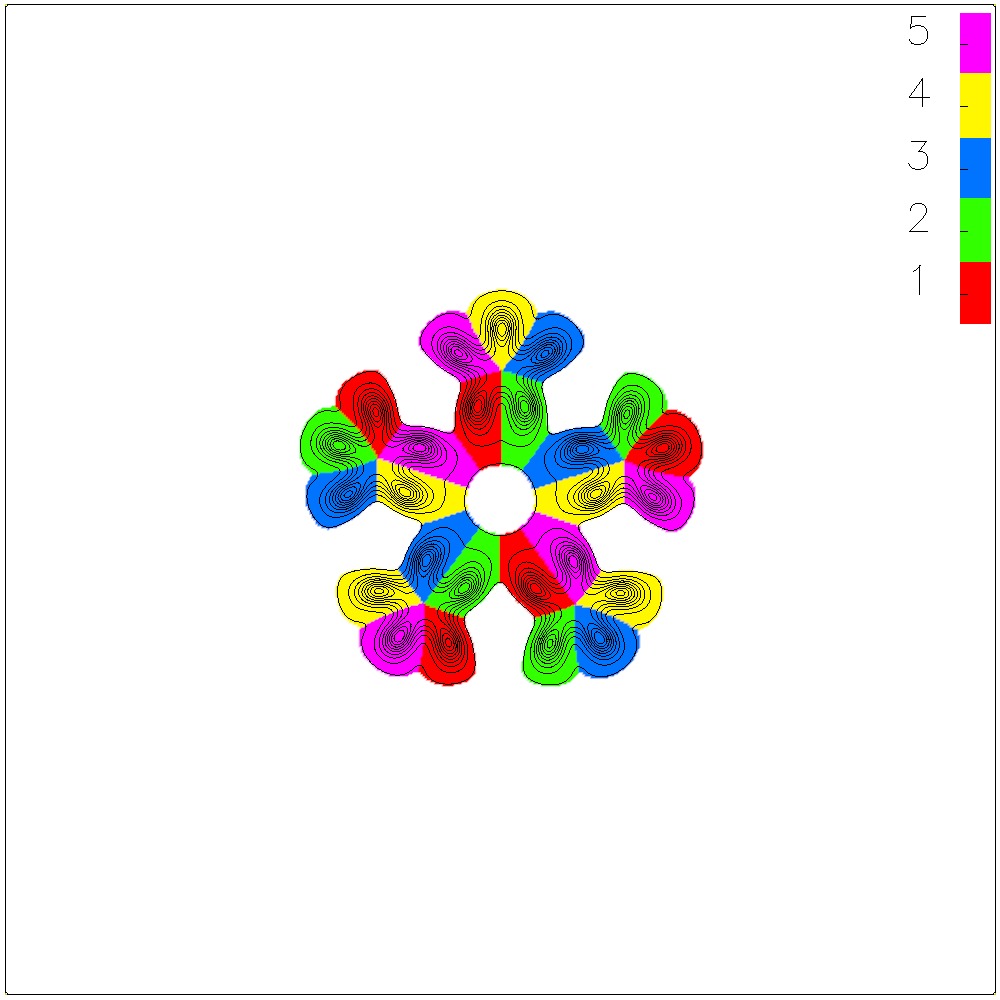} \\
(g) $N=4$, $B=6$ \includegraphics[scale=0.2,natwidth=130,natheight=125]{Images/forms/N4B6hole.jpeg} & (h) $N=5$, $B=5$ \includegraphics[scale=0.14,natwidth=130,natheight=125]{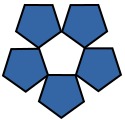} & (i) $N=5$, $B=5$ \includegraphics[scale=0.14,natwidth=130,natheight=125]{Images/forms/pent-5-hole.jpeg}\\
\includegraphics[scale=0.4,natwidth=1000,natheight=1000]{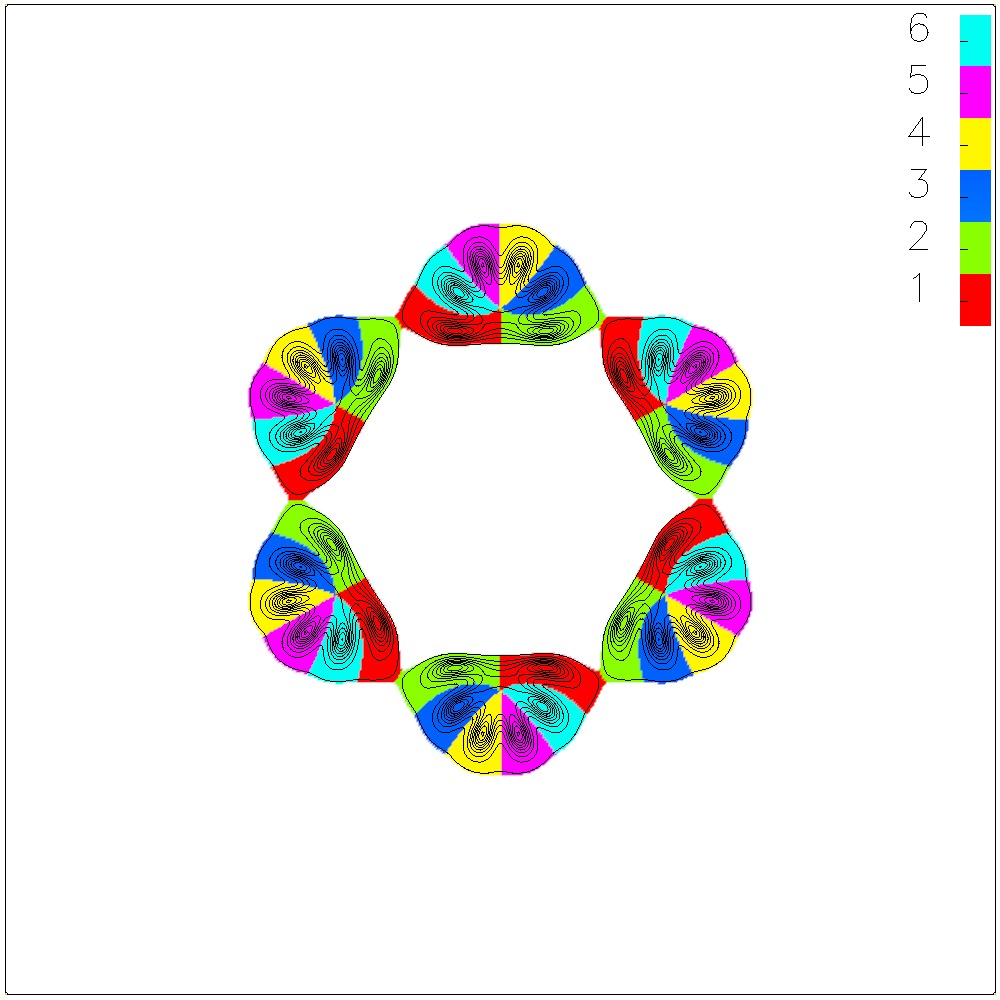} &  \includegraphics[scale=0.4,natwidth=1000,natheight=1000]{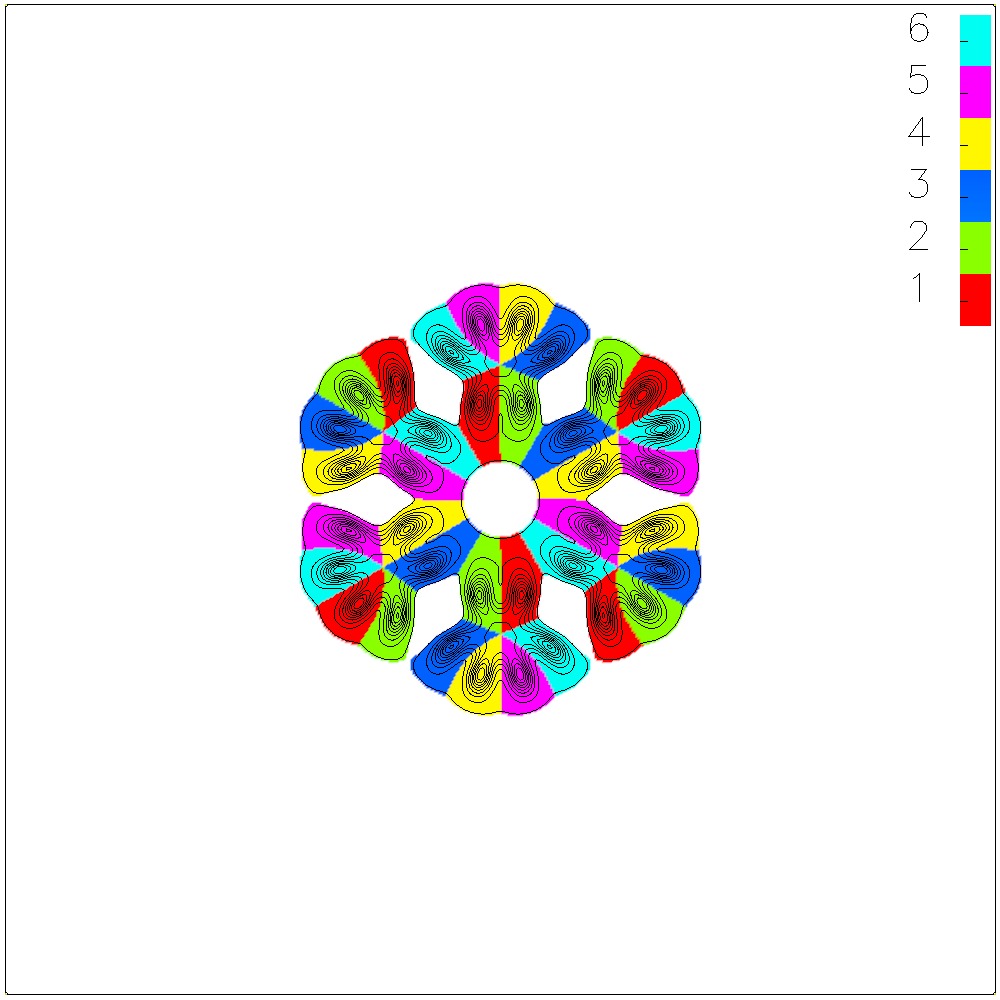} & \includegraphics[scale=0.4,natwidth=1000,natheight=1000]{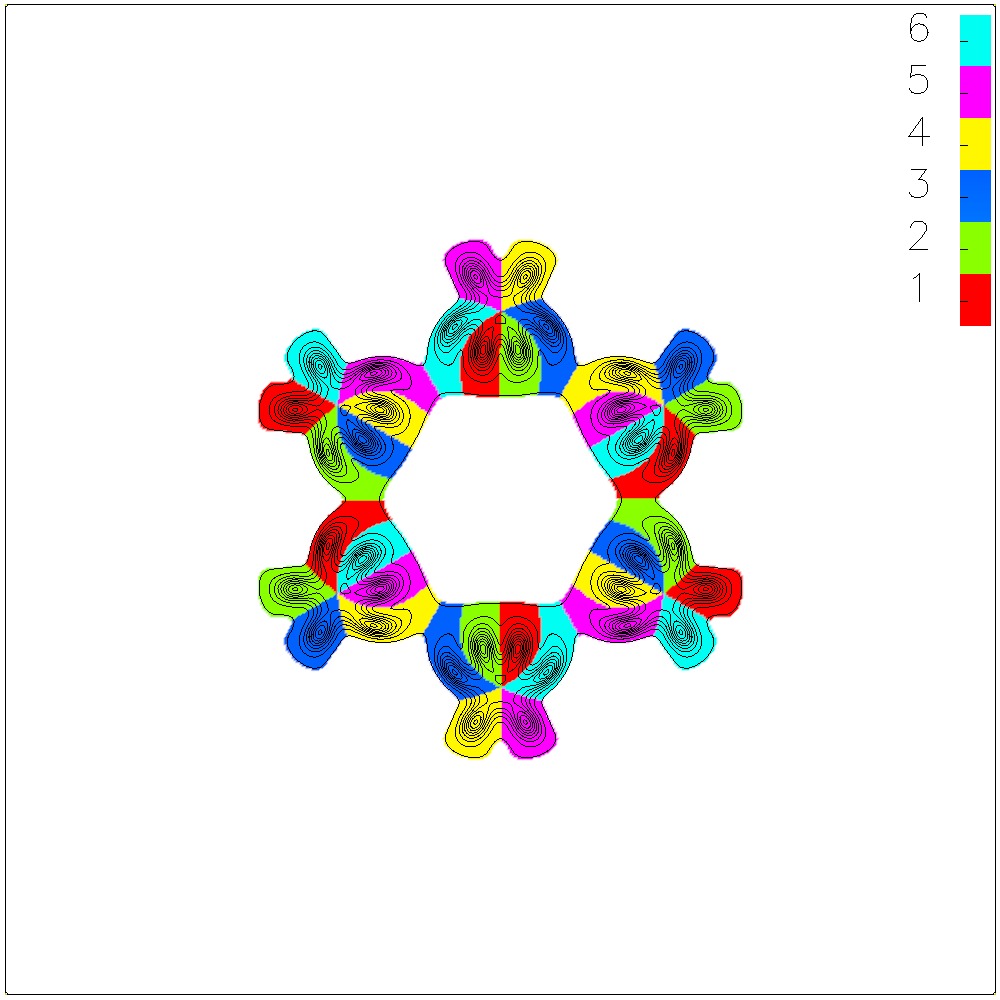}\\
(j) $N=6$, $B=6$ \includegraphics[scale=0.14,natwidth=90,natheight=90]{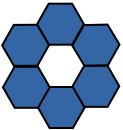} & (k) $N=6$, $B=6$ \includegraphics[scale=0.14,natwidth=90,natheight=90]{Images/forms/hex-6-1.jpeg} & (l) $N=6$, $B=6$ \includegraphics[scale=0.14,natwidth=90,natheight=90]{Images/forms/hex-6-1.jpeg}
\end{tabular}
\caption{Energy density plots detailing the various hole caveats to the predicted polyform structure.}
\label{caveats}
\end{center}
\end{figure}

The first caveat is the formation of hole like structures, which can be seen in figure \ref{caveats}. These hole solutions form higher energy local minima, that break the predicted polyform structure. There is normally only one unique hole solution for each combination of $N \geq 3$ and $B \geq 3$. The solitons have a relative spatial rotation such that the edge contributing to the hole contains alternating colours, as shown in figure \ref{caveats}(a,b,c,d,f,h,j). However if $2B\mod N=0$, we find that additional hole solutions form, with colours going sequentially round the hole as seen in \ref{caveats}(e,g,i,k,l). This can only occur for $2B \mod N= 0$ while retaining the required symmetry to stabilise the hole. Also as $N$ increases, there is no reason why more or less partons can't be contributed to the hole per soliton, as seen in \ref{caveats}(l). Note that the standard hole solutions have significantly higher energies than that of the polyform solutions for the given $B$, while the 
second hole solutions have higher energies still. 

The second caveat is the angle deformations of the $N\geq4$ polyforms. If we consider figure \ref{N4plots}(d), we can see that instead of forming a perfect angle of $\pi/2$, as we might expect for the \includegraphics[scale=0.14,natwidth=82,natheight=82]{Images/forms/poly-3-2.jpeg} shape, the angle is obtuse. This is due to the derivative terms trying to force the phase to change smoothly. This means that segments of the target space next to each other, want to be positioned next to each other spatially. Hence non-neighbouring segments will repel each other as with \includegraphics[scale=0.14,natwidth=82,natheight=82]{Images/forms/poly-3-2.jpeg} in figure \ref{N4plots}(d) and \includegraphics[scale=0.14,natwidth=82,natheight=82]{Images/forms/poly-4-2.jpeg} in figure \ref{N4plots}(g), where green and yellow lie in non-neighbouring segments. This also adds weight to our proposal that the line solutions are the global minima, as this bending pulls the shapes out into more linear structures. The most prominent 
examples of this can be seen for $N \geq 5$, for example \includegraphics[scale=0.14,natwidth=82,natheight=82]{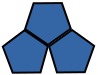} in figure \ref{N5plots}(e) and \includegraphics[scale=0.14,natwidth=82,natheight=82]{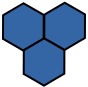} in figure \ref{N6plots}(f). This all stems from the partons themselves being able to move and hence bunch up in the soliton. So our solutions start to look further and further removed from this polyform structure even though they follow the simple rules outlined.

The final caveat occurs with $N=5,6$ only and is denoted \includegraphics[scale=0.14,natwidth=90,natheight=80]{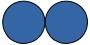}. This additional solution is similar to two warped maximally symmetric $B=2$ solitons joined in a line, as seen in figures \ref{N5plots}(m) and \ref{N6plots}(o). They are of similar energy to the line solution suggesting that as we increase our value for $N$, the lower energy solution does appear to form line structures but not necessarily of standard single solitons.

\section{Dynamics}
The goal of this section is to study the scattering of the various soliton solutions and draw parallels between our results and those of the standard planar Skyrme model \cite{Piette:dbs}. Simulations were performed using a fourth order Runge-Kutta method. These were done on a grid of 751x751 grid points with $\Delta x = 0.04$ and $\Delta t = 0.01$. Our boundary again was fixed to be the vacuum and we included a suitable damping term at the boundary to remove any kinetic energy emitted.  For each simulation we will indicate the initial relative spatial rotations denoted $\psi_0$ and positions denoted $\left(x_0 , y_0 \right)$ of each soliton. We are working in the centre of mass frame, for example for $B=2$ the velocities of the solitons are equal and opposite.

One notional aspect of these scatterings is what we can class as a soliton escaping to infinity. The natural position to take is if the soliton escapes to a point such that the boundary starts to have a significant damping effect on the velocity of the soliton. By slowly moving a soliton we estimate this to be at a distance 5 from the boundary. Hence if a soliton escapes to this line we will class it as having escaped to infinity for all intents and purposes. 

\subsection{$B=2$ scattering}
As you may expect, when given zero velocity the two solitons will attempt to align themselves into the attractive channel. Hence if aligned with $\psi_0=\pi$ the solitons will remain in the attractive channel. However, unlike the standard potential there are additional terms beyond leading order, which cause the solitons to want to be aligned face to face. This only has a significant effect at short range, as shown in figure \ref{anglescattering} in Appendix B.

We are now interested in the head-on collision in the attractive channel with various initial velocities. We place the solitons at $(6.0,0)$ and $(-6.0,0)$, using a range of velocities $0.1 \leq v \leq 0.6$. 

As the solitons collide, they initially form the maximally symmetric solution seen in the $B=2$ static case. They then emerge at $\pi/2$ to their initial direction of motion. This is the same as with the standard potential however what differs is the scattering process itself. As the two solitons collide we can consider the scattering in terms of individual partons. The derivative terms in the energy mean that the change in phase wants to be minimised. Due to this, like colours can in fact overlap, however different colours will have a natural separation, based upon how far away their segments are in the target space. 

Using the above we can predict what will occur in scattering processes, for example in a head-on collision in the attractive channel there are three situations that can occur, based upon the colour of the partons involved in the interaction. 
\begin{itemize}
 \item \emph{like colours} - These partons will cross over each other and scatter at an angle bisecting the incident angles. So for two incident like colours with opposite velocities they will scatter at $\pi/2$.
 \item \emph{sequential colours} - These partons want to lie next to each other, but cannot overlap, leading to the partons approaching each other and then stopping. As they are now the optimal distance apart they will bond together. Assuming the pair of sequential partons can then move off with enough other partons to form an integer charge soliton they will do so. Otherwise they will return to the original soliton they were a part of. 
 \item \emph{non-sequential colours} - These partons do not want to lie next to each other due to a sharper change in phase. Hence they have a larger natural distance and will stop before they approach each other. They will follow the path of the sequential partons they are already bound to when scattering. 
\end{itemize}

\begin{figure}
\centerline{\begin{tabular}{c c c c}
\includegraphics[scale=0.4,natwidth=1000,natheight=1000]{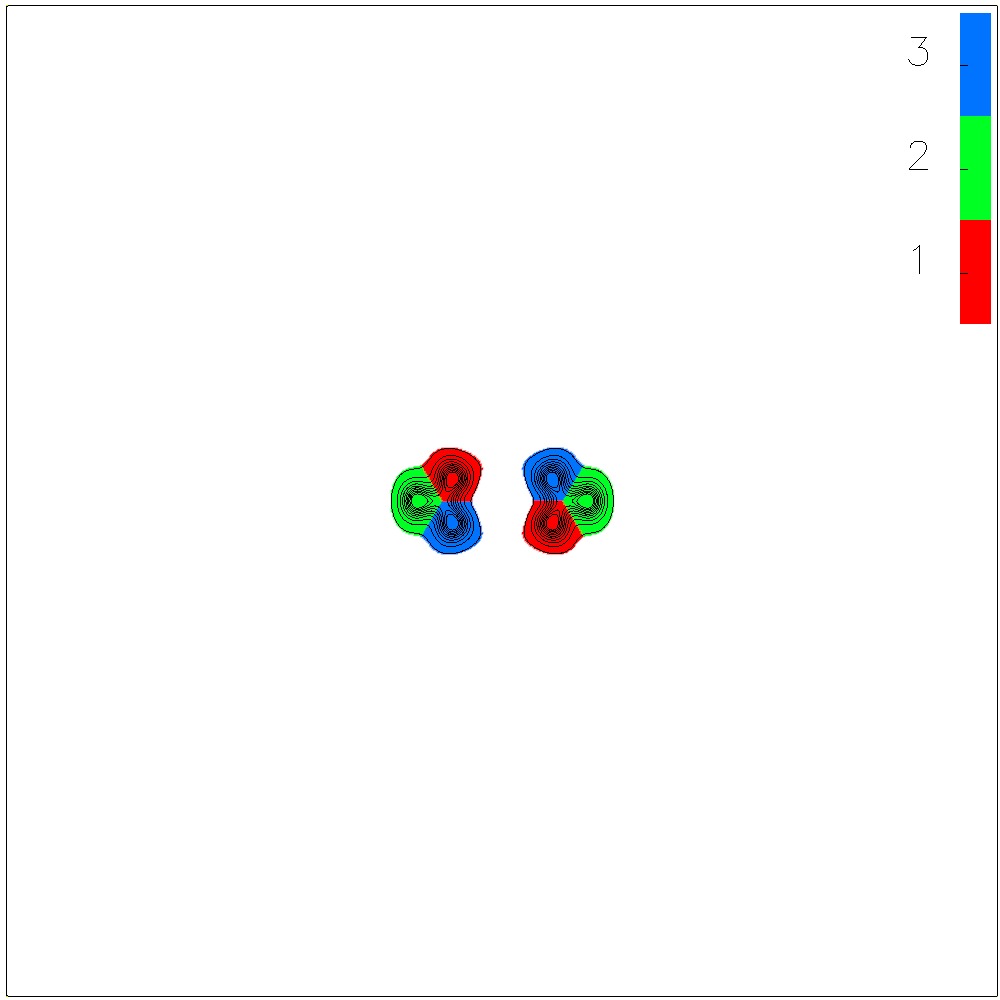} & \includegraphics[scale=0.4,natwidth=1000,natheight=1000]{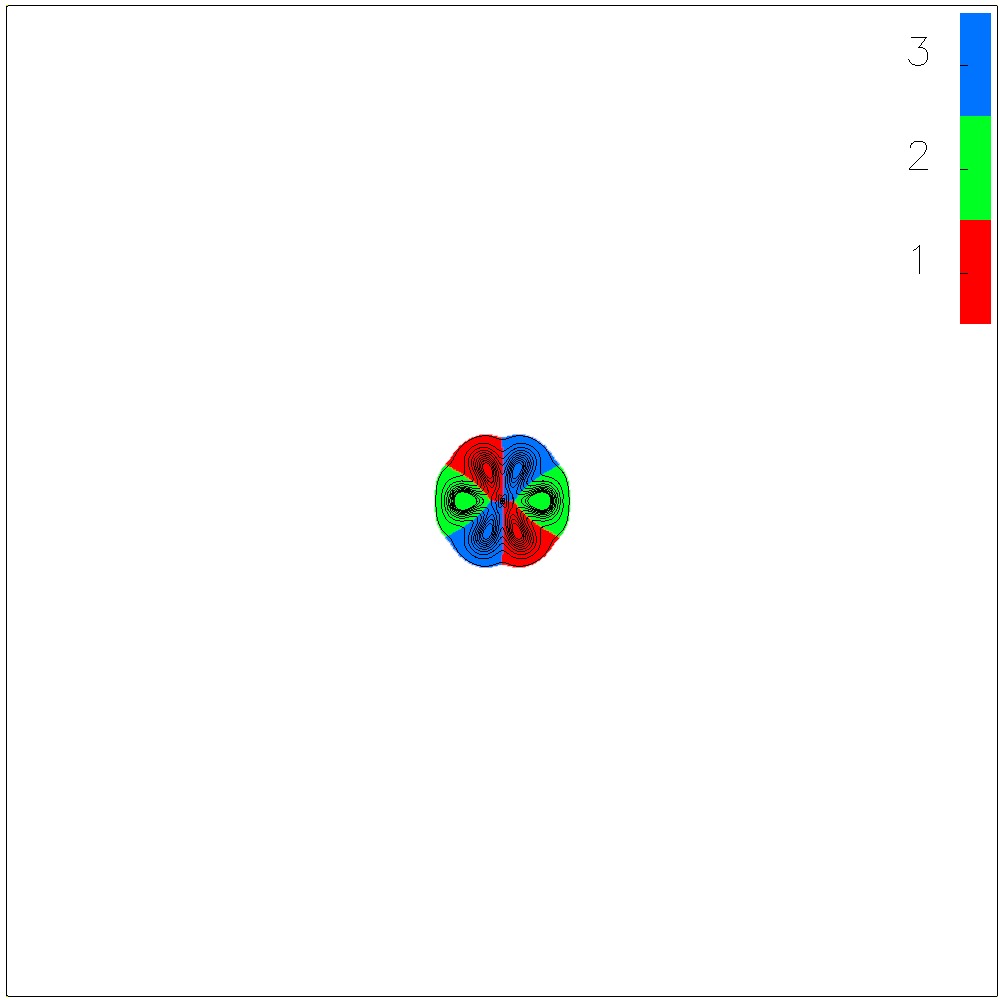} & \includegraphics[scale=0.4,natwidth=1000,natheight=1000]{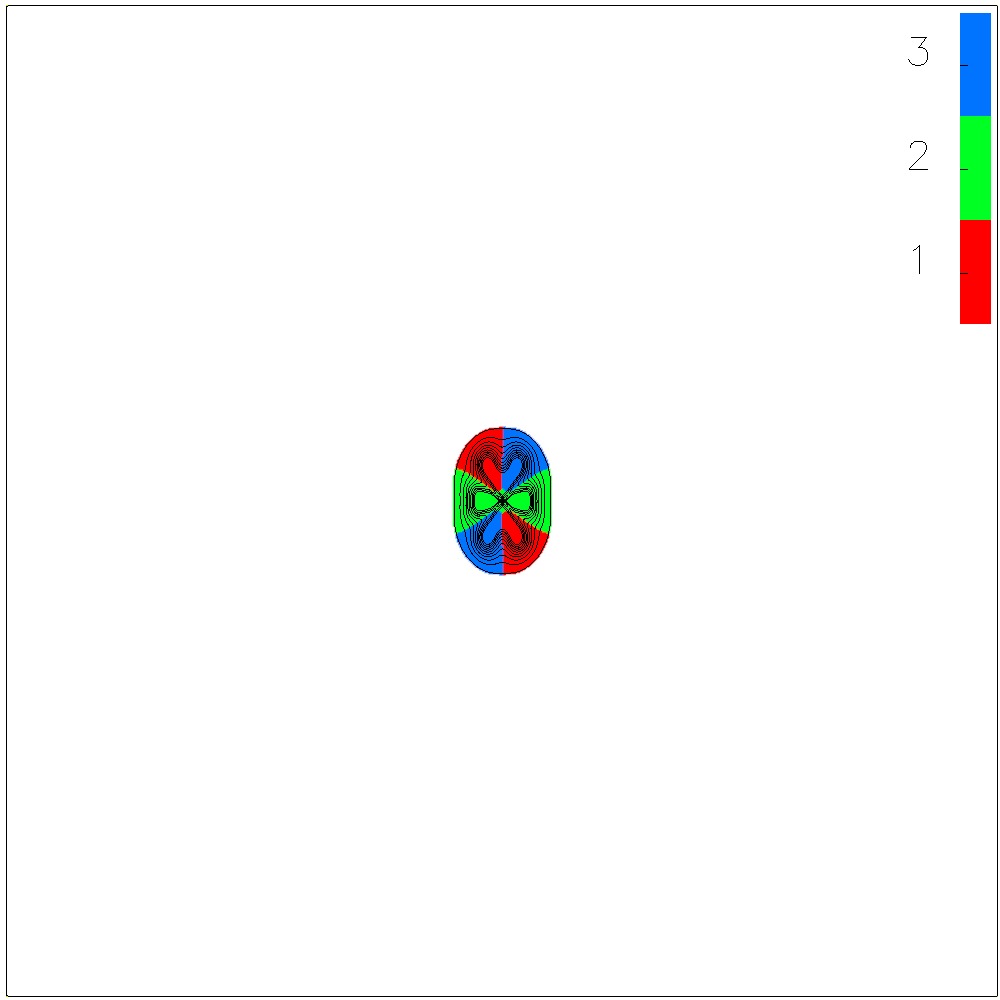} & \includegraphics[scale=0.4,natwidth=1000,natheight=1000]{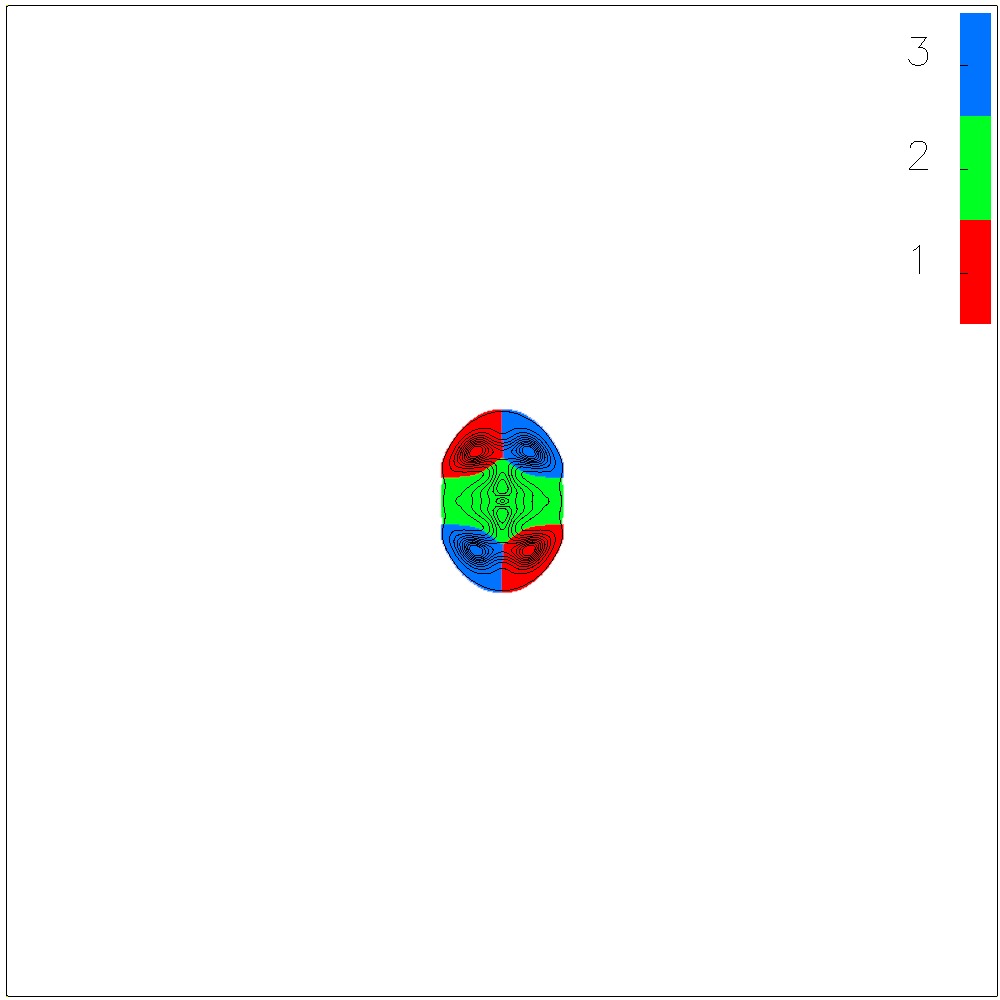}\\
$t = 8.5$ & $t = 11$ & $t = 12$ & $t = 14$\\
\includegraphics[scale=0.4,natwidth=1000,natheight=1000]{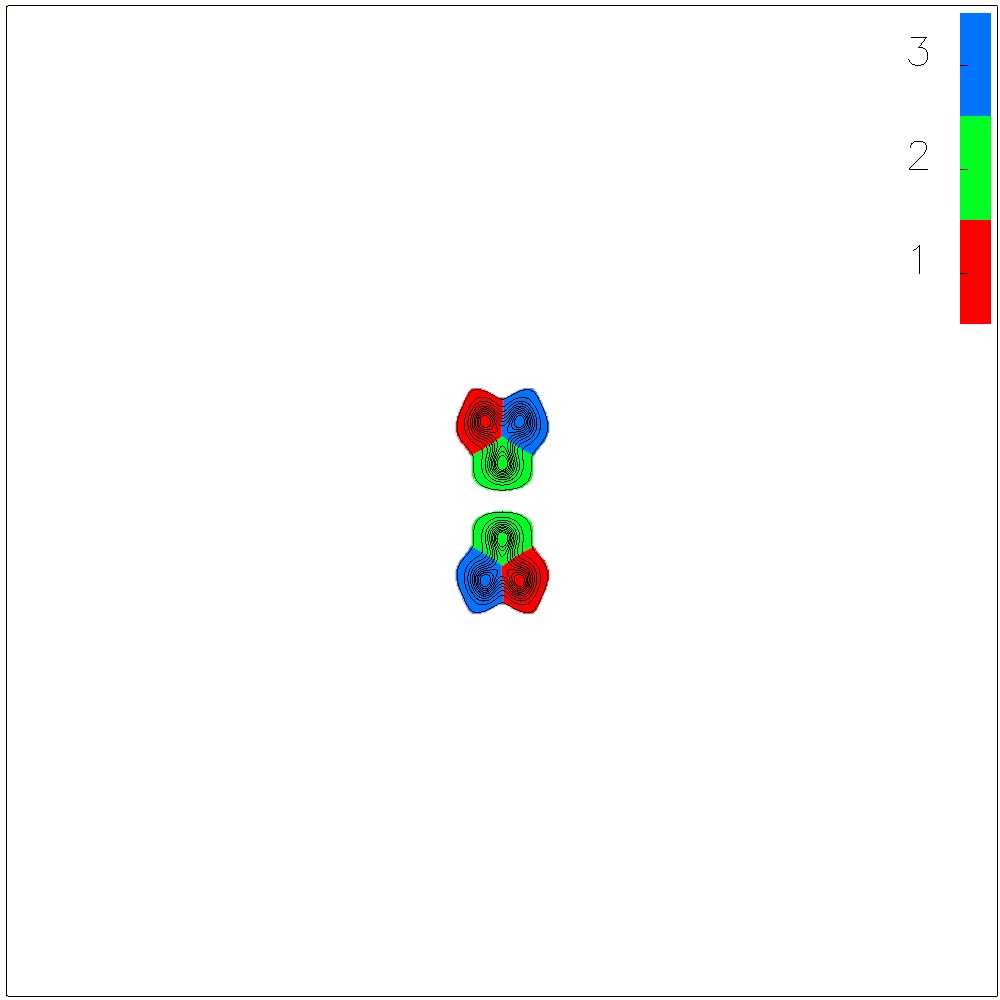} & \includegraphics[scale=0.4,natwidth=1000,natheight=1000]{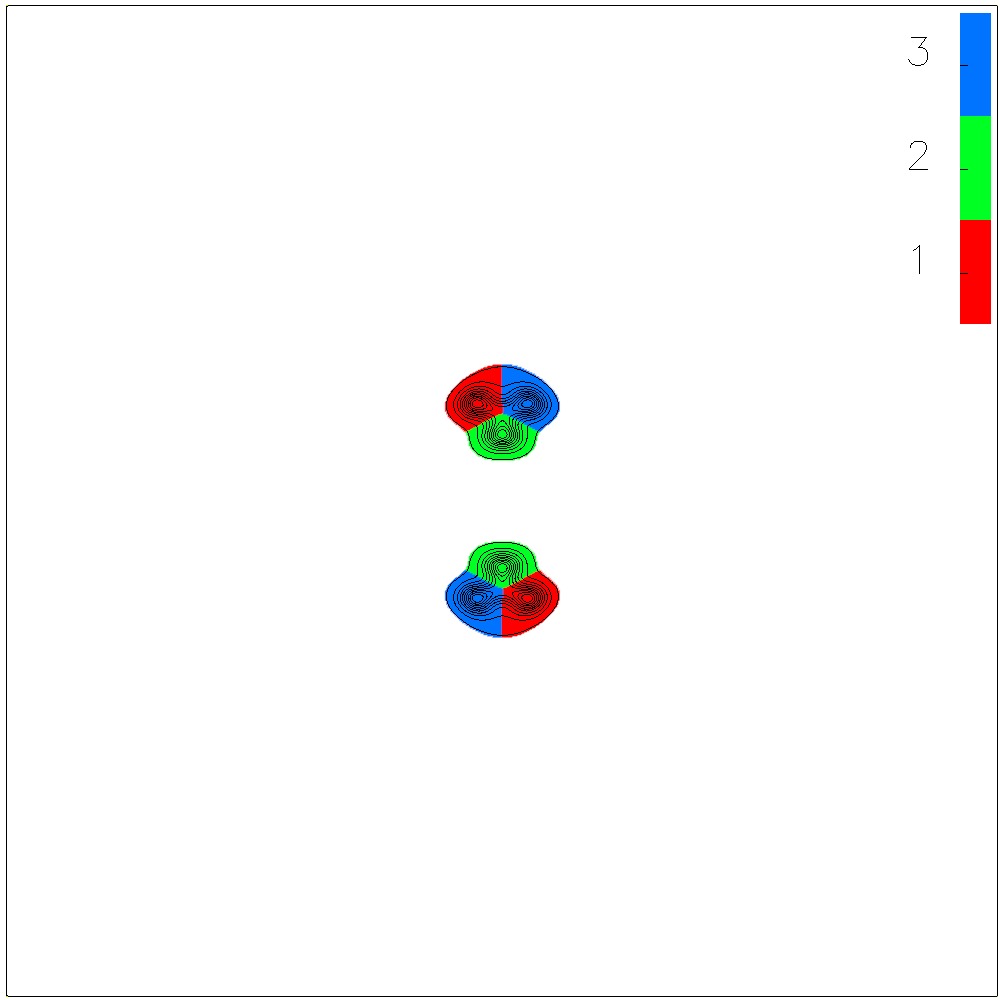} & \includegraphics[scale=0.4,natwidth=1000,natheight=1000]{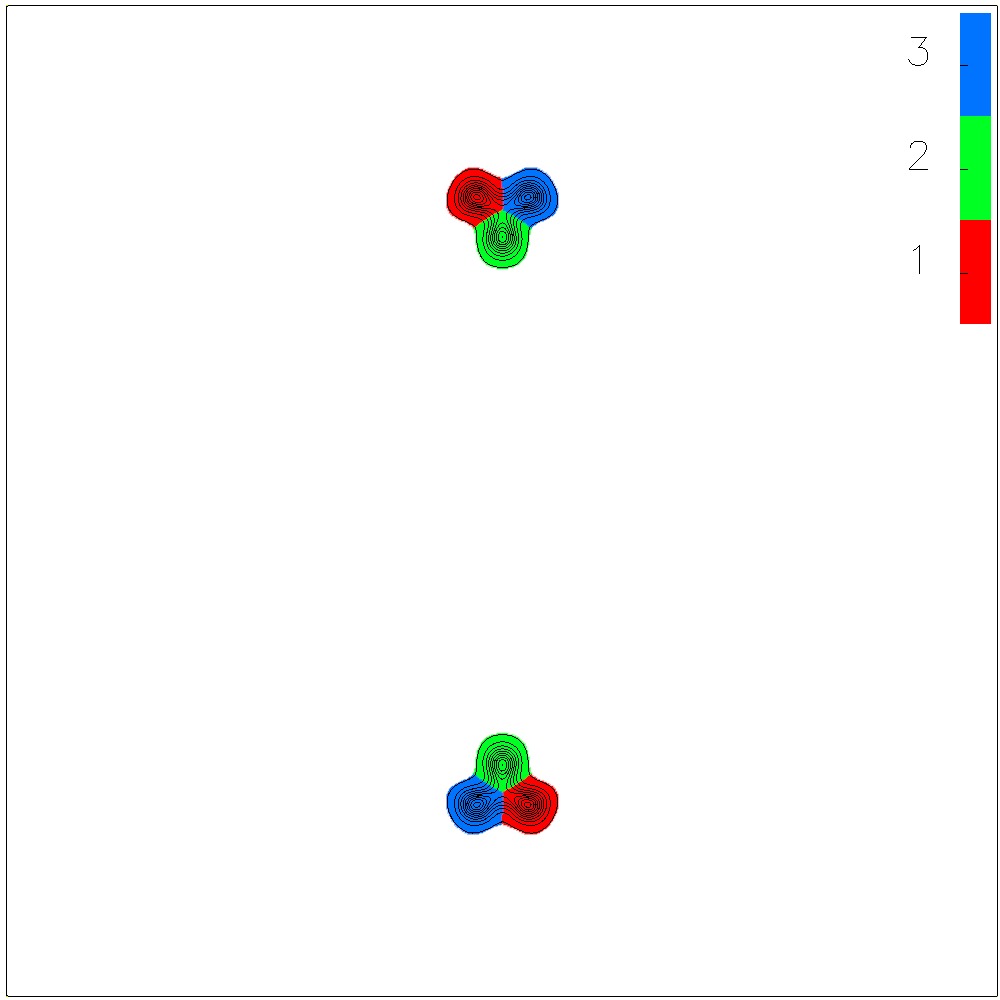} & \includegraphics[scale=0.26,natwidth=1000,natheight=1000]{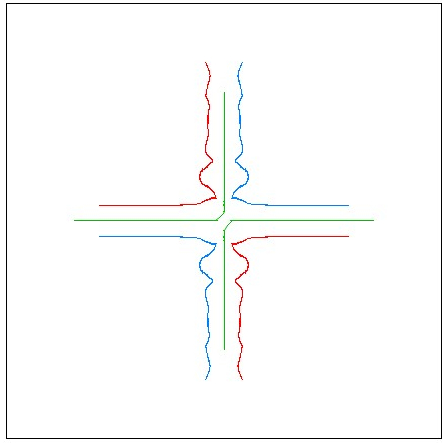}\\
$t = 17.5$ & $t = 20.5$ & $t = 50$ & parton tracks
\end{tabular}}
\caption{Energy density plots at various times during the scattering of two $N=3$ single solitons each with speed 0.4 and with relative spatial rotation of $\pi$}
\label{scattering}
\end{figure}

So scattering processes are determined by the like and sequential colours that meet. If we look at the scattering shown in figure \ref{scattering} we see first the two sets of sequential colours coming together and stopping as predicted. The green partons continue to move, first forming the $B=2$ maximally symmetric solution and then continuing on to overlap and scatter at $\pi/2$. As the sequential colours are currently close enough to be bonded with either of the sequential colours next to it, it is the path of the green partons that will determine which pairs will form the single solitons. Hence the green partons bond with one of the other bonded pairs to form a complete soliton, thus scattering at $\pi/2$.

\begin{figure}
\centerline{\begin{tabular}{c c c c}
\includegraphics[scale=0.4,natwidth=1000,natheight=1000]{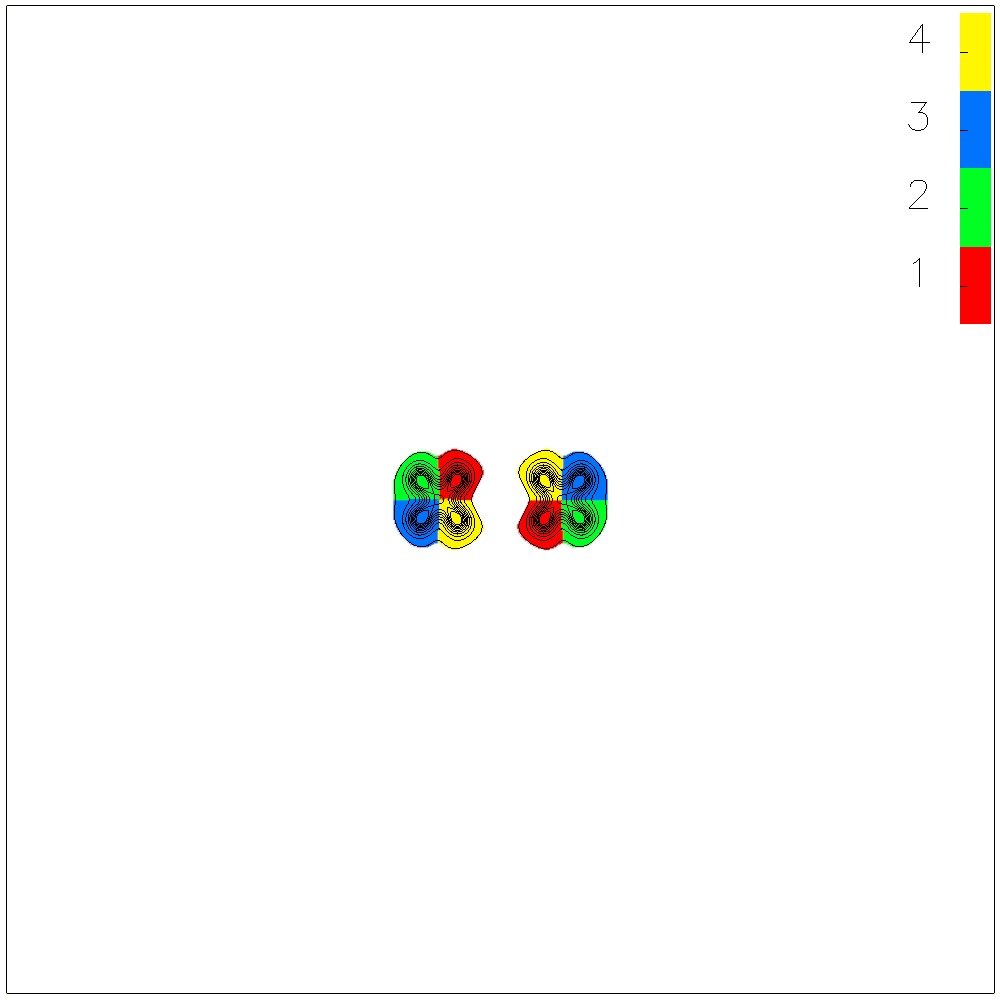} & \includegraphics[scale=0.4,natwidth=1000,natheight=1000]{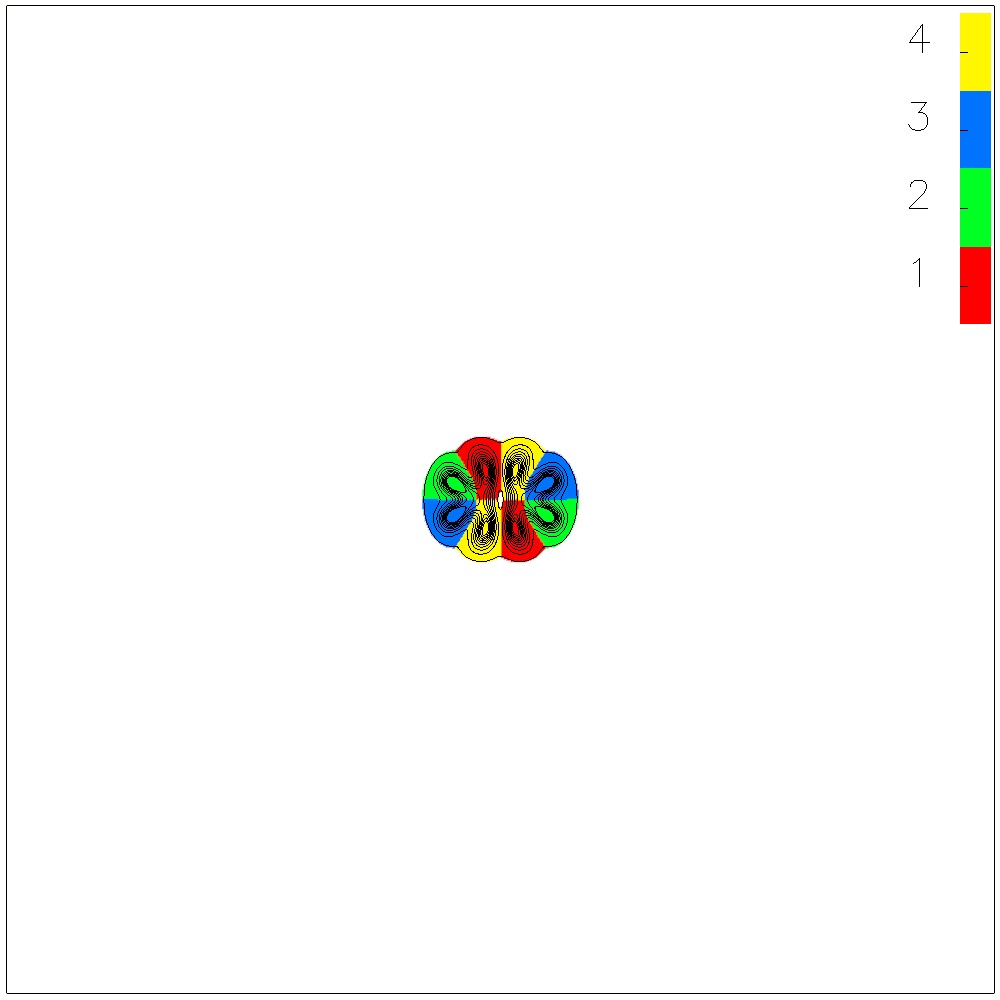} & \includegraphics[scale=0.4,natwidth=1000,natheight=1000]{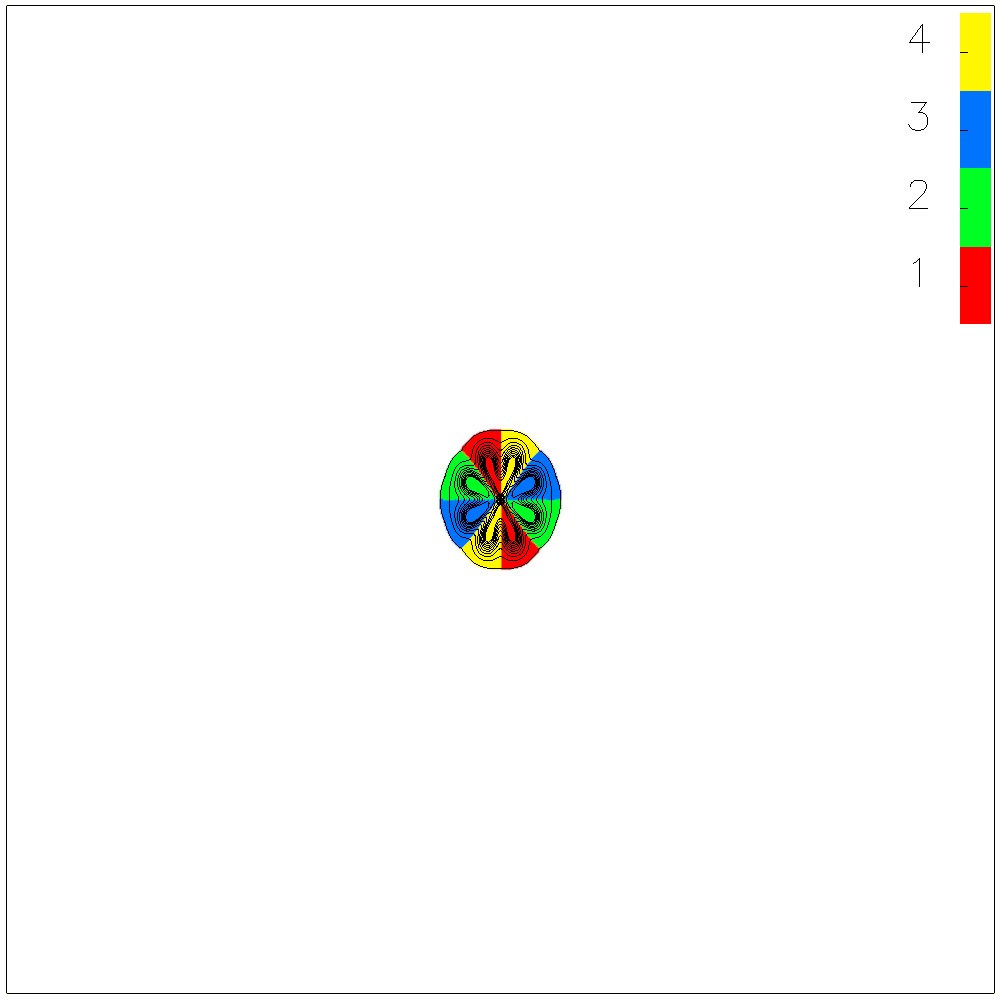} & \includegraphics[scale=0.4,natwidth=1000,natheight=1000]{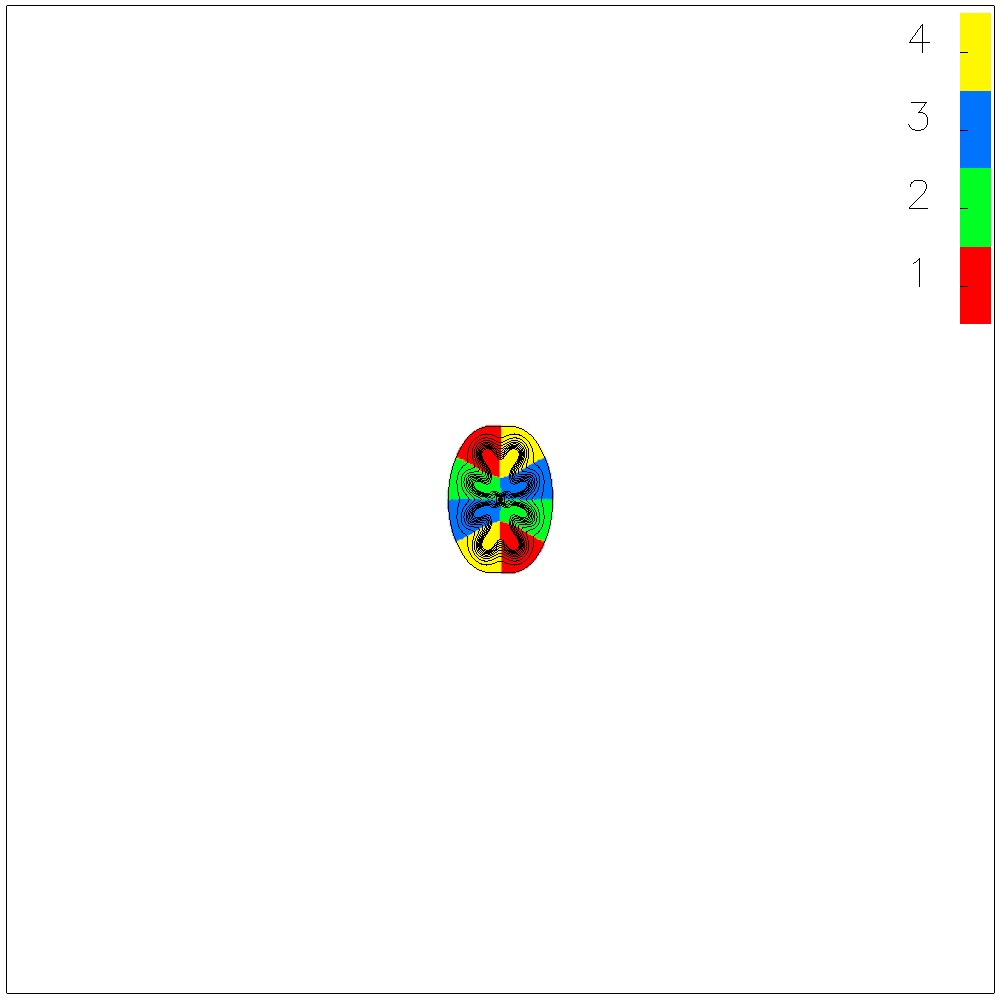}\\
$t = 8.5$ & $t = 10$ & $t = 11$ & $t = 12$\\
\includegraphics[scale=0.4,natwidth=1000,natheight=1000]{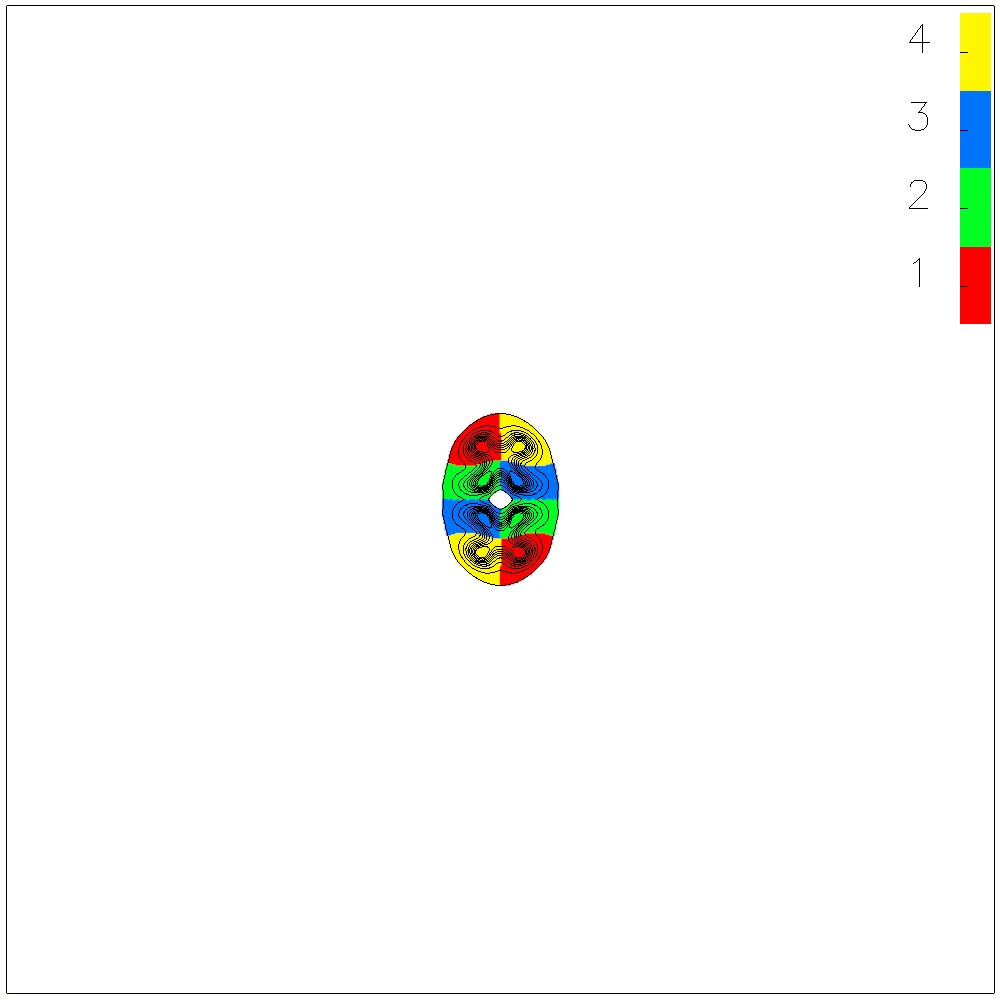} & \includegraphics[scale=0.4,natwidth=1000,natheight=1000]{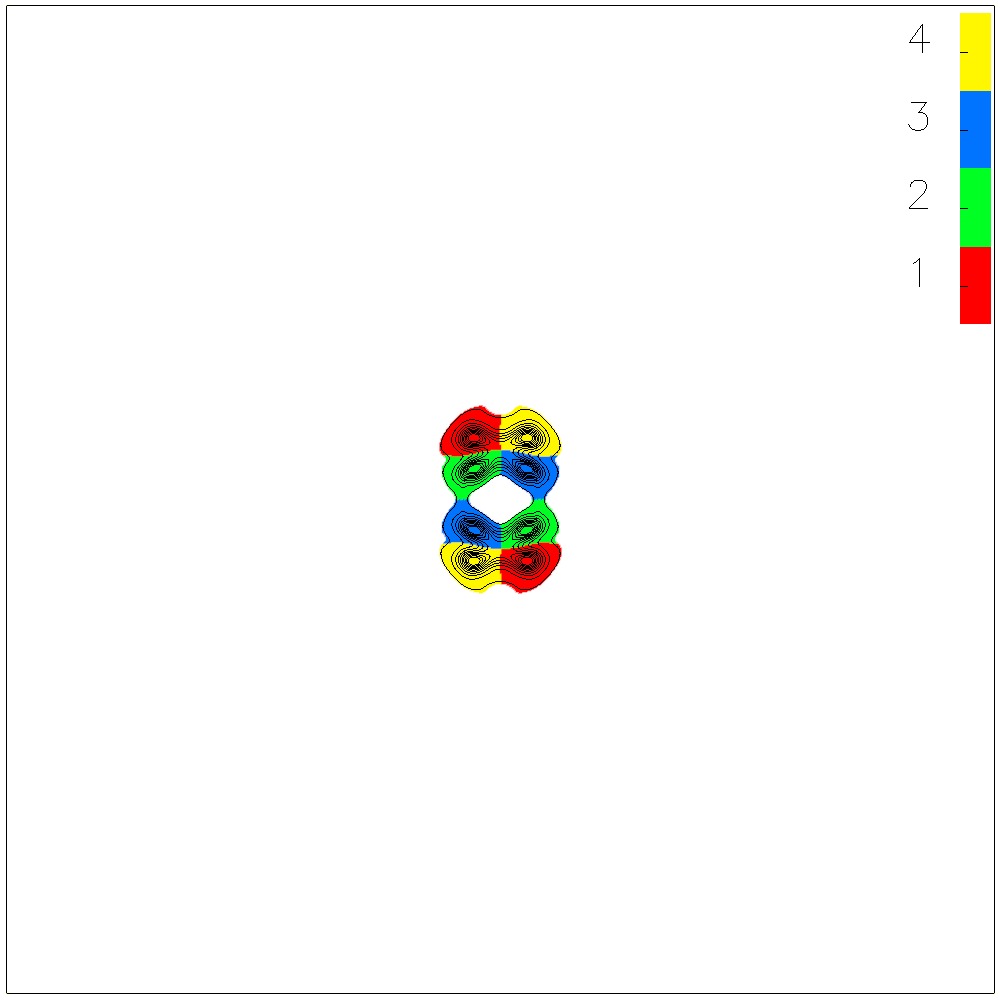} & \includegraphics[scale=0.4,natwidth=1000,natheight=1000]{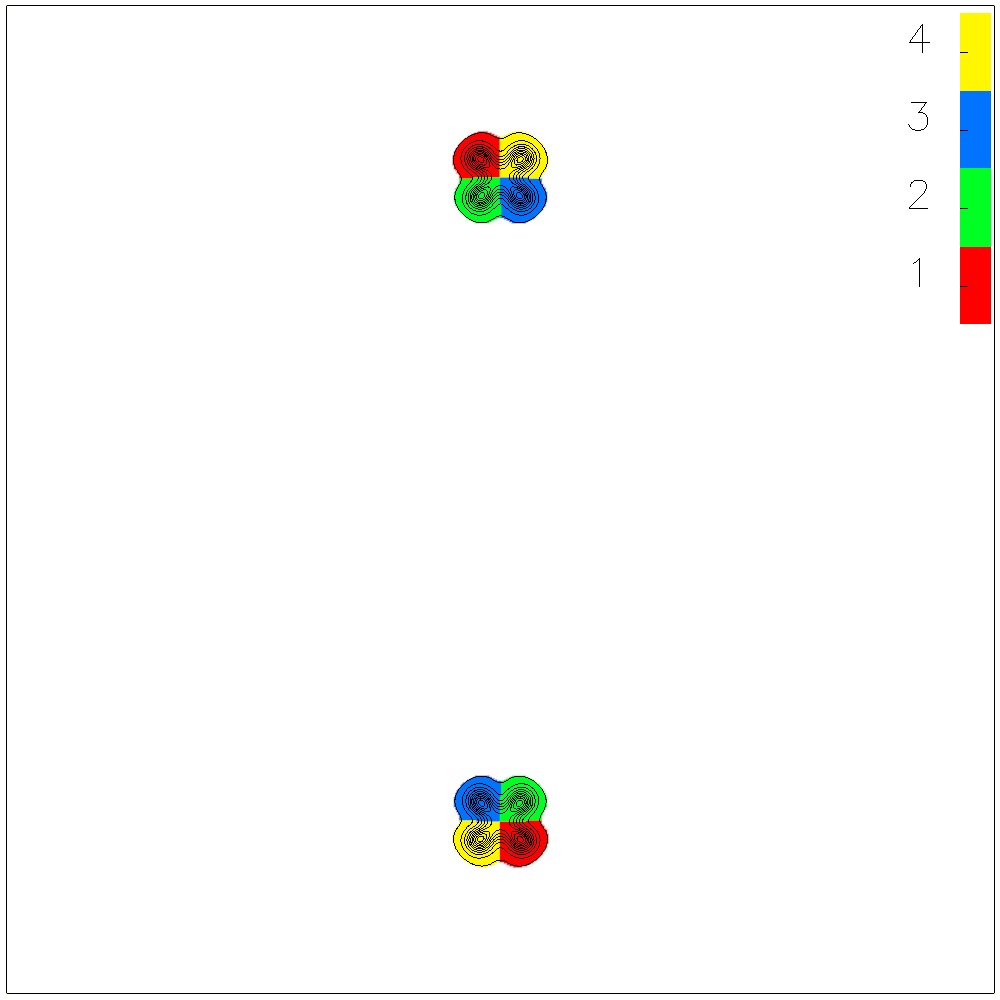} & \includegraphics[scale=0.26,natwidth=1000,natheight=1000]{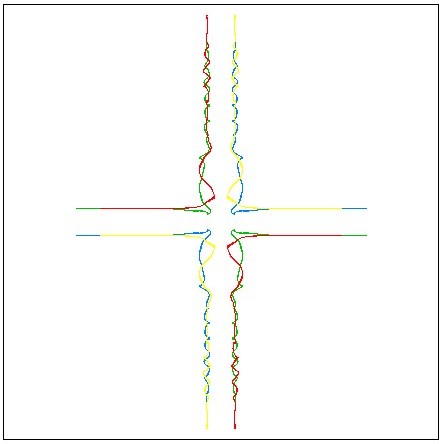}\\
$t = 13$ & $t = 14$ & $t = 30$ & parton tracks
\end{tabular}}
\caption{Energy density plots at various times during the scattering of two $N=4$ single solitons each with speed 0.4 and with relative spatial rotation of $\pi$}
\label{scatteringN4}
\end{figure}

If we now look at the scattering process in figure \ref{scatteringN4} we see only sequential colours meeting. These bond together to form two solitons from different partons. 

In our model we observe a large quantity of kinetic radiation emitted when this intermediate state of the maximally symmetric solution is formed. This radiation significantly reduces the energy from the colliding solitons meaning the escape velocity ($v_e$) of the process is quite high (for the processes we looked at a range of about $0.3$ to $0.5$ was measured). It is also dependent upon the orientation of the solitons in the initial conditions. If we consider the case $v<v_e$, after the collision the attractive forces of the solitons pulls them so they re-collide. The form of this second collision is the time reversal of the original collision however with a smaller velocity. It is also accompanied by the emission of kinetic radiation, and this process will continue until the solitons don't have the kinetic energy to escape the intermediate state of the maximally symmetric solution. The solitons are no longer distinct, and the motion looks more like the excitation of the 2-soliton solution. 

\subsection{$B\geq3$ scattering}
For more than two solitons the scattering processes are a little more complicated but can still be broken down into these simple parton-parton scattering structures discussed above. If we look at the scattering of $N=3$ $B=3$ in figure \ref{3scattering}, we see that it continues to follow the simple rules outlined in the previous subsection. The initial partons meet in the centre scattering at $\frac{2\pi}{3}$, (bisecting the angle of approach relative to each other). The other partons then bond with their neighbour as they sit next to each other in the target space and are dragged off with the blue partons emitted from the centre. Note that a point first scattering is possible, as the attractive asymptotic contribution from edges cancels. This pattern continues for higher values of $N$ and $B$.

\section{Conclusions}
The broken potential breaks the global symmetry to the dihedral group $D_N$. This results in a single soliton composed of $N$ topologically confined partons represented by different colors. We have also extended previous work to demonstrate that multi-soliton solutions take the form of polyforms for all values of $N$. An interesting extension to this would be to consider the soliton lattice formed by tiling these solutions. This was done by J\"{a}ykk\"{a} \emph{et al.} \cite{Jaykka:2011ic} for $N=3$ and as expected the cell was found to be the single soliton which was then tessellated in a cell similar to the standard planar Skyrme model. For those $N$-gons that tessellate (e.g. $N = 4$ or $6$) this is likely to produce similar results as the $N=3$ results, but with some differences due to the corner caveats discussed in section 3.3. Some clues are given in the solutions \includegraphics[scale=0.14,natwidth=82,natheight=82]{Images/forms/poly-4-4.jpeg} and \includegraphics[scale=0.14,natwidth=82,natheight=82]{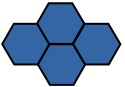} in figures \ref{N4plots}(i) and \ref{N6plots}(m) respectively. For those solutions that don't tessellate, the solution is expected to be more complicated.  

The dynamics of the model was also shown to be classically dependent upon the number of colours $N$. Each scattering process can be understood by considering the separate behaviour of the partons themselves. Additionally  we see that the short range forces differ from the standard model, as it is energetically favourable for edges to be aligned. 

The natural extension to this paper is the analogue in the full $(3+1)$-dimensional Skyrme model. The idea of being able to consider a scattering process by looking at the constituent makeup of the soliton, should transfer to the full model. However if an analogous symmetry breaking potential is constructed in the Skyrme model, we have the physical consequence that isospin symmetry is broken. It is not clear what the physical consequences of this would be.

\section{Acknowledgements}
Paul Jennings would like to thank STFC and Thomas Winyard EPSRC for PhD studentships. We would also like to thank our supervisor Paul Sutcliffe for useful discussions.

\bibliographystyle{prsty}
\bibliography{bbs}
\appendix
\section{Appendix A: Static Solitons for $N=5,6$}
This section contains the static solutions along with their energies for $N=5,6$ upto $B=4$. These results further confirm our predictions but also introduce some interesting caveats which are covered in the caveats section of the paper. 

\begin{table}[h]
\caption{The energy for soliton solutions and their symmetry group $G$ for $B \leq 4$ and (left) $N = 5$ (right) $N=6$}
\begin{center}
\begin{tabular}{c c c}
\begin{tabular}{c c c c c c}
$B$ & form & $E$ & $E/B$ & $G$ & figure \\
\hline
1 & \includegraphics[scale=0.15,natwidth=50,natheight=45]{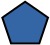} & 34.41 & 34.41 & $D_5$ &\ref{N5plots}(a)\\
2 & \includegraphics[scale=0.15,natwidth=45,natheight=45]{Images/forms/maximal.jpg} & 65.19 & 32.59 & $D_{10}$ &\ref{N5plots}(b)\\
3 & \includegraphics[scale=0.15,natwidth=70,natheight=45]{Images/forms/maximal2.jpg} & 99.23 & 33.23 & $D_{15}$ &\ref{N5plots}(c)\\
3 & \includegraphics[scale=0.15,natwidth=55,natheight=55]{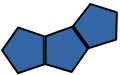} & 97.68 & 32.56 & $C_1$ &\ref{N5plots}(d)\\
3 & \includegraphics[scale=0.15,natwidth=90,natheight=45]{Images/forms/pent-3-2.jpeg} & 98.14 & 32.71 & $C_1$ &\ref{N5plots}(e)\\
4 & \includegraphics[scale=0.15,natwidth=90,natheight=80]{Images/forms/maximal3.jpg} & 137.20 & 34.30 & $D_{20}$ &\ref{N5plots}(f)\\
4 & \includegraphics[scale=0.15,natwidth=90,natheight=80]{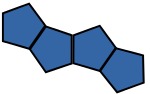} &  129.62 &  32.40 & $D_2$ &\ref{N5plots}(g)\\
4 & \includegraphics[scale=0.15,natwidth=65,natheight=65]{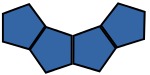} & 129.61 & 32.40 & $C_1$ &\ref{N5plots}(h)\\
4 & \includegraphics[scale=0.15,natwidth=110,natheight=45]{Images/forms/pent-4-3.jpeg} & 130.69 & 32.67 & $C_1$ &\ref{N5plots}(i)\\
4 & \includegraphics[scale=0.15,natwidth=90,natheight=80]{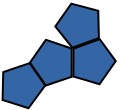} & 130.06 & 32.52 & $C_1$ &\ref{N5plots}(j)\\
4 & \includegraphics[scale=0.15,natwidth=90,natheight=80]{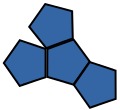} & 131.11 & 32.78 & $C_1$ &\ref{N5plots}(k)\\
4 & \includegraphics[scale=0.15,natwidth=90,natheight=80]{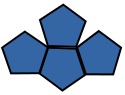} & 131.50 & 32.87 & $C_1$ &\ref{N5plots}(l)\\
4 & \includegraphics[scale=0.15,natwidth=90,natheight=80]{Images/forms/B2.jpeg} & 129.65 & 32.41 & $D_2$ &\ref{N5plots}(m)\\
\hline

\label{N4energytable}
\end{tabular} & ~~ &
\begin{tabular}{c c c c c c}
$B$ & form & $E$ & $E/B$ & $G$ & figure \\
\hline
1 & \includegraphics[scale=0.14,natwidth=43,natheight=44]{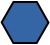} & 34.26 & 34.26 & $D_6$ & \ref{N6plots}(a)\\
2 & \includegraphics[scale=0.14,natwidth=45,natheight=45]{Images/forms/maximal.jpg} & 64.88 & 32.44 & $D_{12}$ & \ref{N6plots}(b)\\
3 & \includegraphics[scale=0.15,natwidth=55,natheight=55]{Images/forms/maximal2.jpg} & 99.23 & 33.08 & $D_{18}$ & \ref{N6plots}(c)\\
3 & \includegraphics[scale=0.14,natwidth=121,natheight=44]{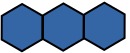} & 97.32 & 32.44 & $D_2$ & \ref{N6plots}(d)\\
3 & \includegraphics[scale=0.14,natwidth=82,natheight=82]{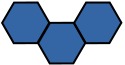} & 97.47 & 32.49 & $D_1$ & \ref{N6plots}(e)\\
3 & \includegraphics[scale=0.15,natwidth=118,natheight=100]{Images/forms/hex-3-3.jpeg} & 97.98 & 32.66 & $D_3$ &\ref{N6plots}(f)\\
4 & \includegraphics[scale=0.15,natwidth=65,natheight=65]{Images/forms/maximal3.jpg} & 136.60 & 34.15 & $D_{24}$ &\ref{N6plots}(g)\\
4 & \includegraphics[scale=0.14,natwidth=160,natheight=43]{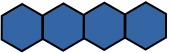}  & 129.11 & 32.28 & $D_2$ & \ref{N6plots}(h)\\
4 & \includegraphics[scale=0.14,natwidth=121,natheight=82]{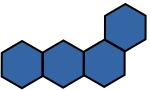} & 129.26 & 32.32 & $C_1$ & \ref{N6plots}(i)\\
4 & \includegraphics[scale=0.14,natwidth=121,natheight=82]{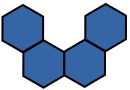} & 129.41 & 32.35 & $D_1$ & \ref{N6plots}(j)\\
4 & \includegraphics[scale=0.14,natwidth=82,natheight=82]{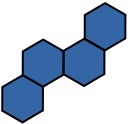} & 129.41 & 32.35 & $D_2$ & \ref{N6plots}(k)\\
4 & \includegraphics[scale=0.14,natwidth=120,natheight=82]{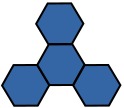} & 130.58 & 32.64 & $D_3$ & \ref{N6plots}(l)\\
4 & \includegraphics[scale=0.14,natwidth=130,natheight=125]{Images/forms/hex-4-6.jpeg} & 130.79 & 32.70 & $D_2$ & \ref{N6plots}(m)\\
4 & \includegraphics[scale=0.14,natwidth=130,natheight=125]{Images/forms/hex-4-7.jpeg} & 130.85 & 32.71 & $D_1$ & \ref{N6plots}(n)\\
4 & \includegraphics[scale=0.15,natwidth=90,natheight=80]{Images/forms/B2.jpeg} & 129.09 & 32.27 & $D_2$ & \ref{N6plots}(o)\\
\hline
\label{N6energytable}
\end{tabular}
\end{tabular}
\end{center}
\end{table}

\begin{figure}
\begin{center}
\begin{tabular}{c c c}
\includegraphics[scale=0.35,natwidth=1000,natheight=1000]{Images/N5B1.jpeg} & \includegraphics[scale=0.35,natwidth=1000,natheight=1000]{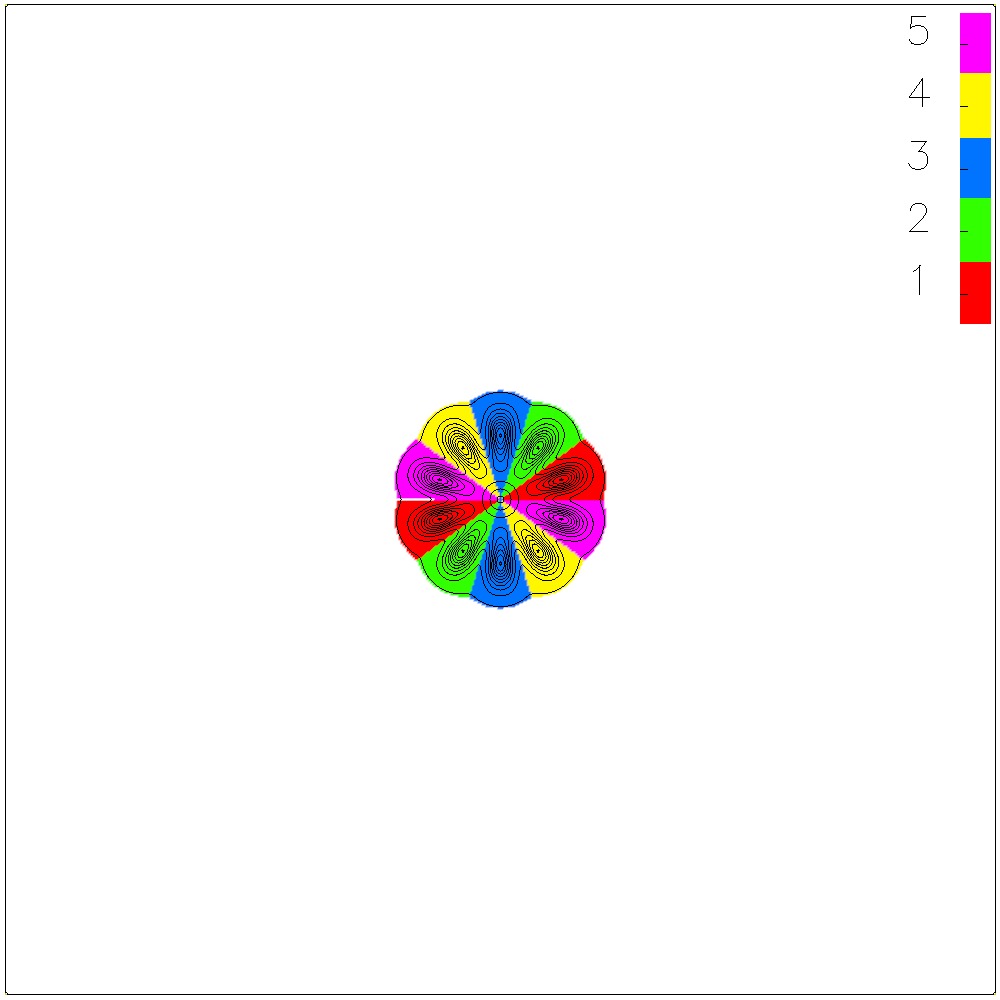} & \includegraphics[scale=0.35,natwidth=1000,natheight=1000]{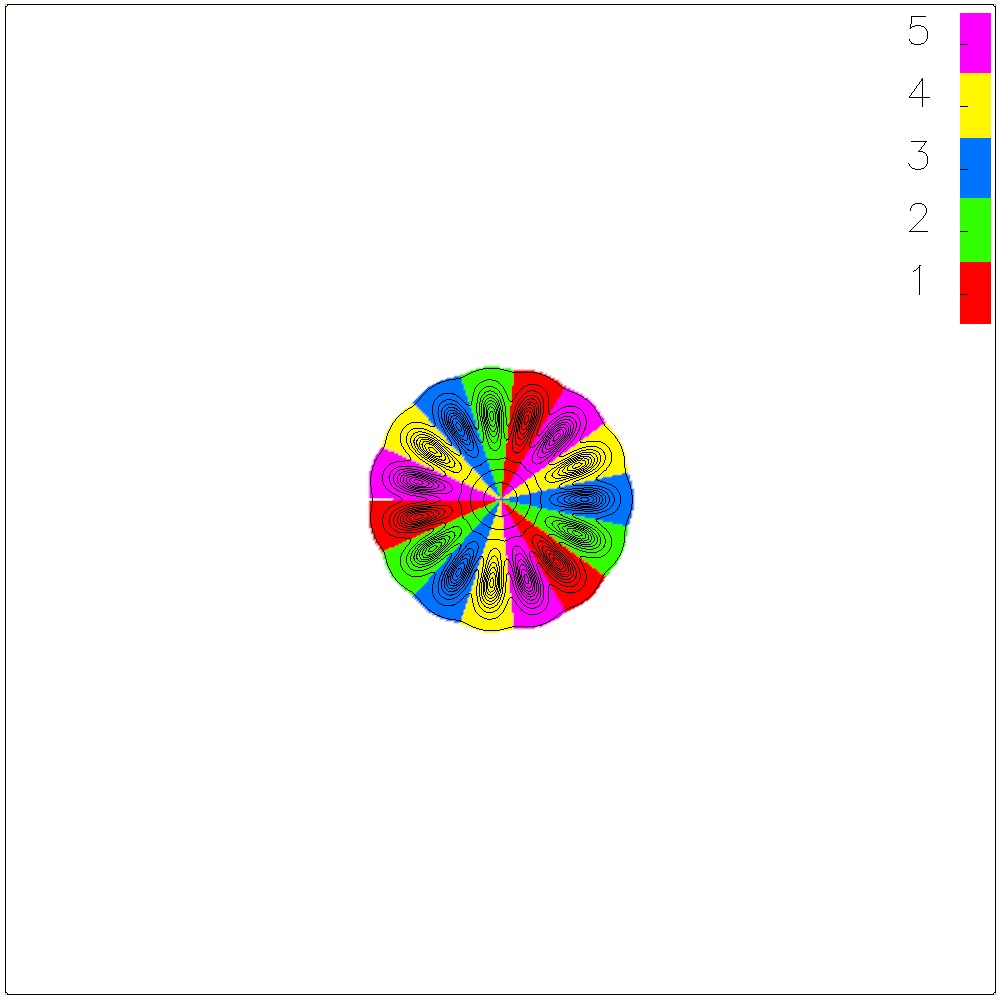}\\
(a) $B=1$ form = \includegraphics[scale=0.14,natwidth=45,natheight=45]{Images/forms/pent-1-1.jpeg} & (b) $B=2$ form = \includegraphics[scale=0.15,natwidth=55,natheight=55]{Images/forms/maximal.jpg} & (c) $B=3$ form = \includegraphics[scale=0.14,natwidth=121,natheight=44]{Images/forms/maximal2.jpg} \\
\includegraphics[scale=0.35,natwidth=1000,natheight=1000]{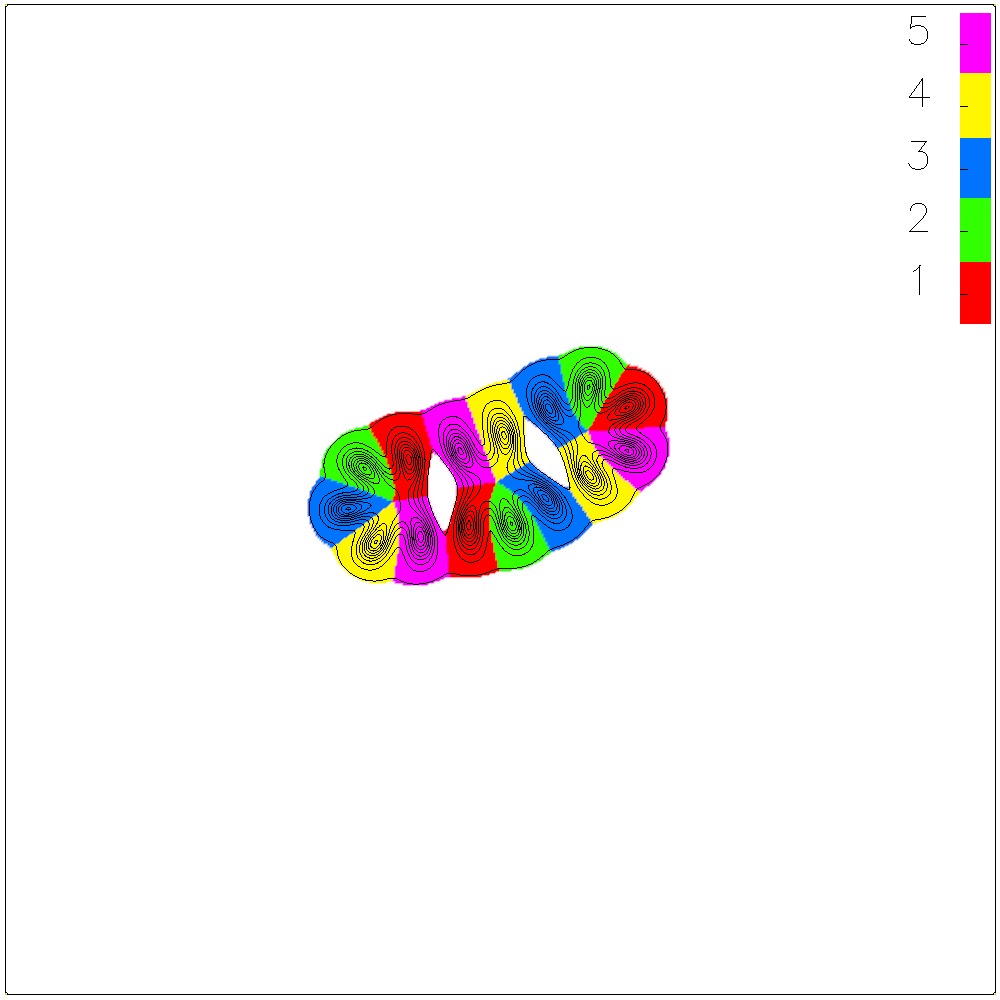} & \includegraphics[scale=0.35,natwidth=1000,natheight=1000]{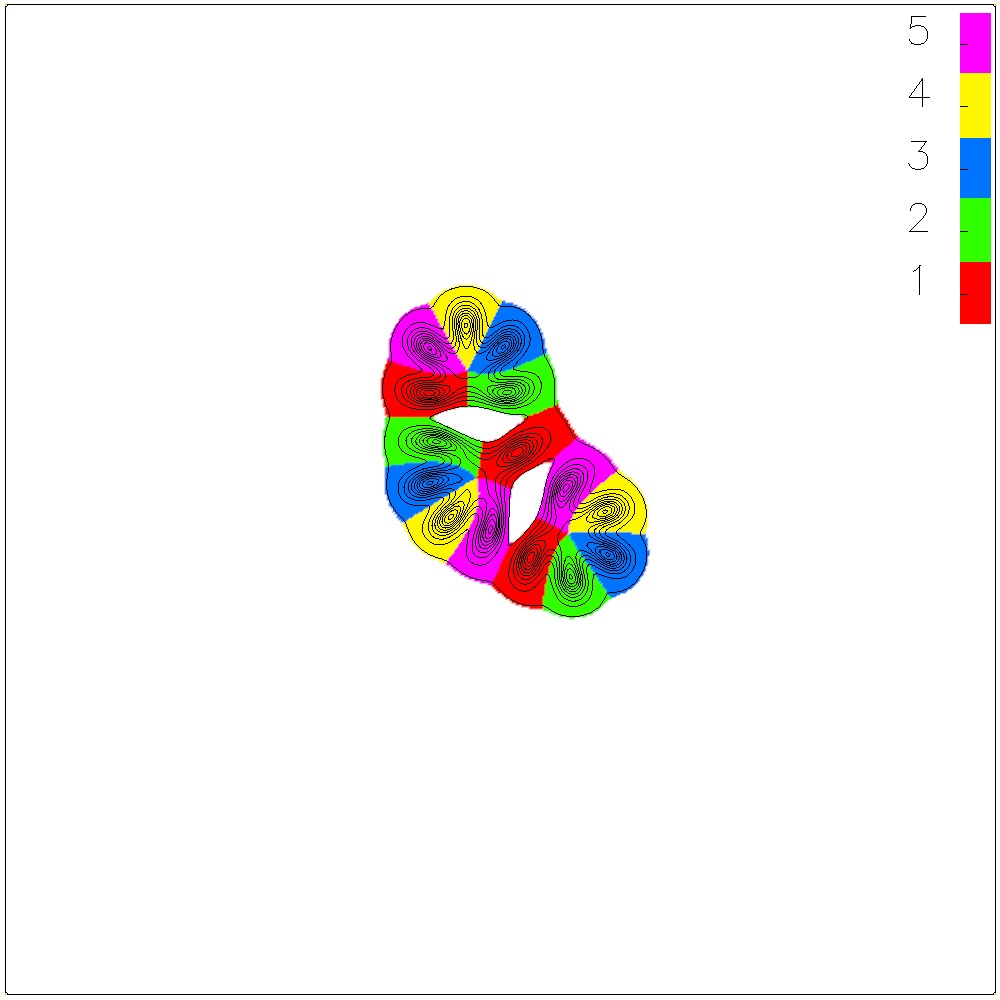} & \includegraphics[scale=0.35,natwidth=1000,natheight=1000]{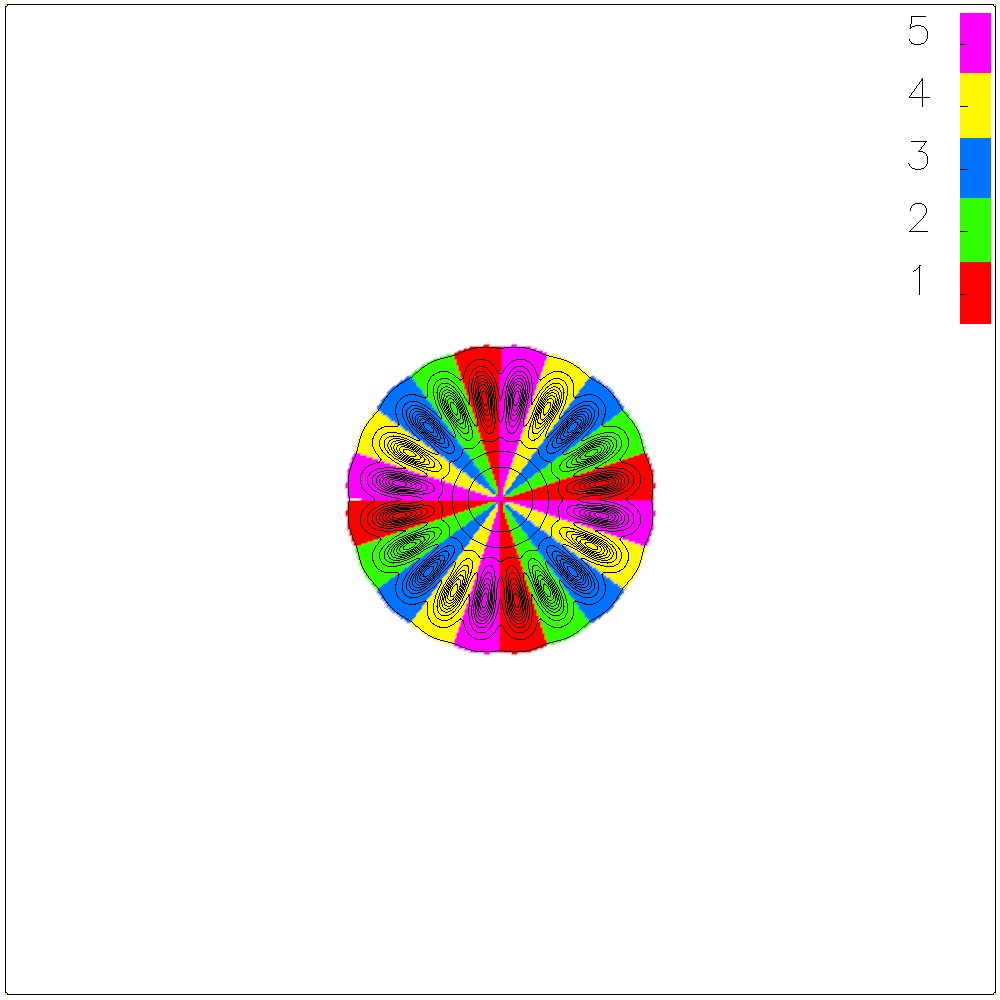}\\
(d) $B=3$ form = \includegraphics[scale=0.14,natwidth=82,natheight=82]{Images/forms/pent-3-1.jpeg} & (e) $B=3$ form = \includegraphics[scale=0.15,natwidth=65,natheight=65]{Images/forms/pent-3-2.jpeg} & (f) $B=4$ form = \includegraphics[scale=0.14,natwidth=121,natheight=82]{Images/forms/maximal3.jpg} \\
\includegraphics[scale=0.35,natwidth=1000,natheight=1000]{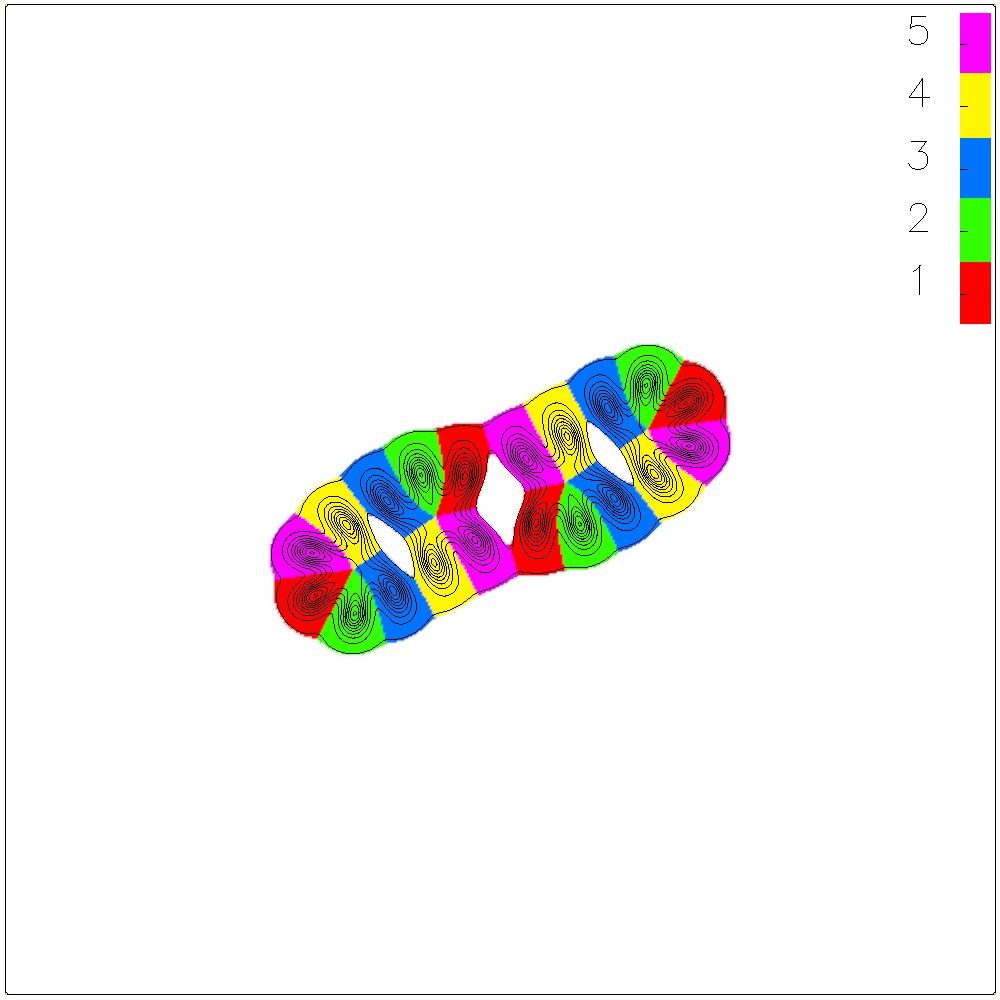} & \includegraphics[scale=0.35,natwidth=1000,natheight=1000]{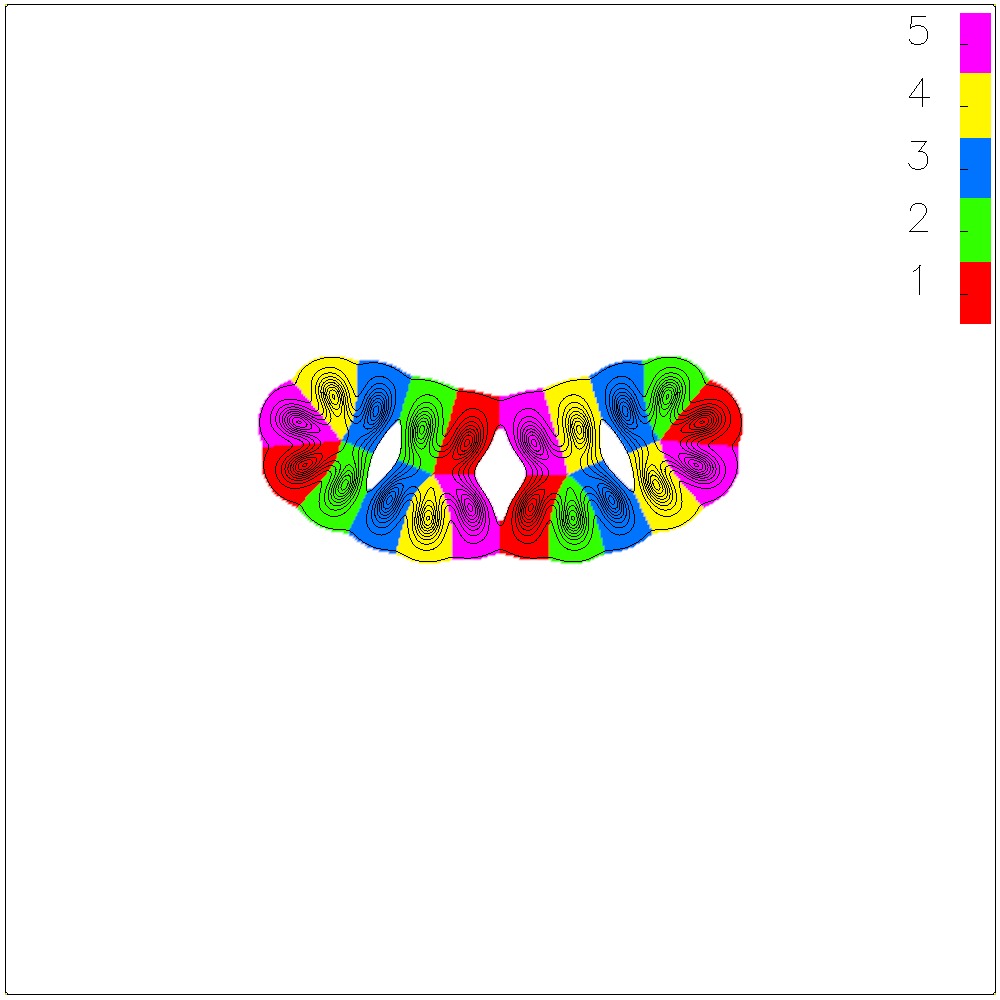} & \includegraphics[scale=0.35,natwidth=1000,natheight=1000]{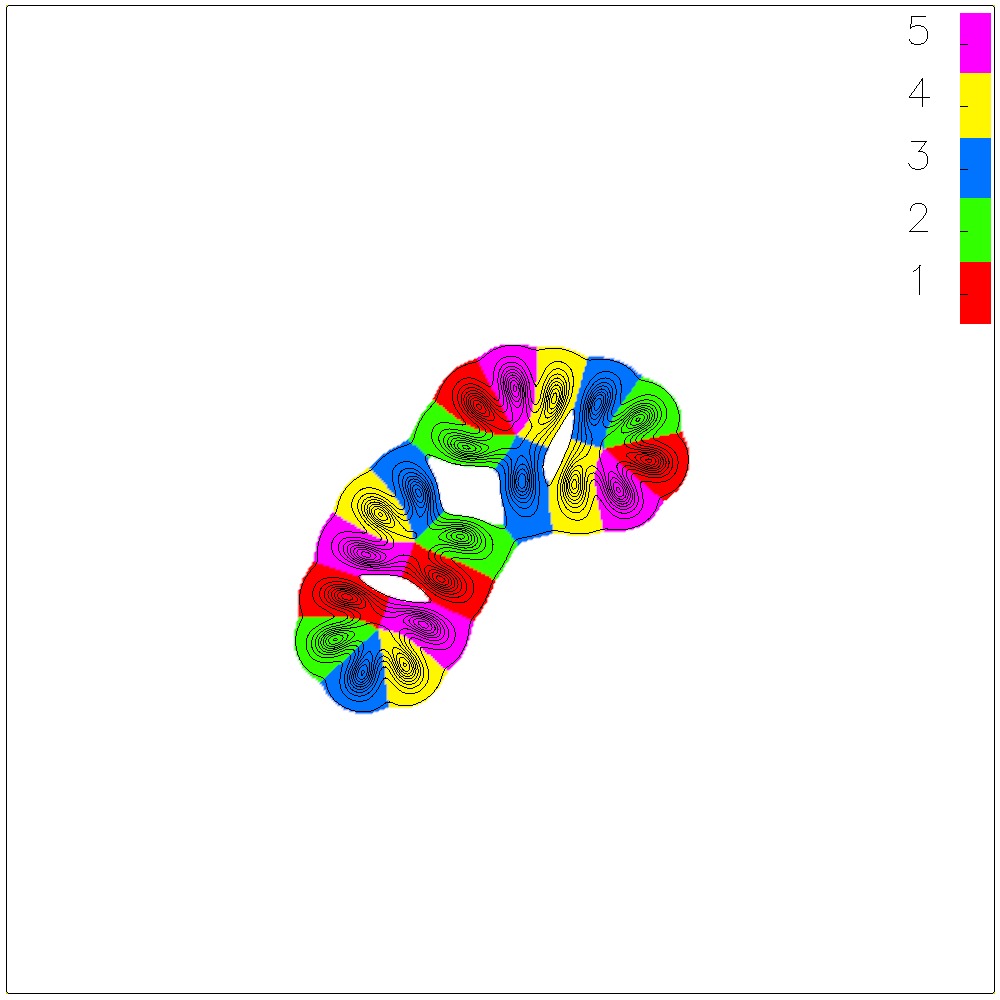}\\
(g) $B=4$ form = \reflectbox{\includegraphics[scale=0.14,natwidth=121,natheight=82]{Images/forms/pent-4-1.jpeg}} & (h) $B=4$ form = \includegraphics[scale=0.14,natwidth=82,natheight=82]{Images/forms/pent-4-2.jpeg} & (i) $B=4$ form = \reflectbox{\includegraphics[scale=0.14,natwidth=121,natheight=82]{Images/forms/pent-4-3.jpeg}} \\
\includegraphics[scale=0.35,natwidth=1000,natheight=1000]{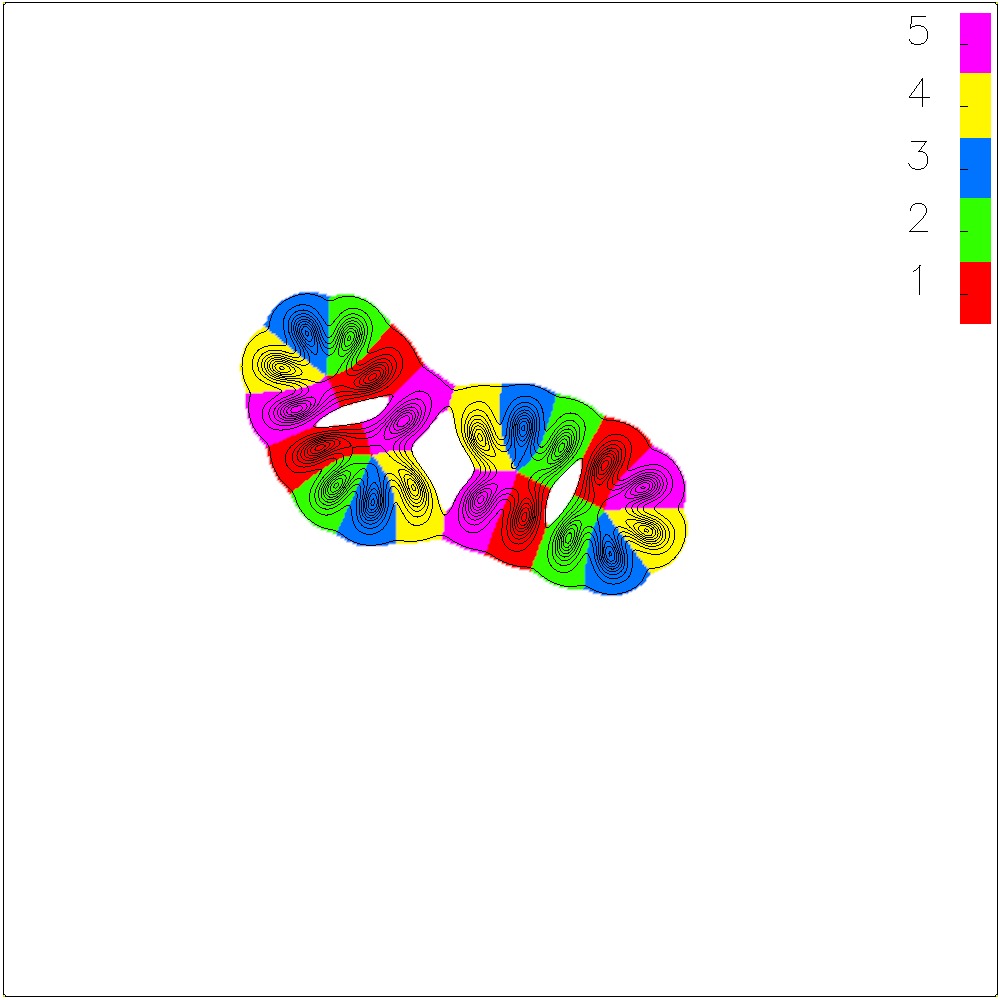} & \includegraphics[scale=0.35,natwidth=1000,natheight=1000]{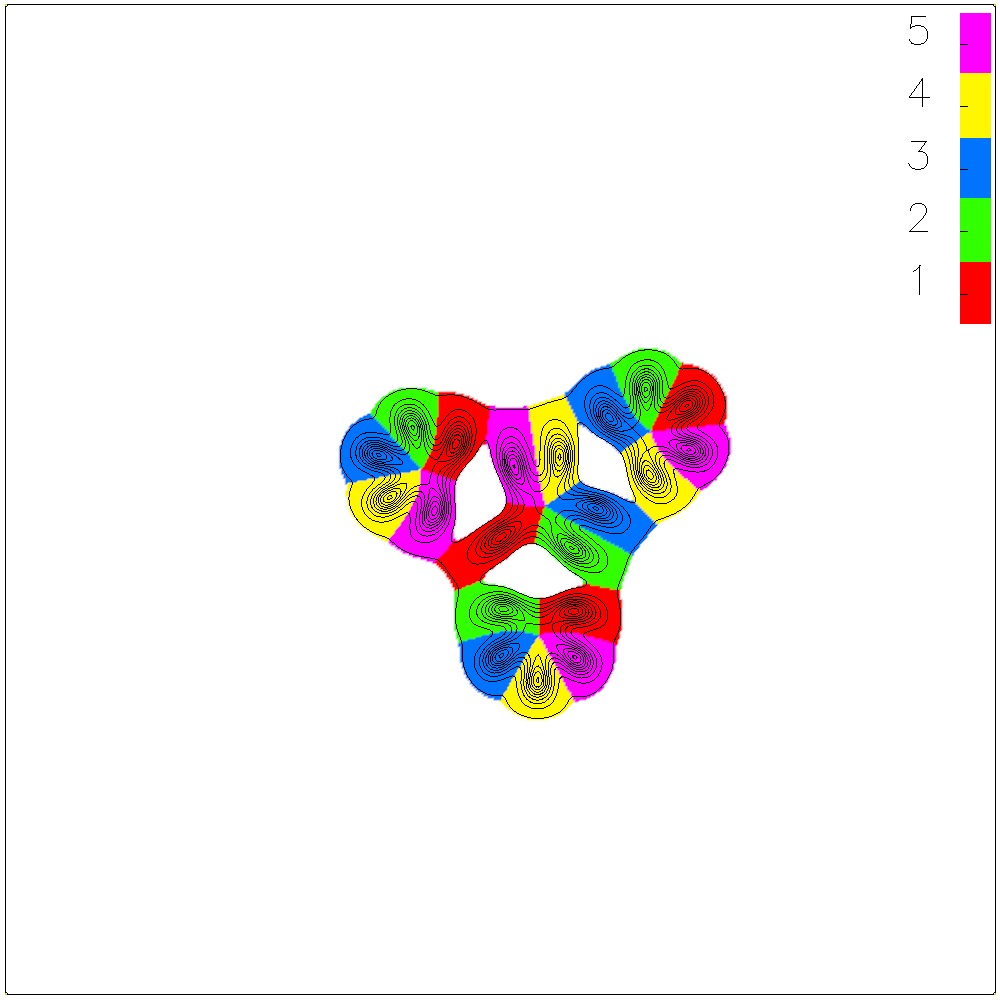} & \includegraphics[scale=0.35,natwidth=1000,natheight=1000]{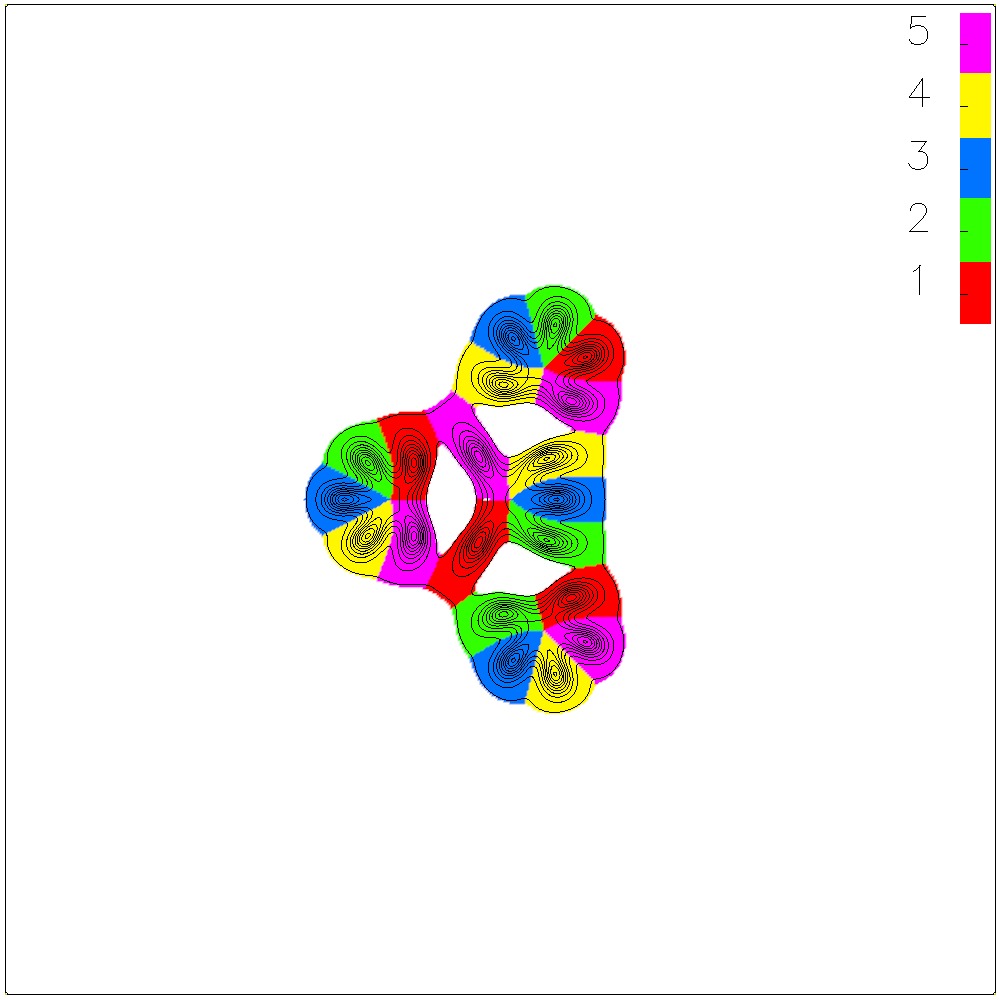}\\
(j) $B=4$ form = \includegraphics[scale=0.14,natwidth=121,natheight=82]{Images/forms/pent-4-4.jpeg} & (k) $B=4$ form = \includegraphics[scale=0.14,natwidth=82,natheight=82]{Images/forms/pent-4-5.jpeg} & (l) $B=4$ form = \reflectbox{\includegraphics[scale=0.14,natwidth=121,natheight=82]{Images/forms/pent-4-6.jpeg}} \\
\end{tabular}
\begin{tabular}{c}
 \includegraphics[scale=0.35,natwidth=1000,natheight=1000]{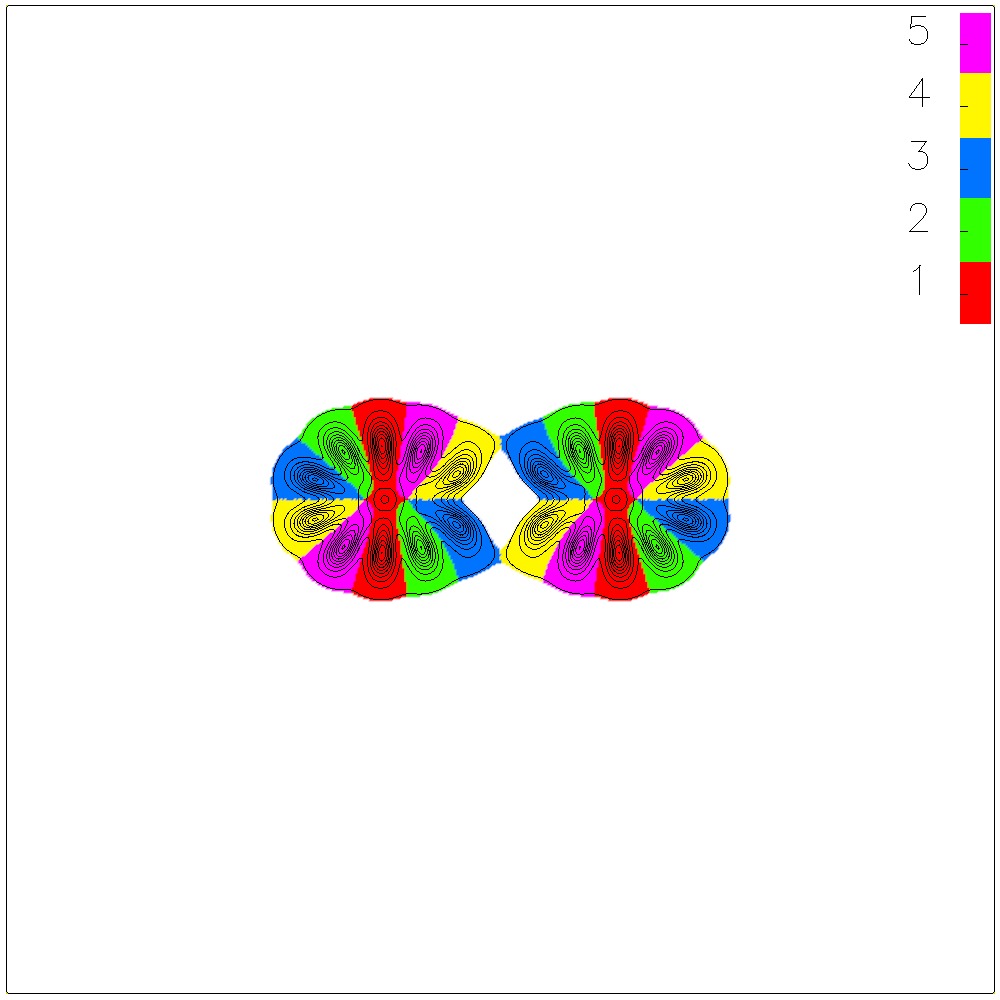}\\
(m) $B=4$ form = \includegraphics[scale=0.14,natwidth=90,natheight=80]{Images/forms/B2.jpeg}
\end{tabular}

\caption{Energy density plots of the multi-soliton solutions for $N=5$ and $B\leq4$ (colouring is based on the segment in which the point lies in the target space). Note that the \protect\includegraphics[scale=0.12,natwidth=121,natheight=82]{Images/forms/pent-4-7.jpeg} solution was not obtained, although we still expect this solution to exist. It was very similar to the \protect\includegraphics[scale=0.12,natwidth=90,natheight=80]{Images/forms/B2.jpeg} caveat in (m) meaning it was difficult to pick out initial conditions that would relax to the desired solution rather than this lower energy caveat form.}
\label{N5plots}
\end{center}
\end{figure}

\begin{figure}
\begin{center}
\begin{tabular}{c c c}
\includegraphics[scale=0.35,natwidth=1000,natheight=1000]{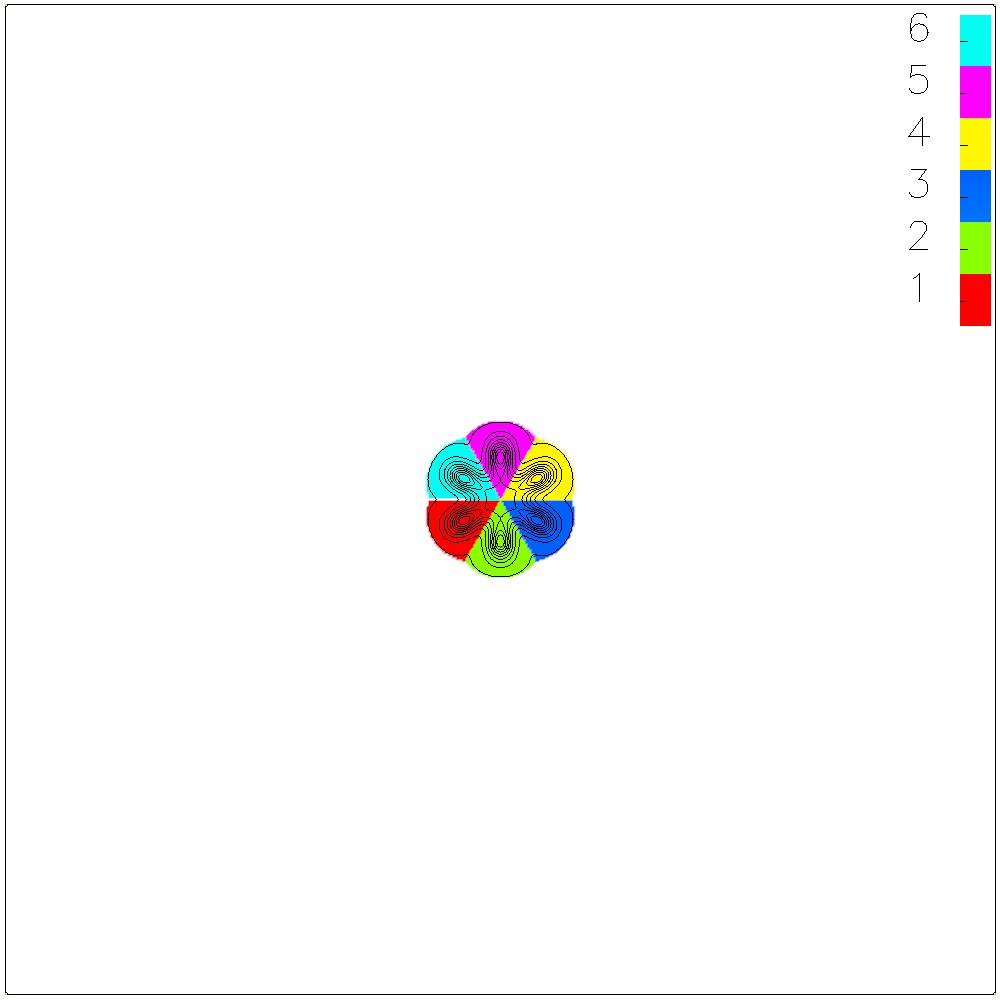} & \includegraphics[scale=0.35,natwidth=1000,natheight=1000]{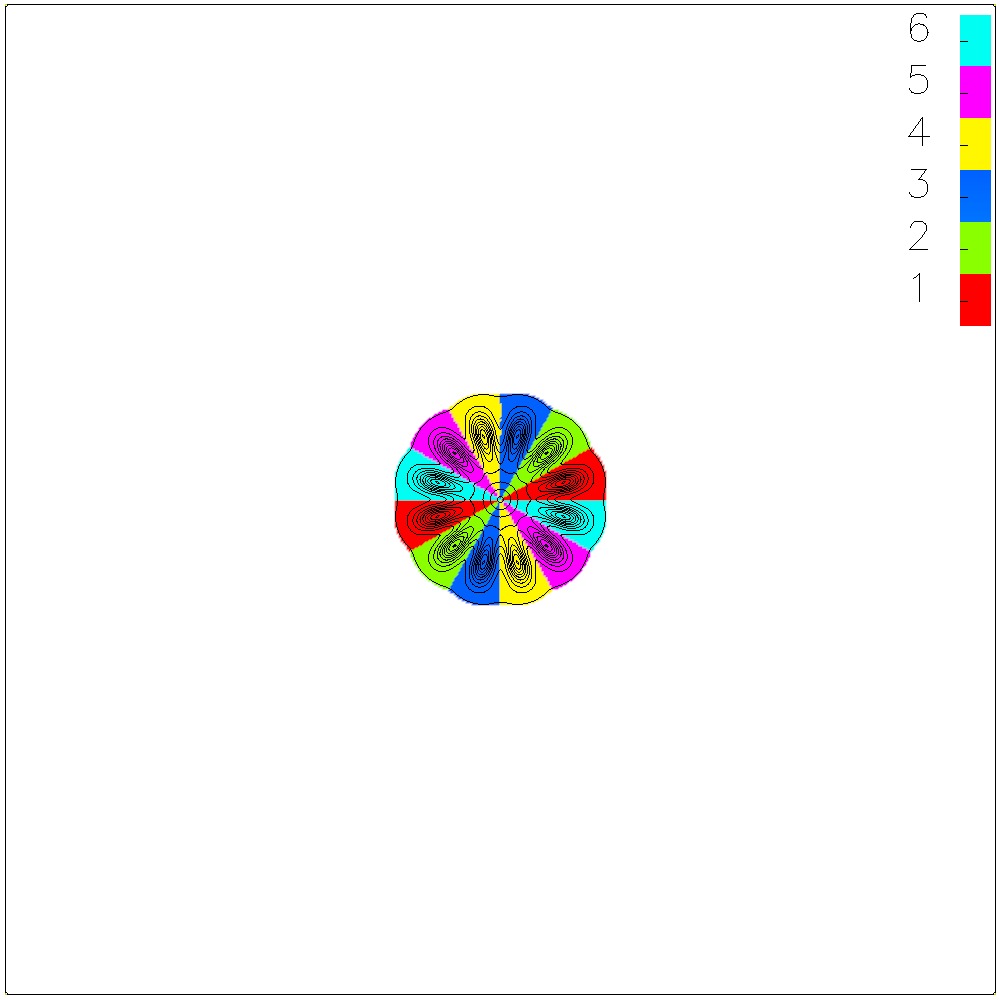} & \includegraphics[scale=0.35,natwidth=1000,natheight=1000]{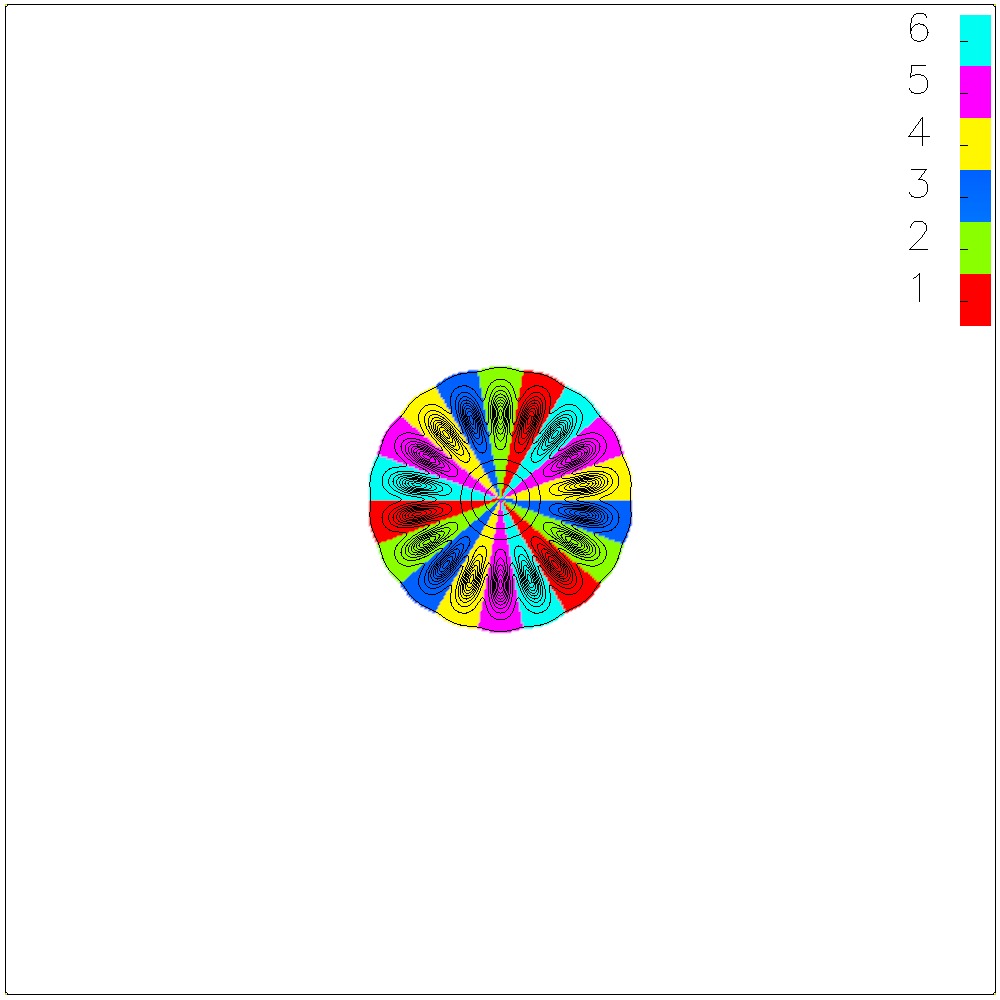}\\
(a) $B=1$ form = \includegraphics[scale=0.14,natwidth=45,natheight=45]{Images/forms/hex-1-1.jpeg} & (b) $B=2$ form = \includegraphics[scale=0.15,natwidth=55,natheight=55]{Images/forms/maximal.jpg} & (c) $B=3$ form = \includegraphics[scale=0.14,natwidth=121,natheight=44]{Images/forms/maximal2.jpg} \\
\includegraphics[scale=0.35,natwidth=1000,natheight=1000]{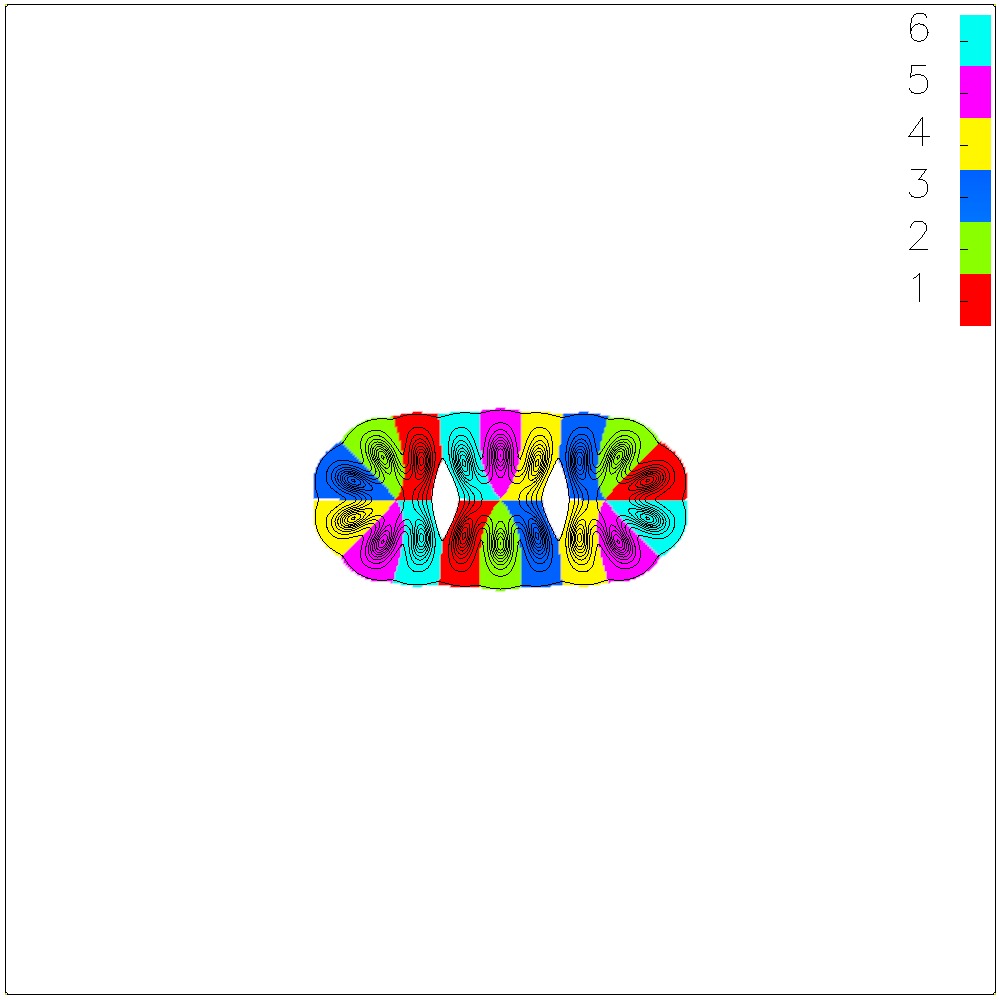} & \includegraphics[scale=0.35,natwidth=1000,natheight=1000]{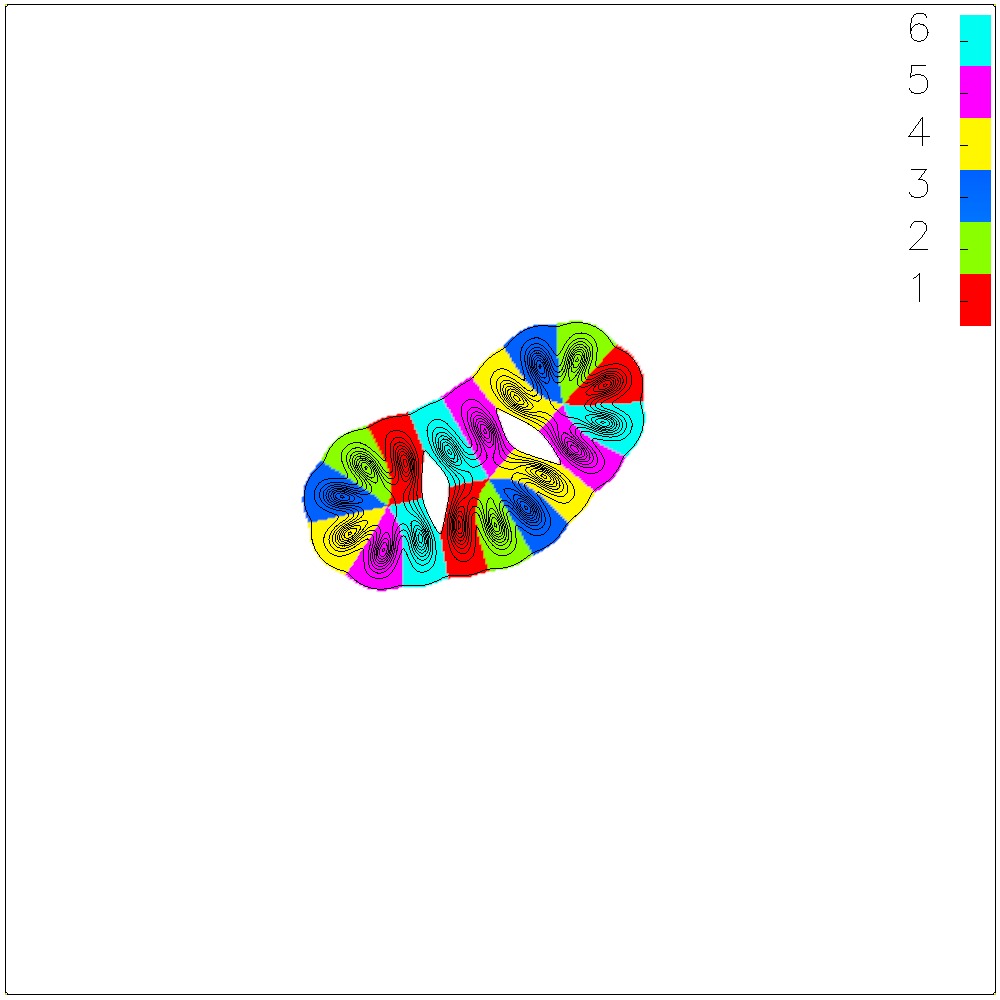} & \includegraphics[scale=0.35,natwidth=1000,natheight=1000]{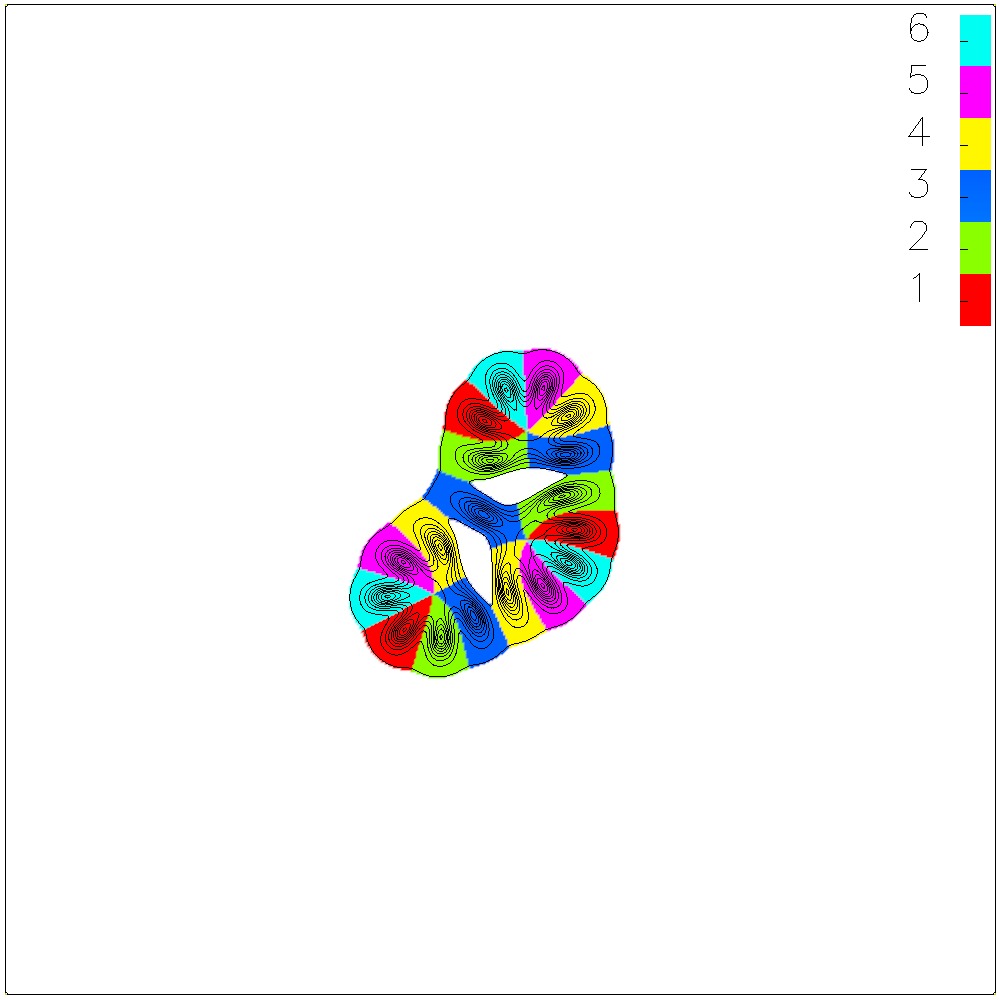}\\
(d) $B=3$ form = \includegraphics[scale=0.14,natwidth=82,natheight=82]{Images/forms/hex-3-1.jpeg} & (e) $B=3$ form = \includegraphics[scale=0.15,natwidth=65,natheight=65]{Images/forms/hex-3-2.jpeg} & (f) $B=4$ form = \includegraphics[scale=0.14,natwidth=121,natheight=82]{Images/forms/hex-3-3.jpeg} \\
\includegraphics[scale=0.35,natwidth=1000,natheight=1000]{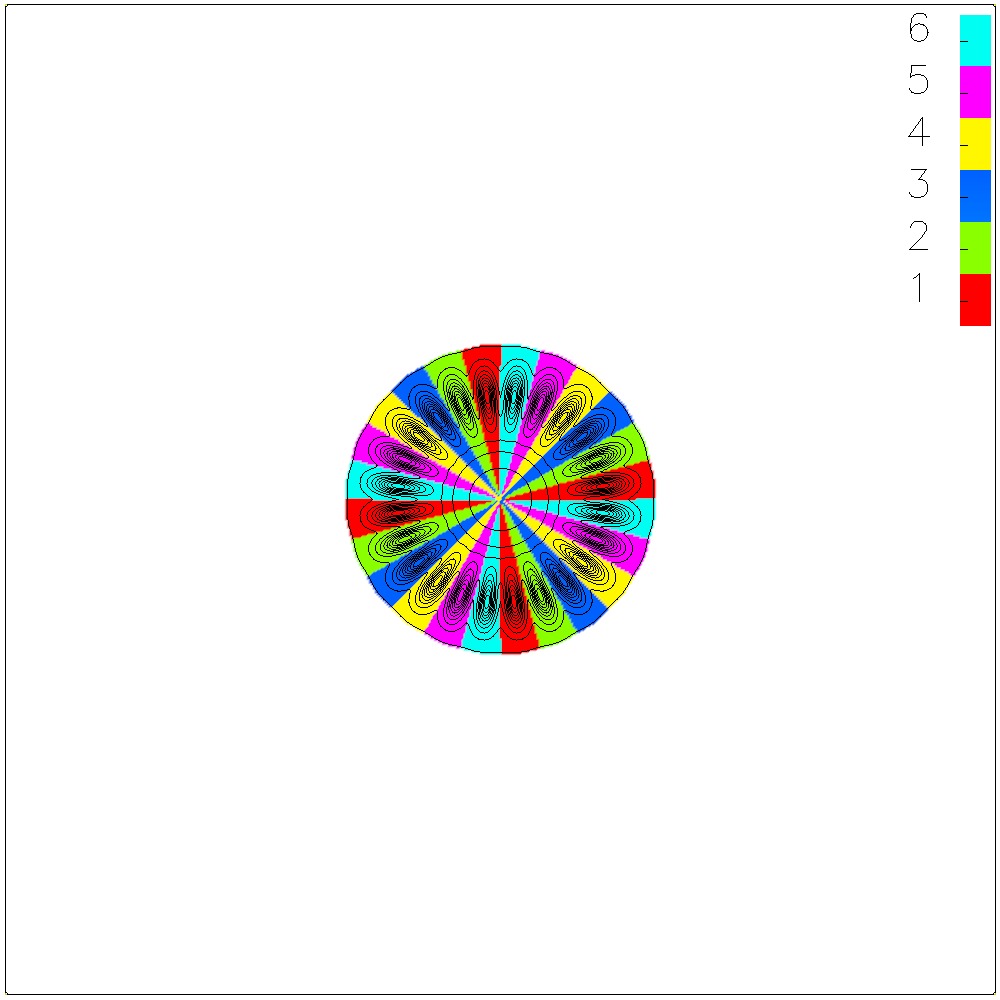} & \includegraphics[scale=0.35,natwidth=1000,natheight=1000]{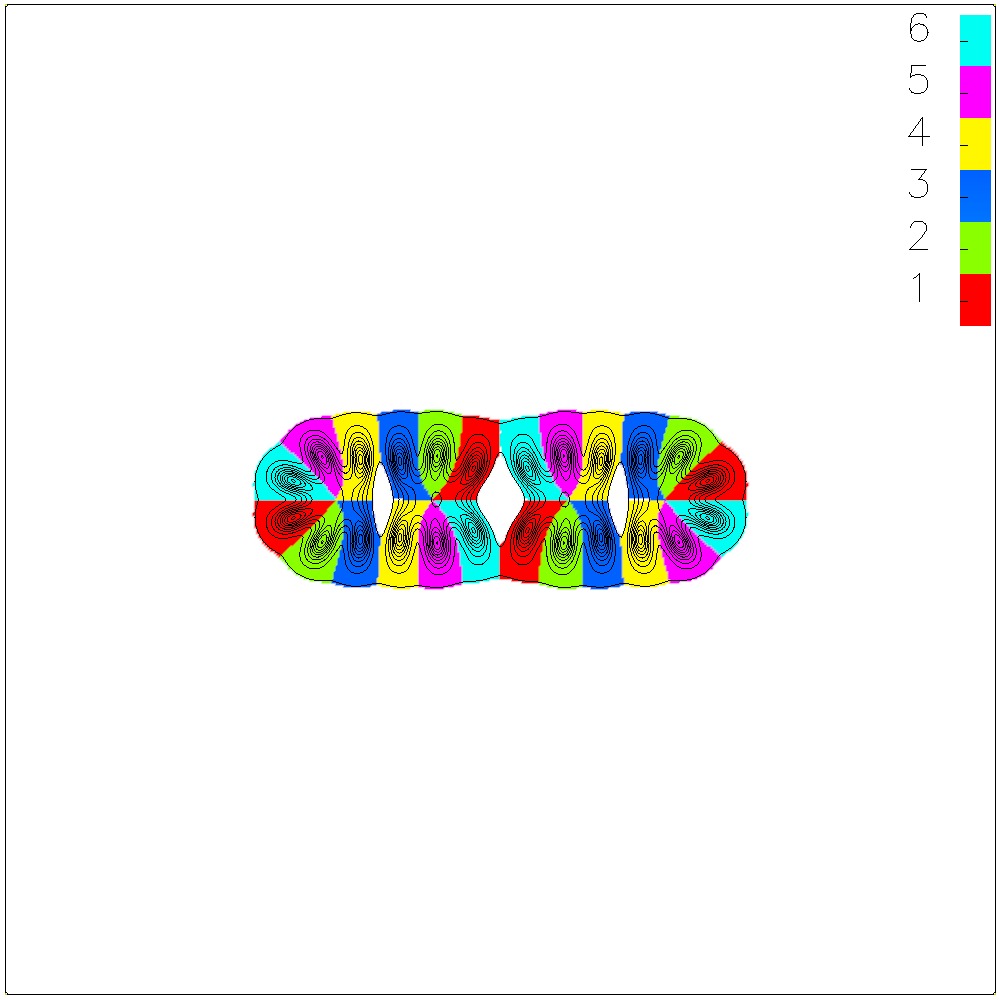} & \includegraphics[scale=0.35,natwidth=1000,natheight=1000]{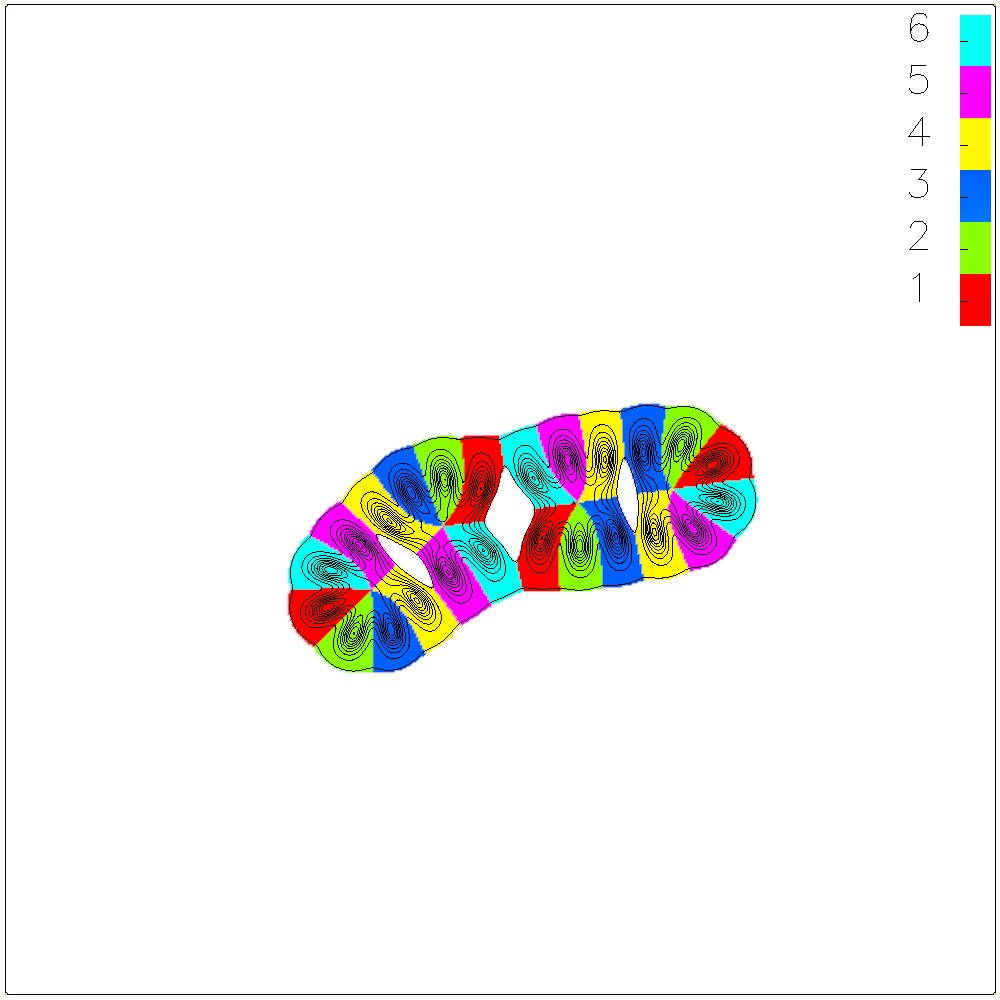}\\
(g) $B=4$ form = \reflectbox{\includegraphics[scale=0.14,natwidth=121,natheight=82]{Images/forms/maximal3.jpg}} & (h) $B=4$ form = \includegraphics[scale=0.14,natwidth=82,natheight=82]{Images/forms/hex-4-1.jpeg} & (i) $B=4$ form = \reflectbox{\includegraphics[scale=0.14,natwidth=121,natheight=82]{Images/forms/hex-4-2.jpeg}} \\
\includegraphics[scale=0.35,natwidth=1000,natheight=1000]{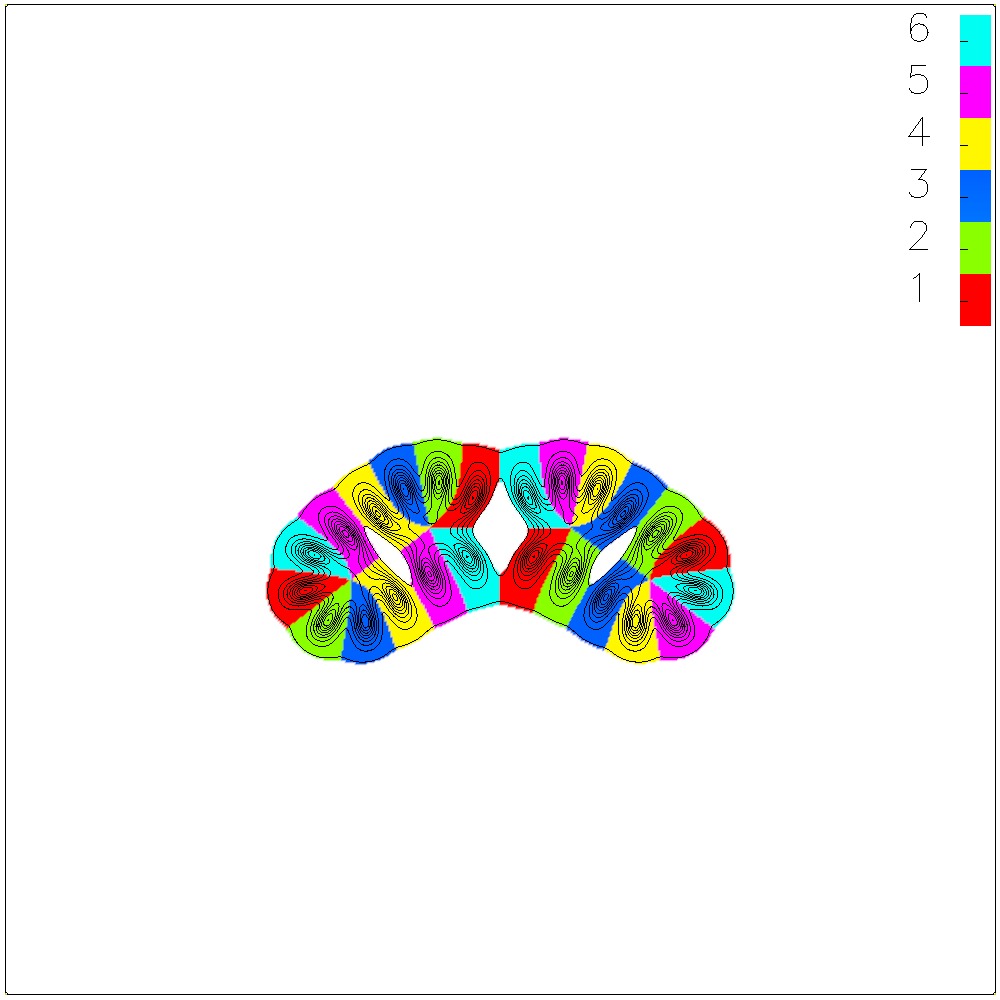} & \includegraphics[scale=0.35,natwidth=1000,natheight=1000]{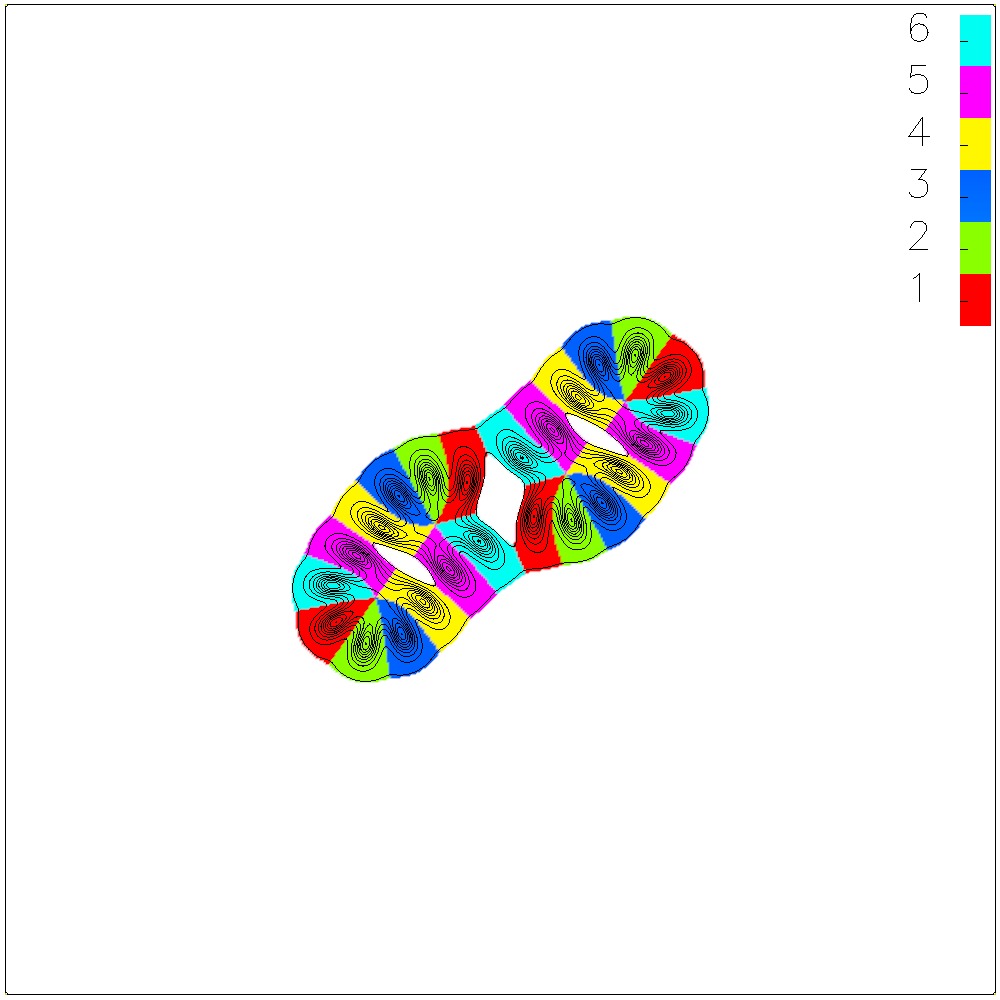} & \includegraphics[scale=0.35,natwidth=1000,natheight=1000]{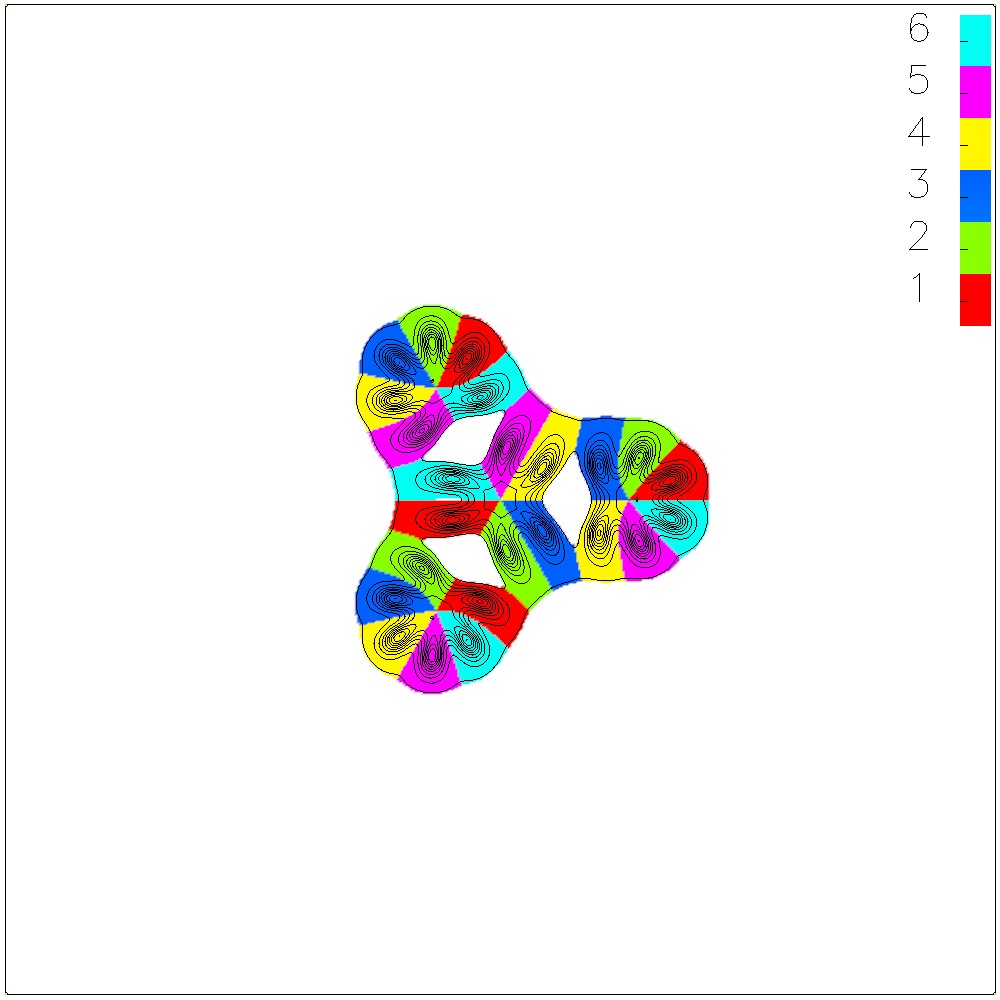}\\
(j) $B=4$ form = \includegraphics[scale=0.14,natwidth=121,natheight=82]{Images/forms/hex-4-3.jpeg} & (k) $B=4$ form = \includegraphics[scale=0.14,natwidth=82,natheight=82]{Images/forms/hex-4-4.jpeg} & (l) $B=4$ form = \reflectbox{\includegraphics[scale=0.14,natwidth=121,natheight=82]{Images/forms/hex-4-5.jpeg}} \\
 \includegraphics[scale=0.35,natwidth=1000,natheight=1000]{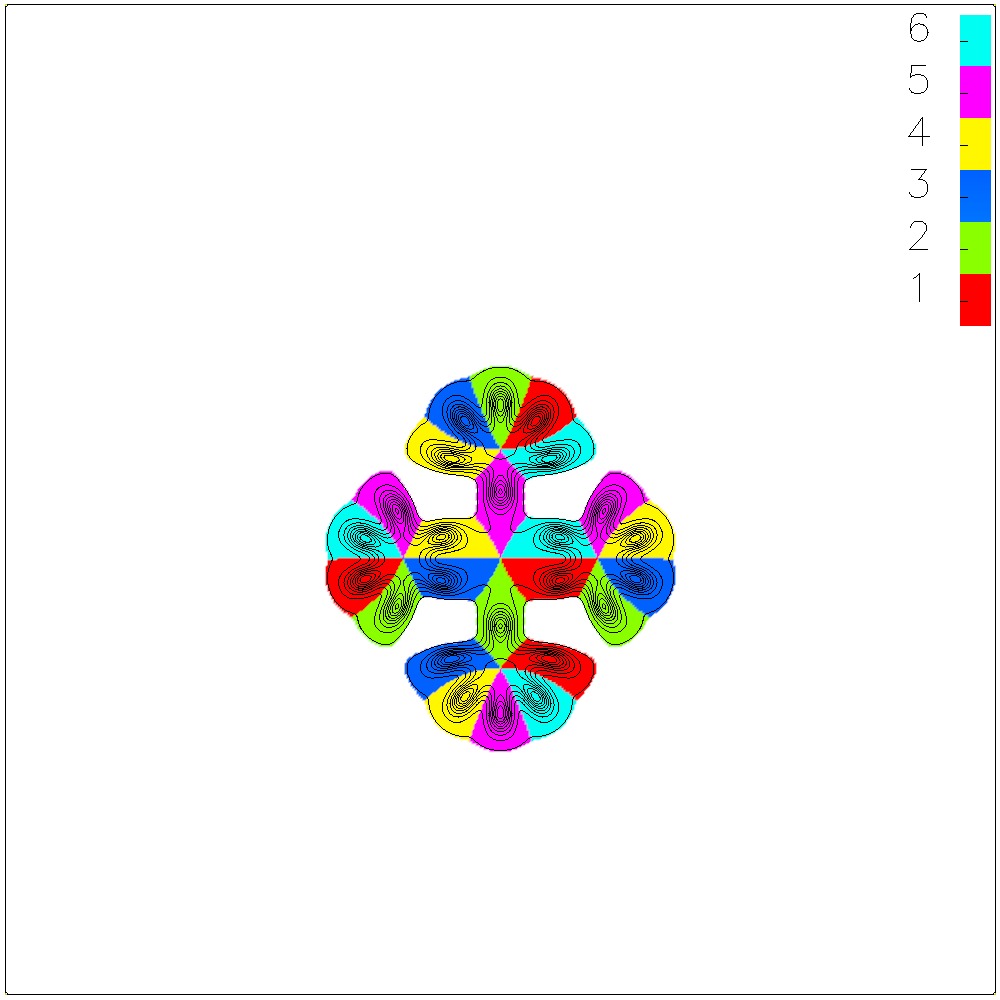} & \includegraphics[scale=0.35,natwidth=1000,natheight=1000]{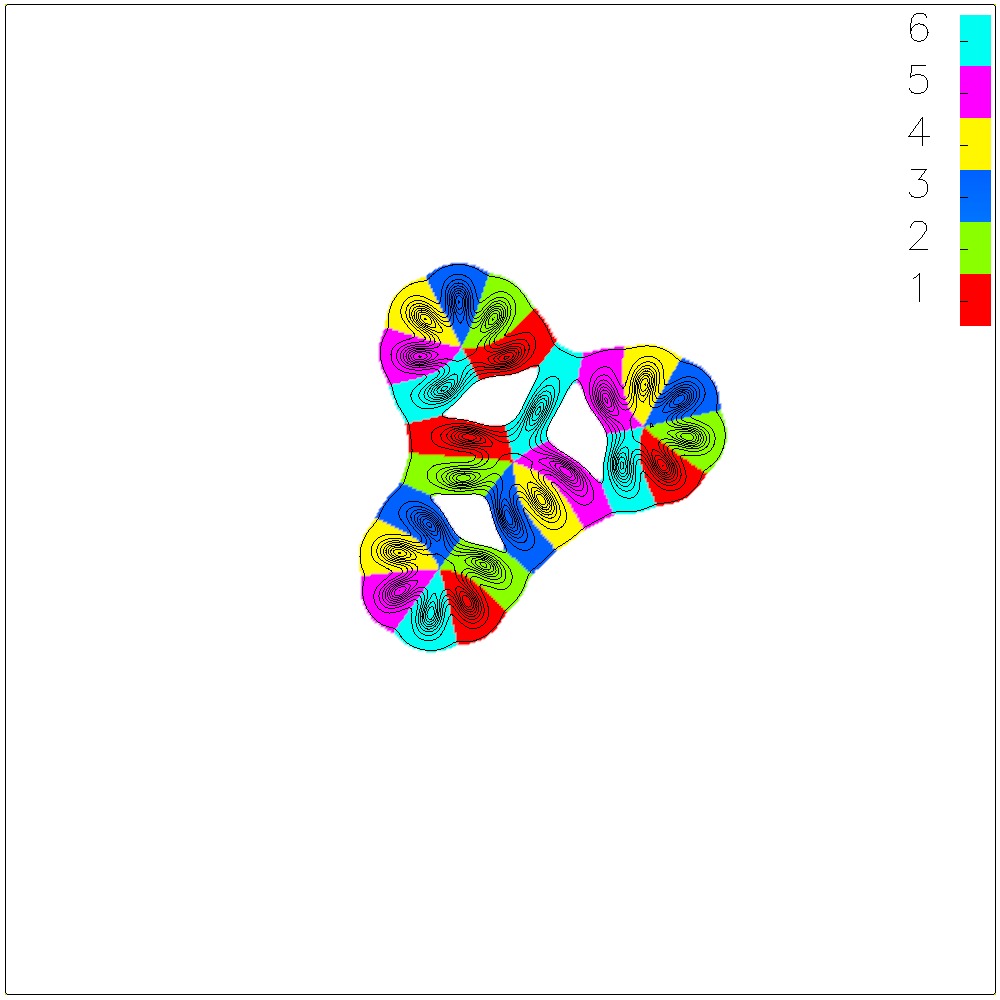} & \includegraphics[scale=0.35,natwidth=1000,natheight=1000]{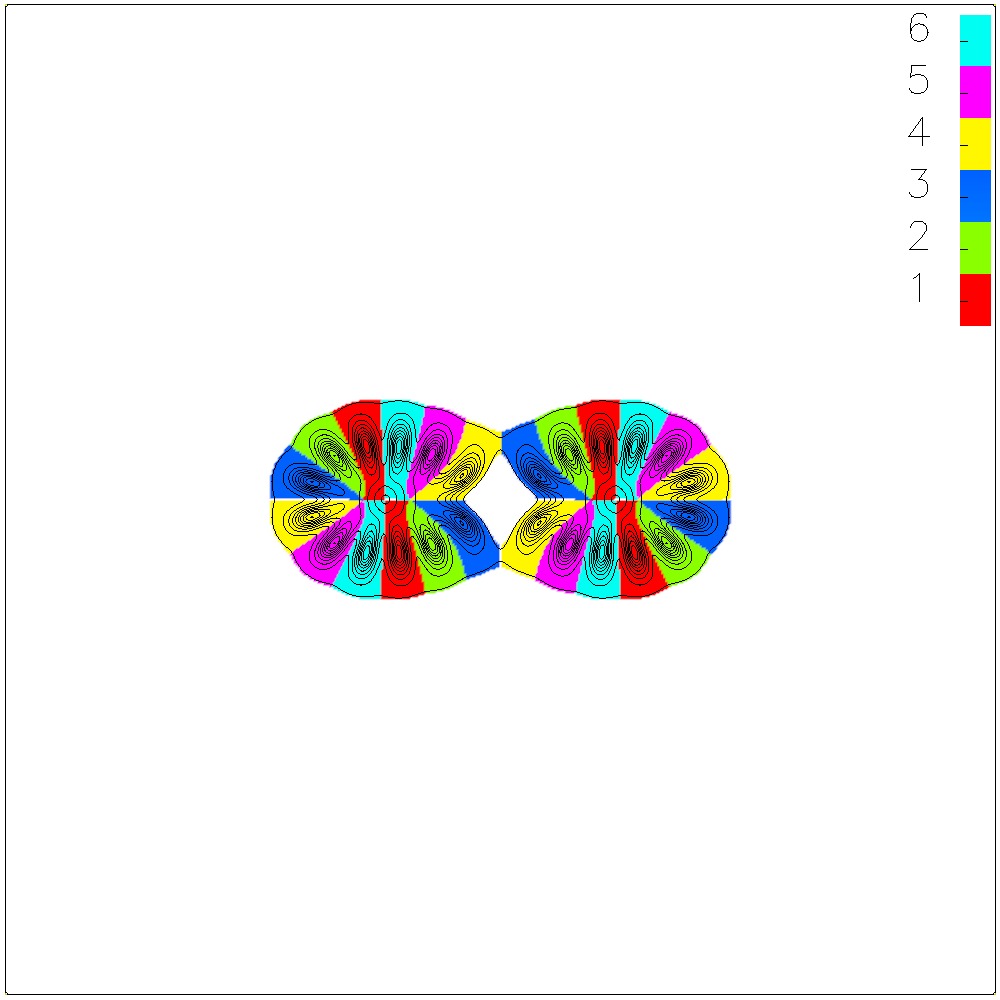} \\
(m) $B=4$ form = \includegraphics[scale=0.14,natwidth=121,natheight=82]{Images/forms/hex-4-6.jpeg} & (n) $B=4$ form = \includegraphics[scale=0.14,natwidth=120,natheight=82]{Images/forms/hex-4-7.jpeg} & (o) $B=4$ form = \includegraphics[scale=0.14,natwidth=90,natheight=80]{Images/forms/B2.jpeg}
\end{tabular}

\caption{Energy density plots of the multi-soliton solutions for $N=6$ and $B\leq4$ (colouring is based on the segment in which the point lies in the target space).}
\label{N6plots}
\end{center}
\end{figure}
\clearpage
\section{Appendix B: Additional Scatterings}
\begin{figure}[h]
\centerline{\begin{tabular}{c c c c}
\includegraphics[scale=0.4,natwidth=1000,natheight=1000]{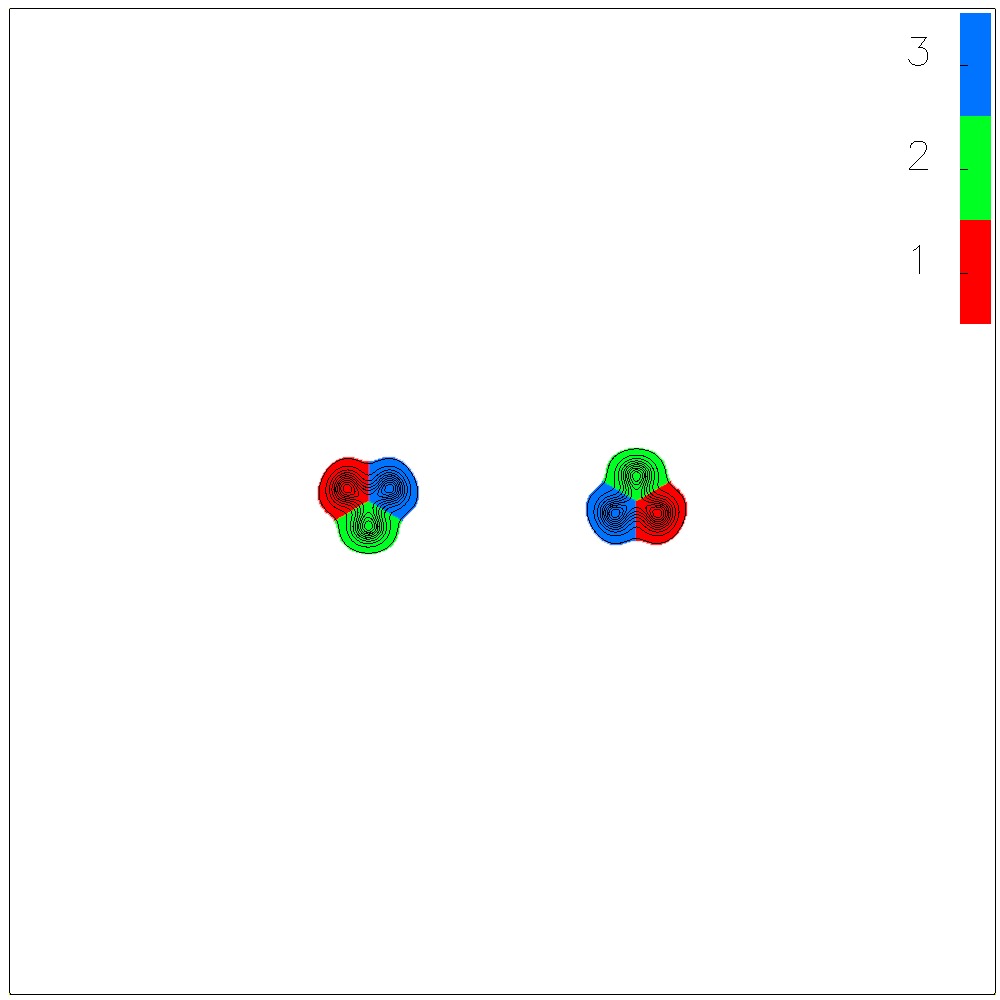} & \includegraphics[scale=0.4,natwidth=1000,natheight=1000]{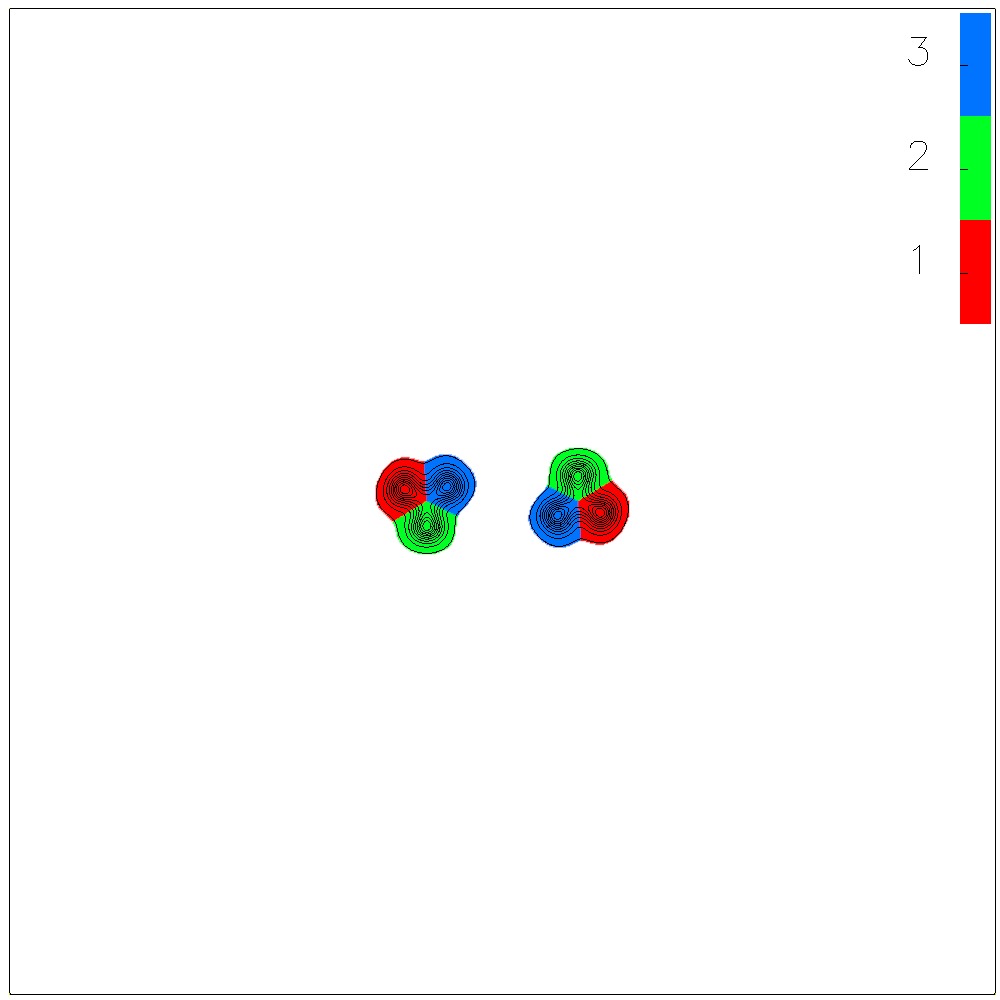} & \includegraphics[scale=0.4,natwidth=1000,natheight=1000]{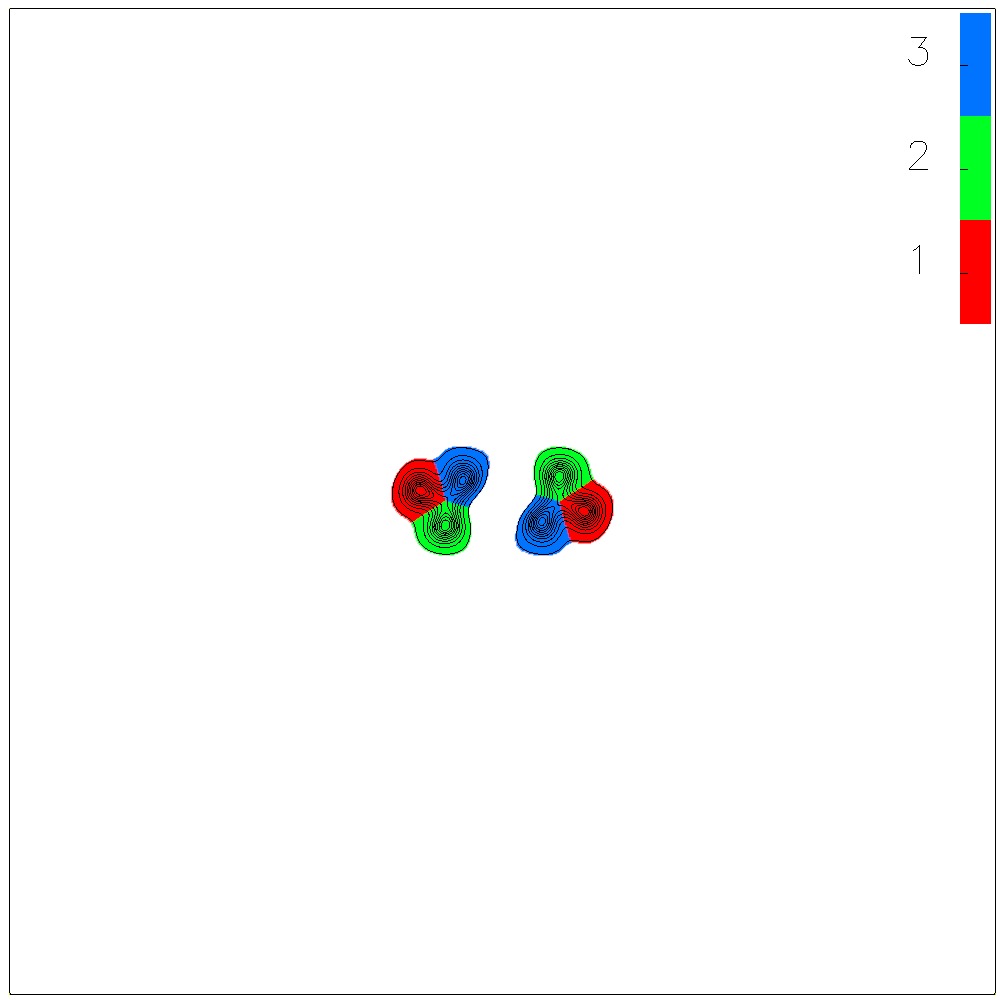} & \includegraphics[scale=0.4,natwidth=1000,natheight=1000]{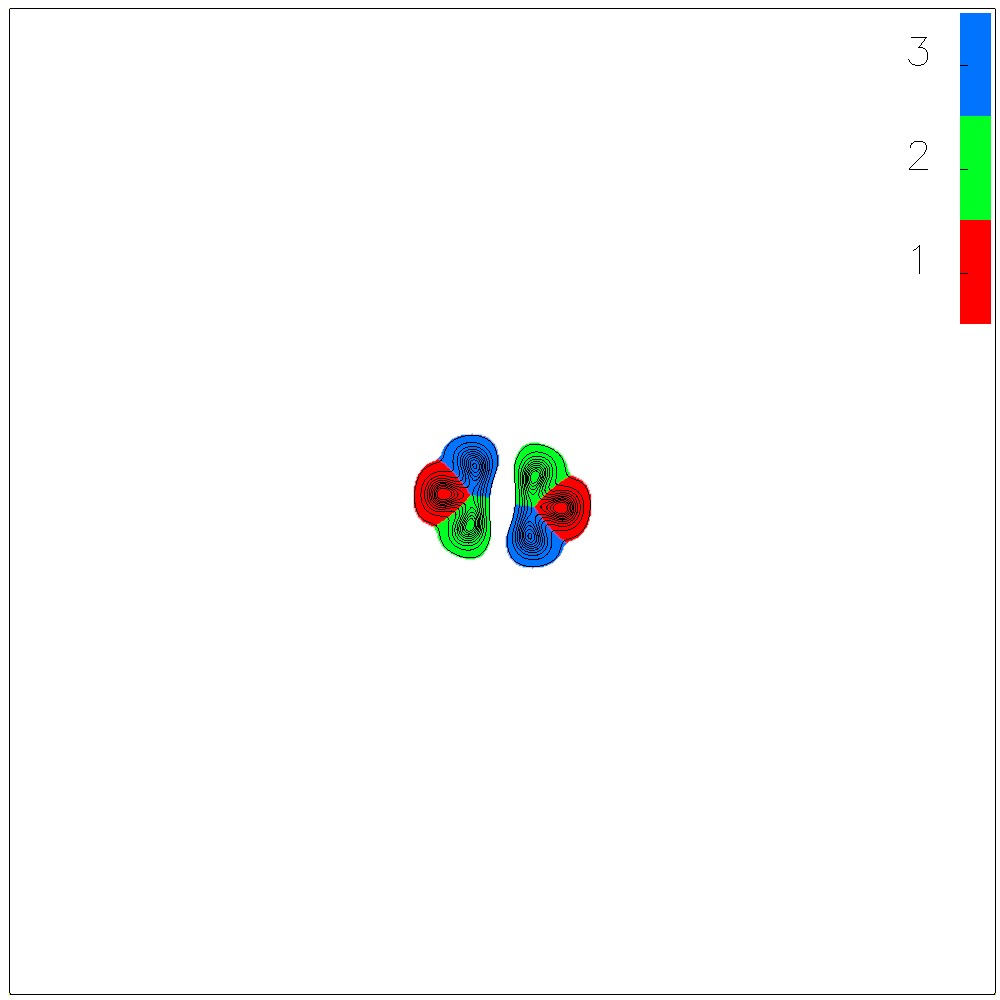}\\
$t = 8.5$ & $t = 17$ & $t = 19$ & $t = 21$\\
\includegraphics[scale=0.4,natwidth=1000,natheight=1000]{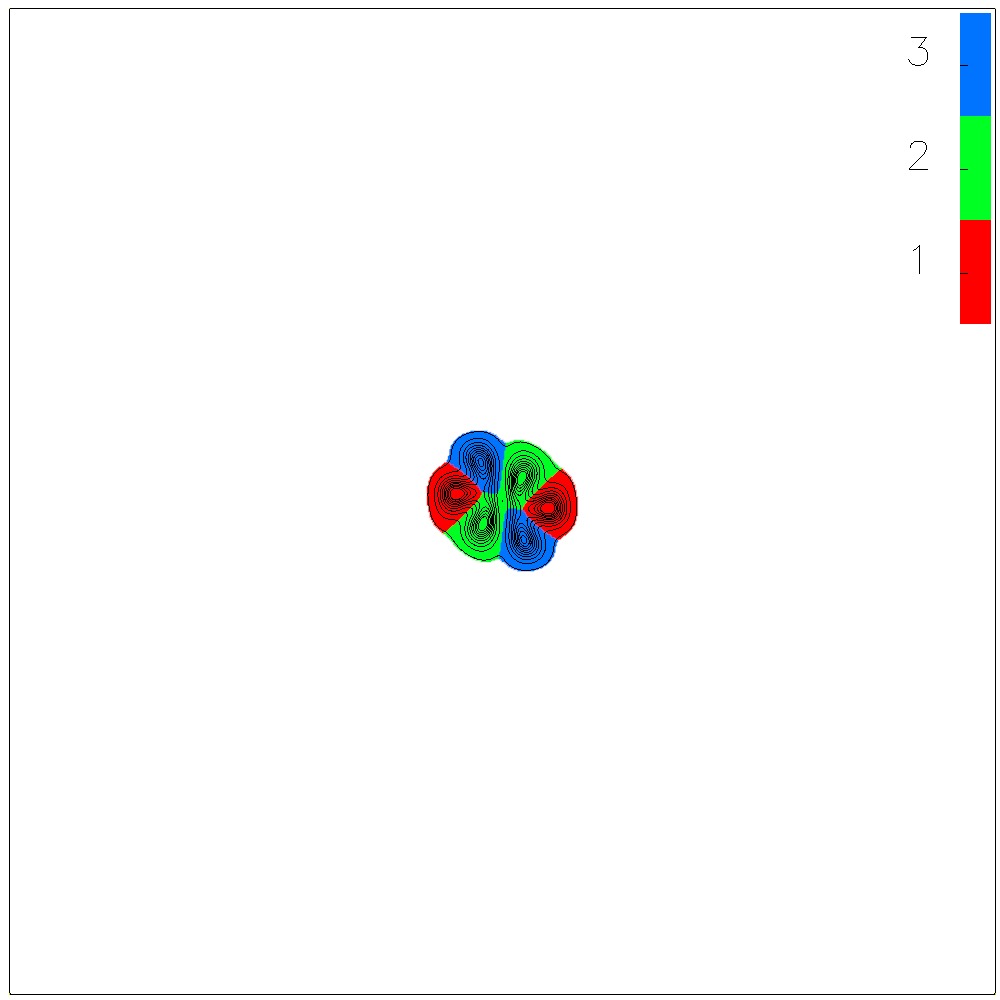} & \includegraphics[scale=0.4,natwidth=1000,natheight=1000]{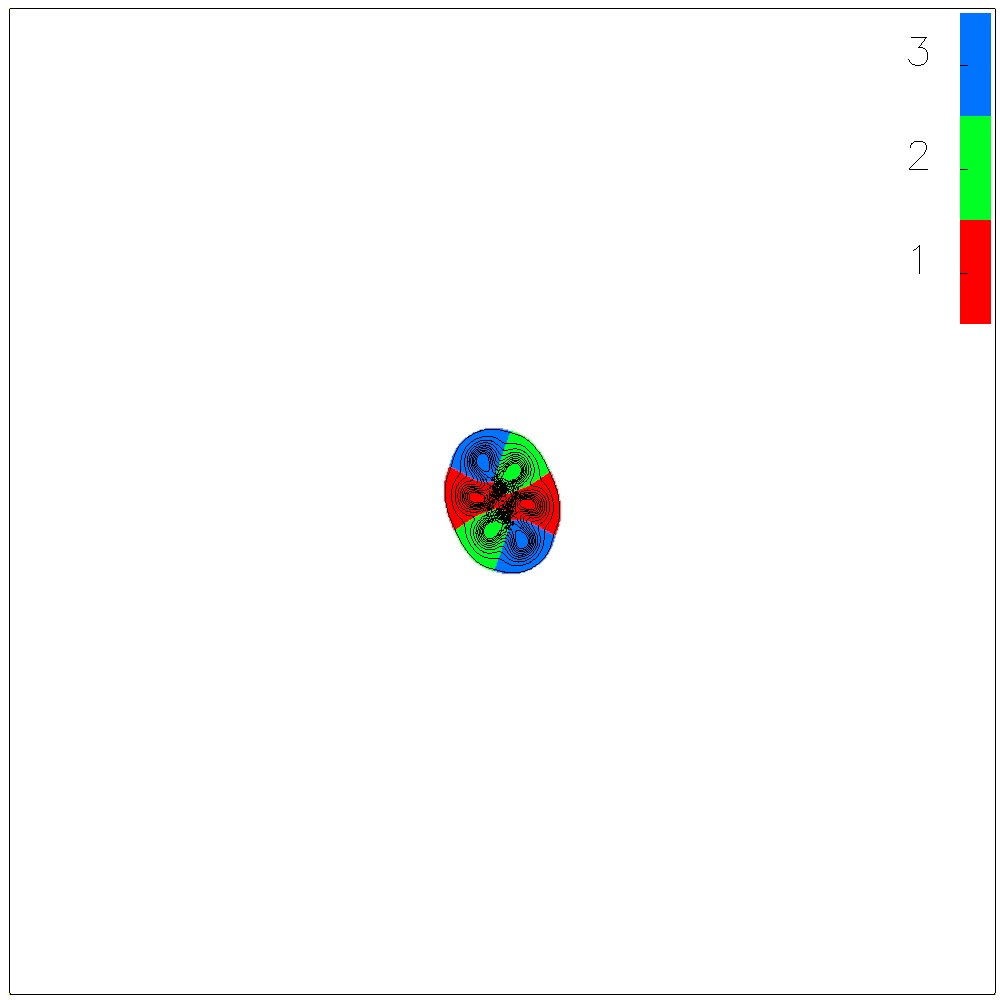} & \includegraphics[scale=0.4,natwidth=1000,natheight=1000]{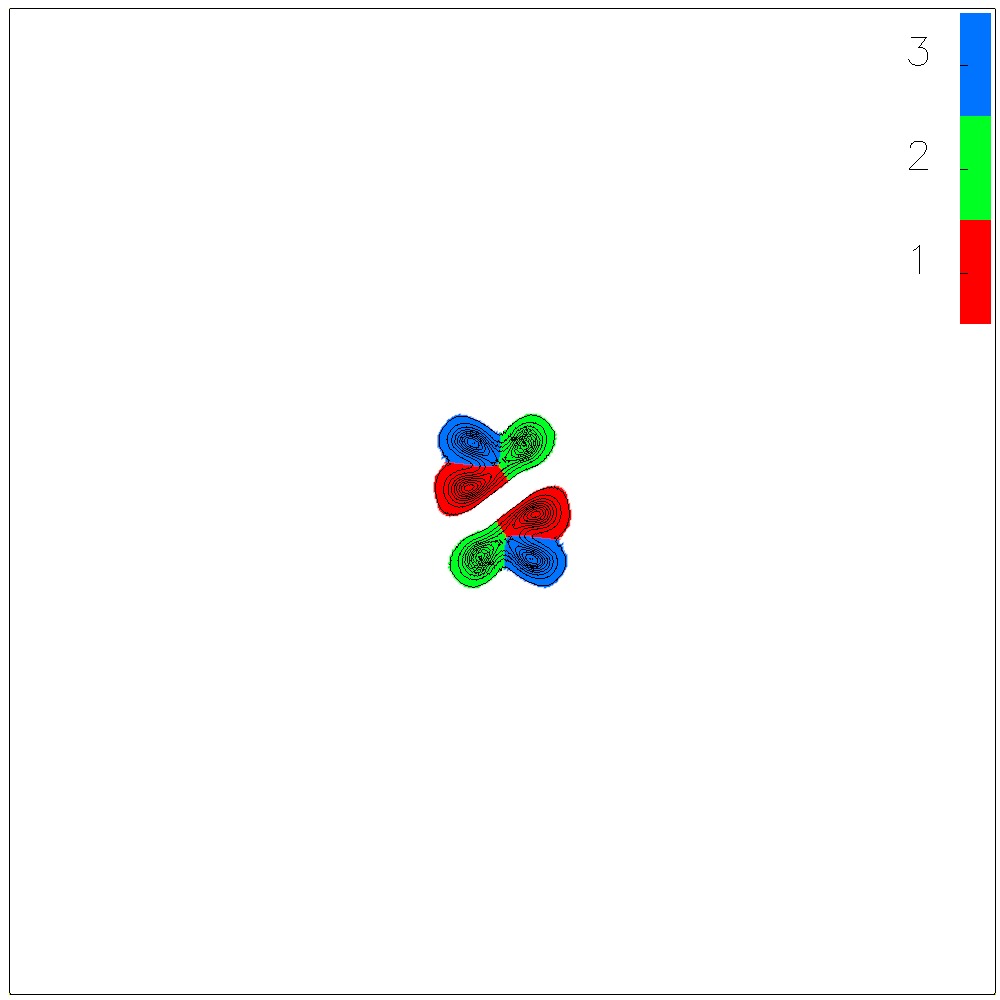} & \includegraphics[scale=0.4,natwidth=1000,natheight=1000]{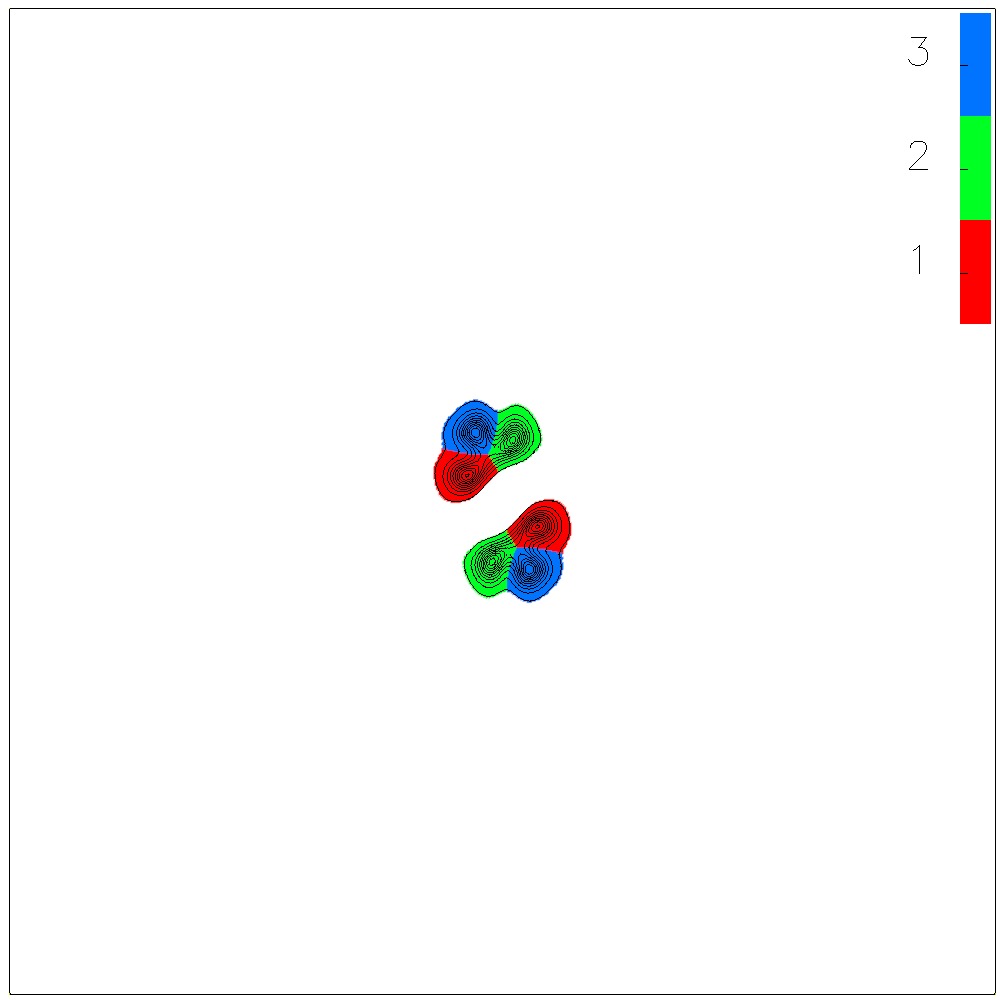}\\
$t = 22$ & $t = 23$ & $t = 27$ & $t = 29$
\end{tabular}}
\caption{Energy density plots at various times during the scattering of two $N=3$ single solitons each with speed 0.4 and with relative spatial rotation of $\pi$. The solitons' edges however, are not aligned.}
\label{anglescattering}
\end{figure}

In this section we present a few additional scatterings that demonstrate that the simple rules outlined in the scattering section apply to more complicated systems. Figure \ref{anglescattering} demonstrates how the broken potential introduces additional terms making edges wanting to come together. Hence the solitons rotate into the maximally attractive channel before they scatter, then continuing to rotate after the scattering while preserving the symmetry of the system.

\begin{figure}[h]
\centerline{\begin{tabular}{c c c c}
\includegraphics[scale=0.4,natwidth=1000,natheight=1000]{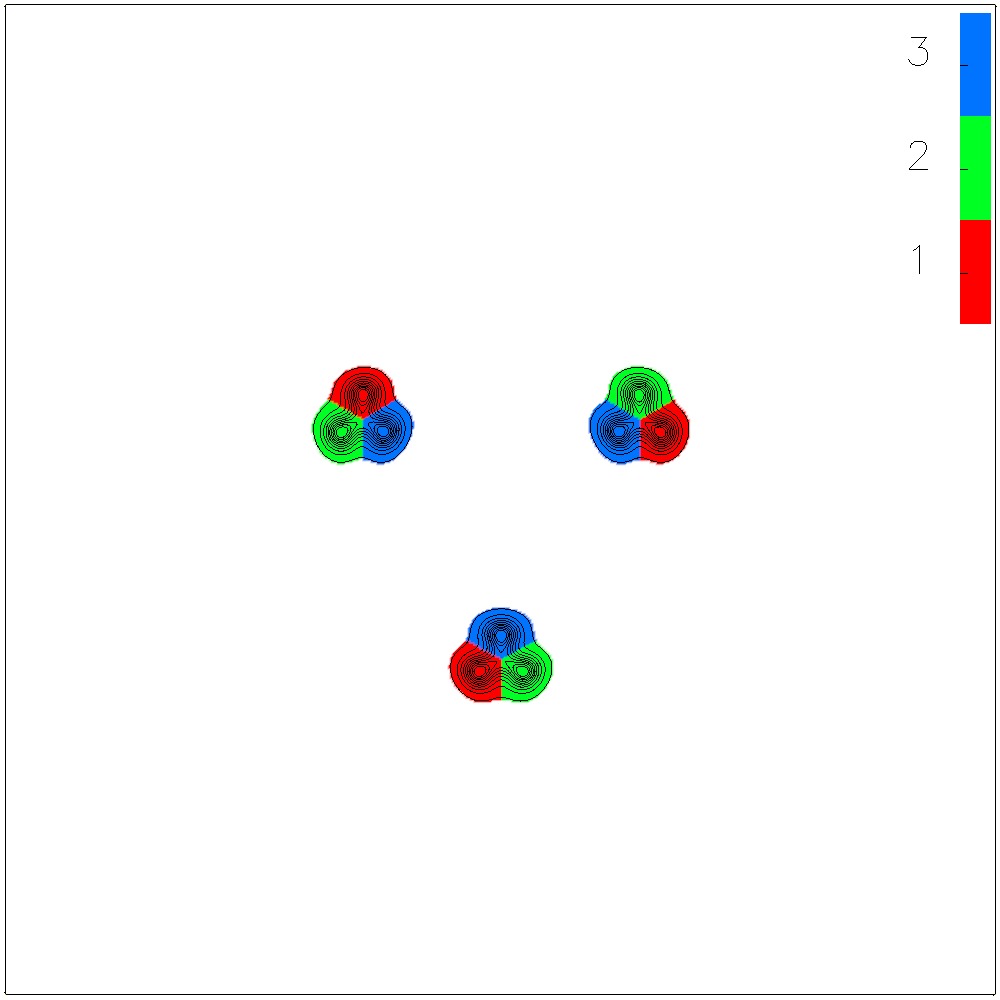} & \includegraphics[scale=0.4,natwidth=1000,natheight=1000]{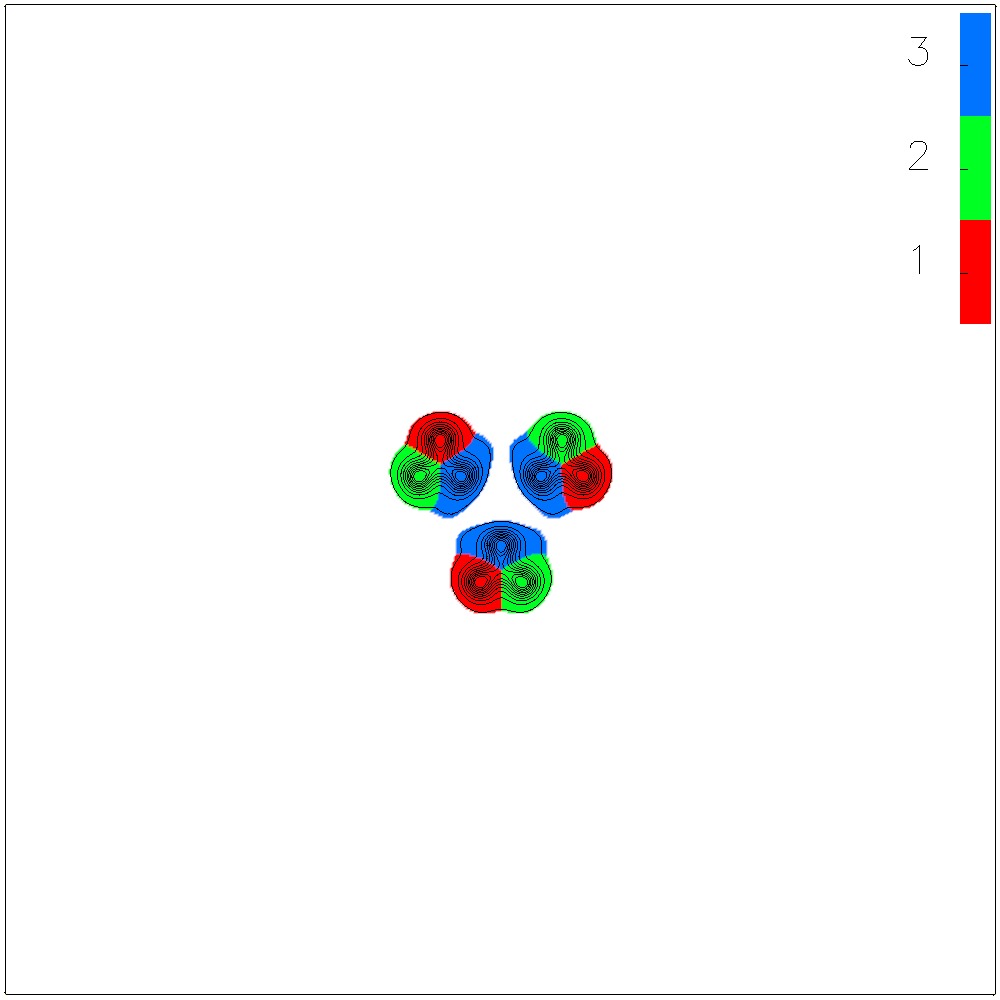} & \includegraphics[scale=0.4,natwidth=1000,natheight=1000]{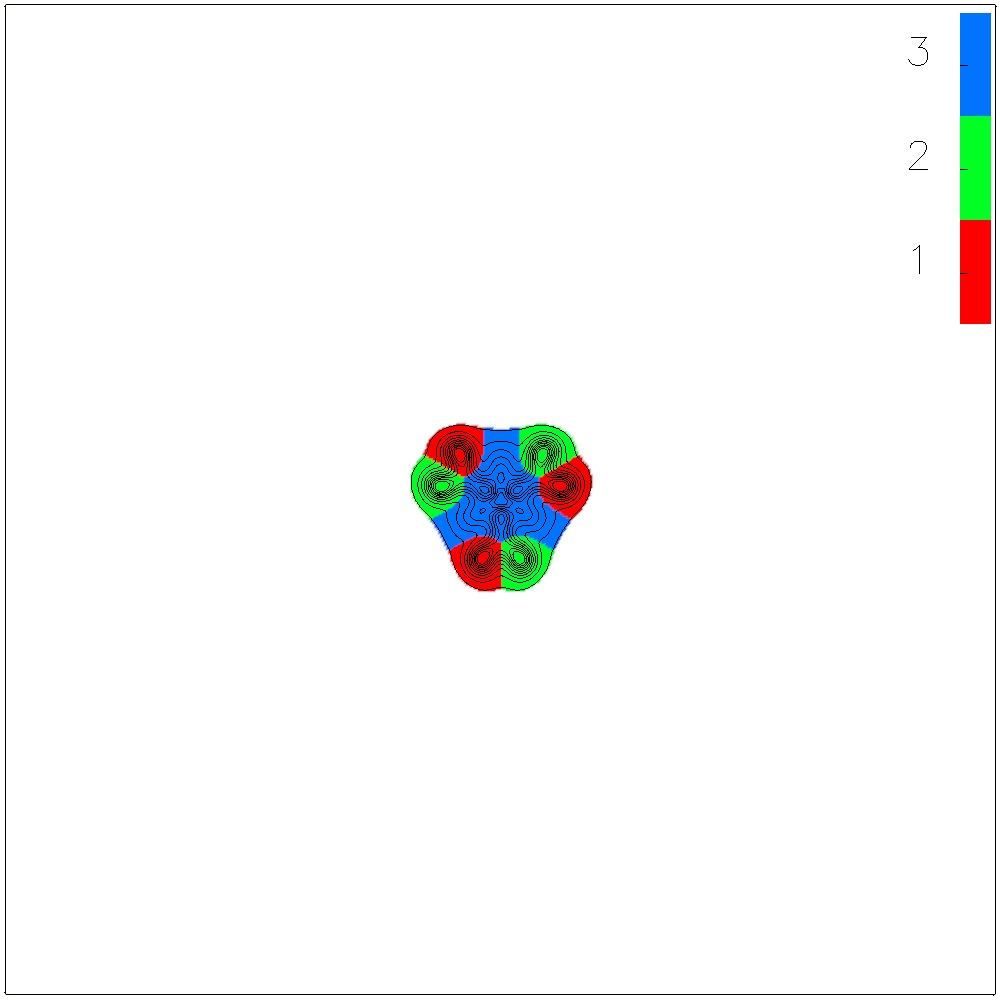} & \includegraphics[scale=0.4,natwidth=1000,natheight=1000]{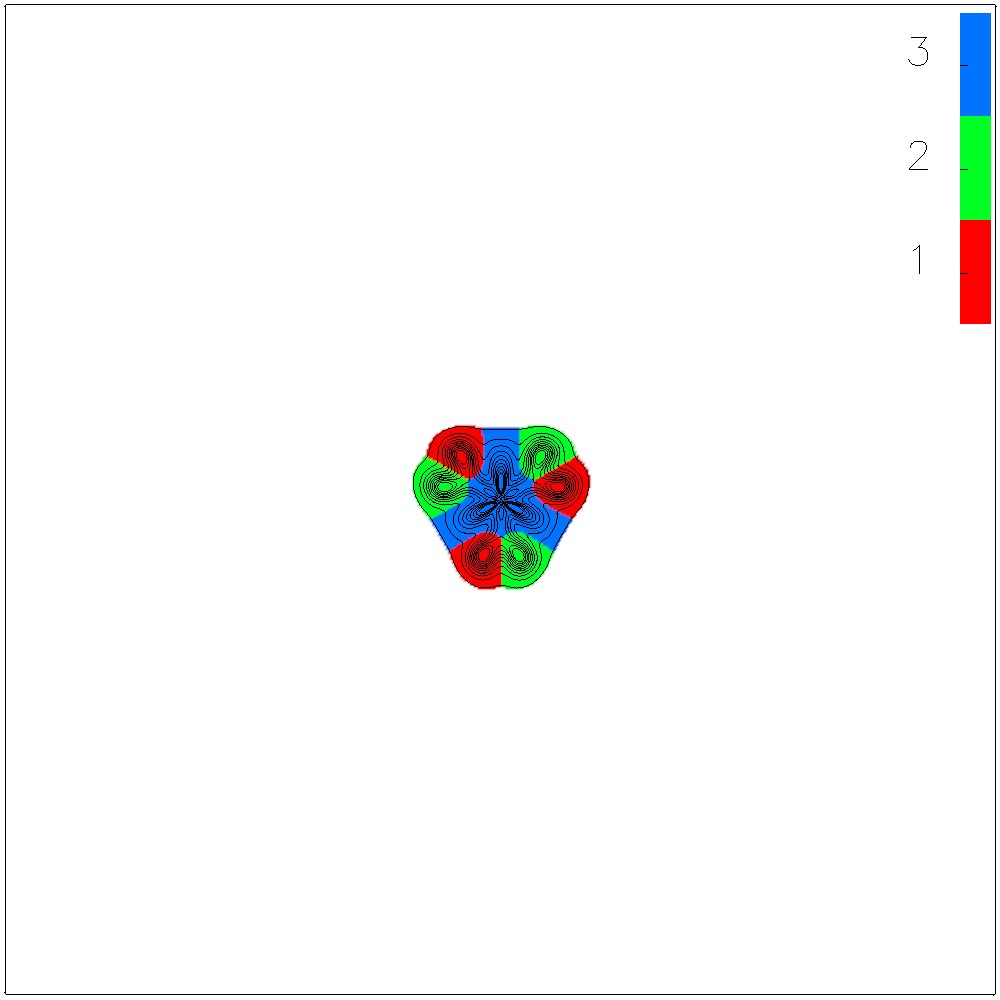}\\
$t = 36$ & $t = 72$ & $t = 84$ & $t = 86.4$\\
\includegraphics[scale=0.4,natwidth=1000,natheight=1000]{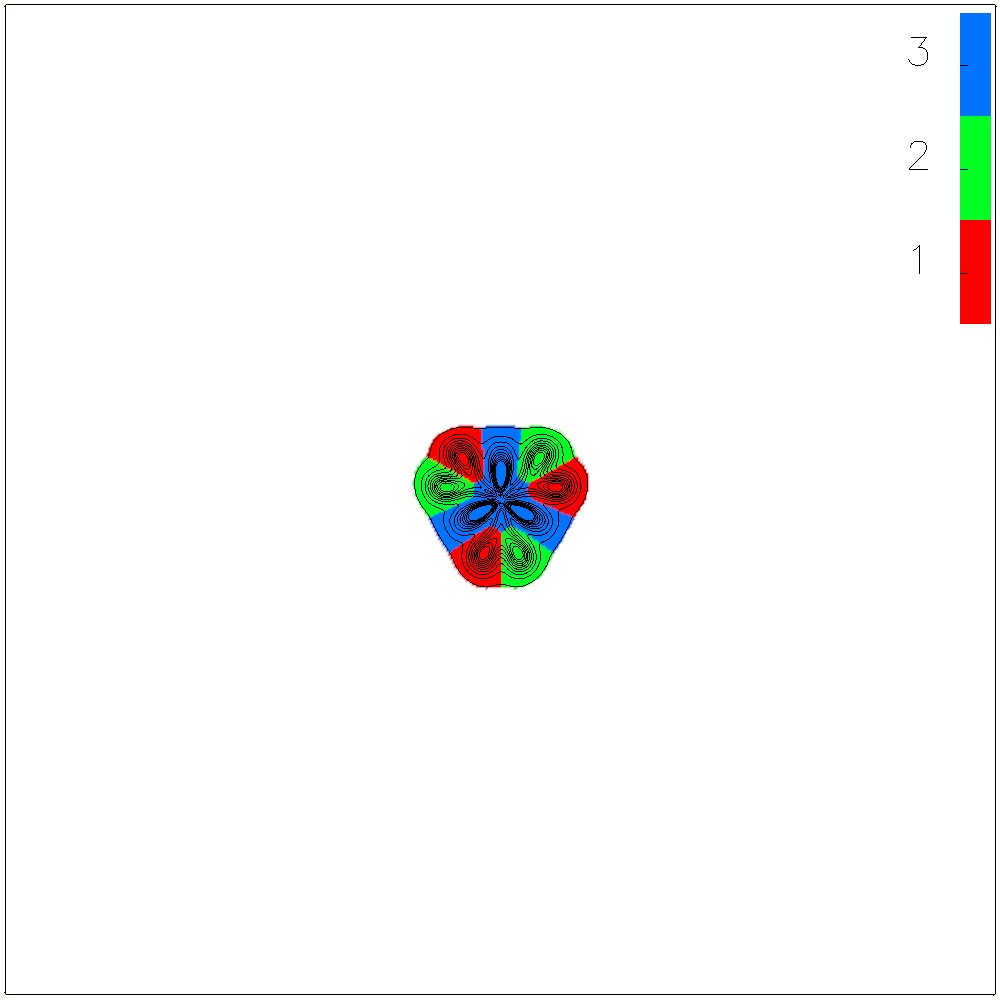} & \includegraphics[scale=0.4,natwidth=1000,natheight=1000]{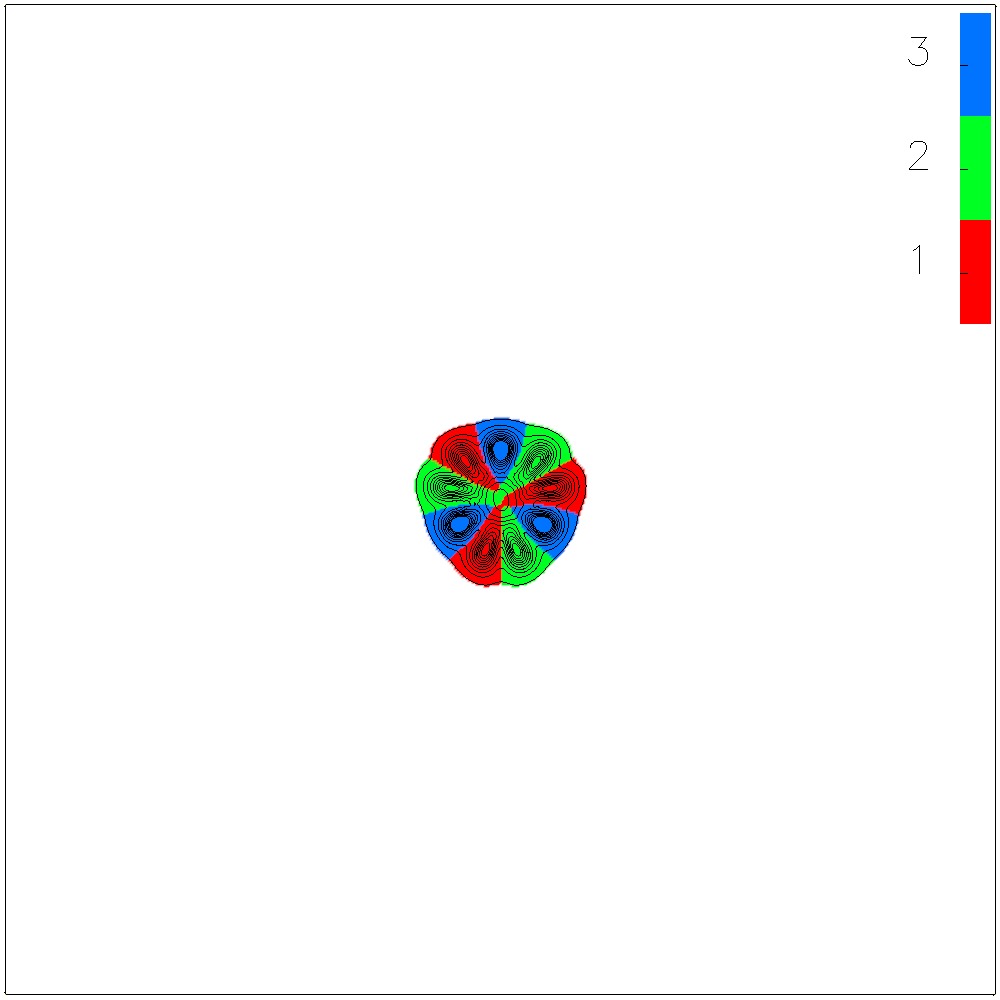} & \includegraphics[scale=0.4,natwidth=1000,natheight=1000]{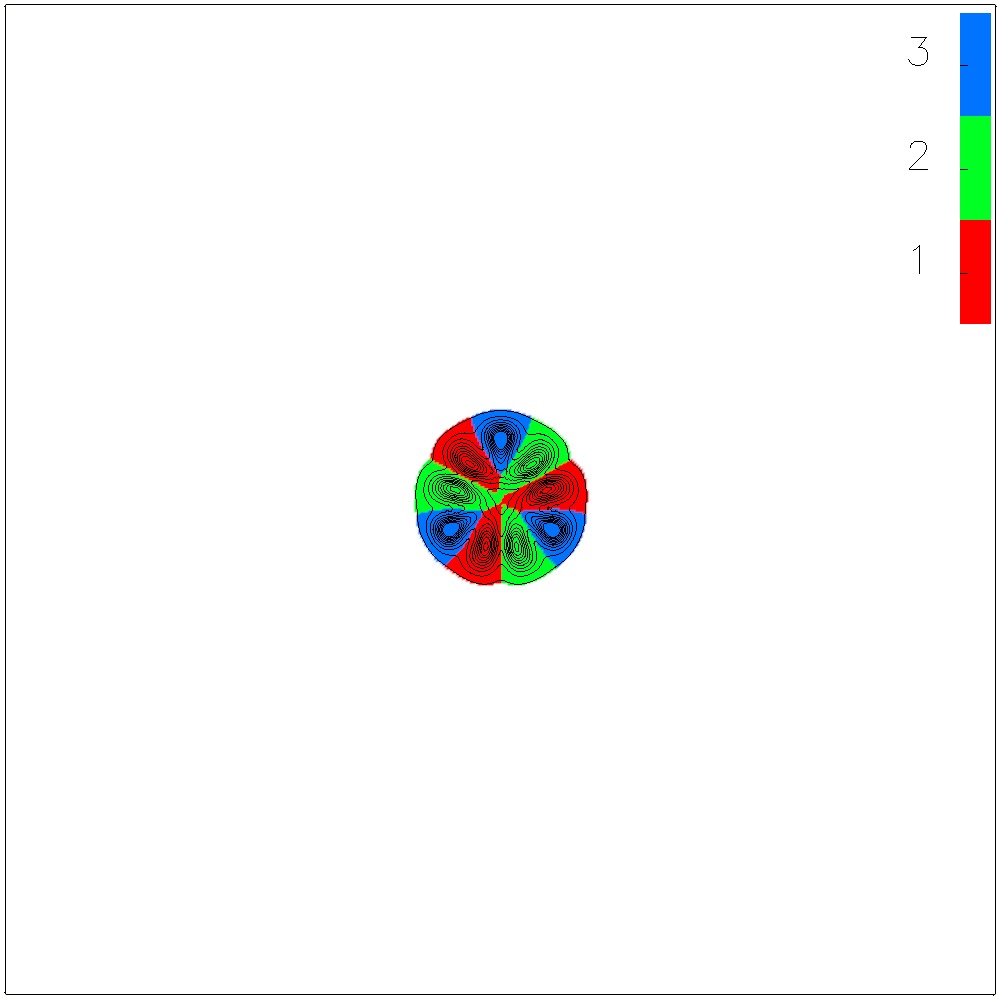} & \includegraphics[scale=0.4,natwidth=1000,natheight=1000]{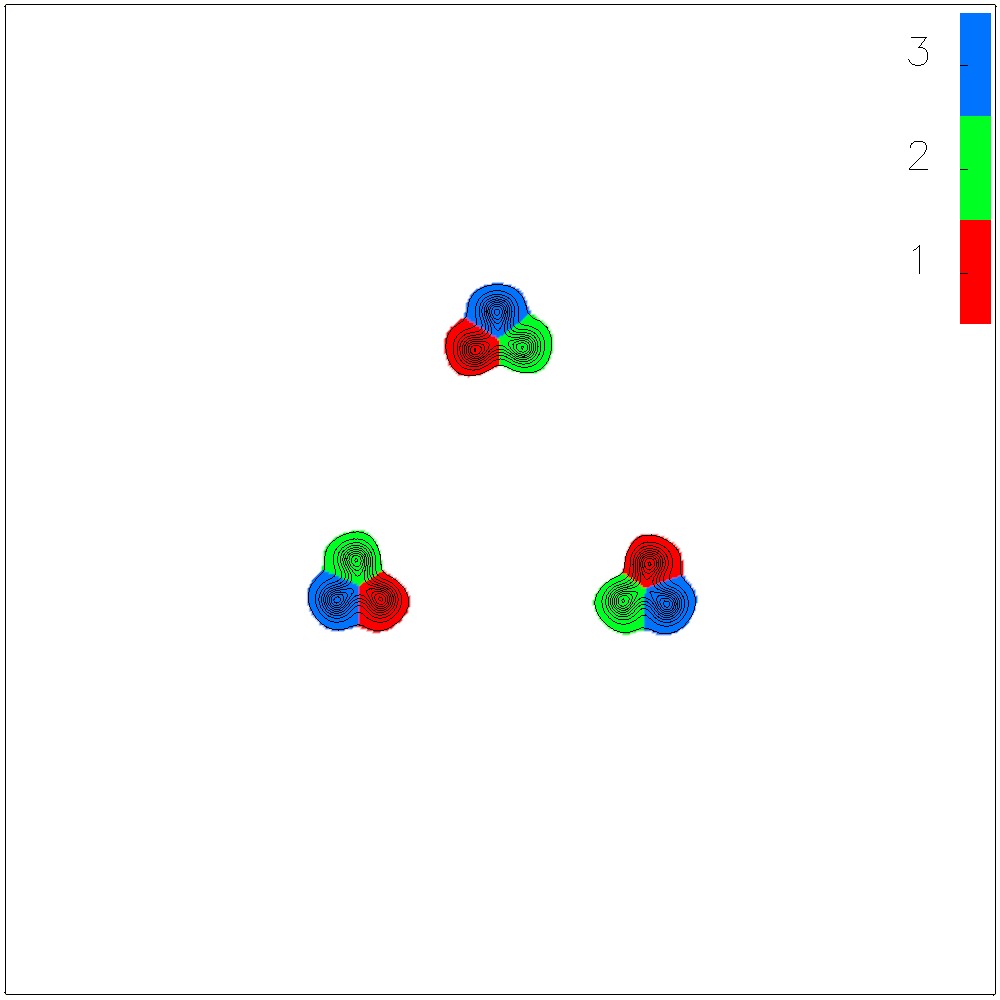}\\
$t = 88.8$ & $t = 93.6$ & $t = 96$ & $t = 168$
\end{tabular}}
\caption{Energy density plots at various times during the scattering of three $N=3$ single solitons each with speed 0.3 and with relative spatial rotation of $\frac{2\pi}{3}$.}
\label{3scattering}
\end{figure}

In figure \ref{3scattering} we see an example of scattering for a higher value of $B$, specifically $B=3$. This demonstrates that the standard rules still apply for a higher number of solitons. The like colours scattering in the centre, bisecting the angles on which they approached. With colours linked to neighbouring segments in the target space then bonding together. 

\end{document}